\documentclass[aps,prb,reprint,superscriptaddress,showpacs,amsmath,amssymb,floatfix,tighten]{revtex4-1}

\usepackage{graphicx}
\usepackage{bm}
\usepackage[table]{xcolor}
\usepackage{multirow}
\usepackage{mathptmx}
\usepackage[breaklinks=true,colorlinks=true,urlcolor=blue,
citecolor=blue,linkcolor=blue,bookmarks=false]{hyperref}

\newcommand{\ltsim}{\protect\raisebox{-0.5ex}{$\:\stackrel{\textstyle <}{\sim}\:$}}

\newcommand{\eV}{\,\,\rm eV} 

\begin{document}

\title{Electronic structure and correlations of vitamin B$_{\bf 12}$ studied within \\
the Haldane-Anderson impurity model}

\author{Zafer Kandemir$^{1,2}$, Selma Mayda Bacaksiz$^{1,2}$, and Nejat Bulut}


\affiliation{Department of Physics, Izmir Institute of Technology, 
Urla 35430, Turkey\\
$^2$ Department of Materials Science and Engineering, 
Izmir Institute of Technology, Urla 35430, Turkey}

\date{August 26, 2015}

\begin{abstract}
We study the electronic structure and correlations of vitamin B$_{12}$ (cyanocobalamine)
by using the framework of the multi-orbital single-impurity 
Haldane-Anderson model of a transition-metal impurity in a semiconductor host. 
The parameters of the effective Haldane-Anderson model are obtained 
within the Hartree-Fock (HF) approximation. 
The quantum Monte Carlo (QMC) technique is then used to calculate
the one-electron and magnetic correlation functions of this effective model. 
We observe that new states form inside the semiconductor gap found by HF
due to the intra-orbital Coulomb interaction at the impurity $3d$ orbitals.
In particular, 
the lowest unoccupied states correspond to an 
impurity bound state,
which consists of states from mainly the CN axial ligand 
and the corring ring as well as the Co $e_g$-like orbitals.
We also observe that the Co($3d$) orbitals can develop 
antiferromagnetic correlations with the surrounding atoms
depending on the filling of the impurity bound states.
In addition, we make comparisons of the HF+QMC data with 
the density functional theory calculations. 
We also discuss the photoabsorption spectrum of cyanocobalamine. 

\end{abstract}


\maketitle

\section{Introduction}
 
\subsection{Metalloproteins and metalloenzymes} 
 
Proteins are important building blocks of organisms.
About one third of proteins have a metal atom such as iron,
cobalt or zinc, and also about half of enzymes contain a metal atom 
\cite{Pratt,Krautler,Banerjee,Brown}.
The metal atom in these organometallic molecules is usually coordinated 
with nitrogen, oxygen or sulphur atoms. 
This kind of proteins and enzymes are called as metalloproteins and
metalloenzymes. 
An example for metalloproteins is hemoglobin which contains 
iron, and is involved in the transportation of oxygen and 
carbon dioxide in blood. 
An example for metalloenzymes is vitamin B$_{12}$, which
contains a cobalt atom and plays an important role in 
the production of red blood cells and in the 
functioning of the nervous system and the brain. 

Metalloenzymes and metalloproteins 
form an unusual class of organometallic molecules;
they enable various chemical reactions to take place
in the biological environment of the organisms
instead of requiring, for example, 
higher values for pressure or temperature.
The mechanisms for the 
functioning of metalloenzymes and metalloproteins 
have been studied for more than half a century. 
However, 
much still remains to be understood about 
these organometallic molecules.
In particular,
we think that it is important to know 
whether their unique chemical functions have a common
underlying electronic mechanism.

As an example for metalloenzymes, 
in this paper we study vitamin B$_{12}$, 
which is also called cobalamin (Cbl).
We have chosen to study Cbl 
because it contains relatively few atoms and 
is suitable for numerical calculations 
in addition to its biological importance.
The molecular structure of Cbl was determined by X-ray
measurements \cite{Hodgkin},
and consists of a ligand and a nucleotide attached 
to a corrin ring as shown in Fig. 1(a).
In the corrin ring, four nitrogen atoms 
are attached to the cobalt 
atom which is located at the center. 
Here, we concentrate on the cofactor cyanocobalamin 
(CNCbl) with the chemical formula 
C$_{63}$H$_{88}$CoN$_{14}$O$_{14}$P,
in which a CN molecule is attached to Co as the axial ligand.
The other two cofactors of Cbl are adenosylcobalamine (AdoCbl)
and methylcobalamin (MeCbl),
in which either an adenosyl or a methyl CH$_3$ group is attached to
Co instead of CN as the axial ligand. 

\subsection{Comparison with the dilute magnetic semiconductors}

The CNCbl exhibits an energy gap of about 2.2 eV
in its electronic spectrum \cite{Firth}
and contains a transition metal atom.
In this respect, 
it is similar to an entirely different class of compounds, 
the dilute magnetic semiconductors (DMS) \cite{Ohno92,Ohno96}.
The DMS materials such as (Ga,Mn)As are obtained 
by substituting transition metal impurities into 
a semiconducting host material. 
They exhibit interesting physical properties
such as
long-range ferromagnetic correlations,
which are found to exist
between the Mn magnetic moments at elevated temperatures.  
The transport measurements have shown that, 
in the dilute limit,
an impurity bound state exists near the top of the 
valence band in the semiconductor gap\cite{Jungwirth}. 
The impurity bound state,
which is a sharp resonant state in the single-particle spectrum
consisting of spectral weight from the Mn impurity and the host,
plays a crucial role in 
determining the electronic and magnetic properties
of (Ga,Mn)As \cite{Ichimura,Bulut,Tomoda}
when studied within the framework of the 
Haldane-Anderson\cite{Haldane} model. 
For example,
the long-range ferromagnetic correlations among the Mn impurities 
disappear rapidly as the impurity bound state 
becomes occupied with electrons. 

\begin{figure}[t]
\includegraphics[width=6.5cm]{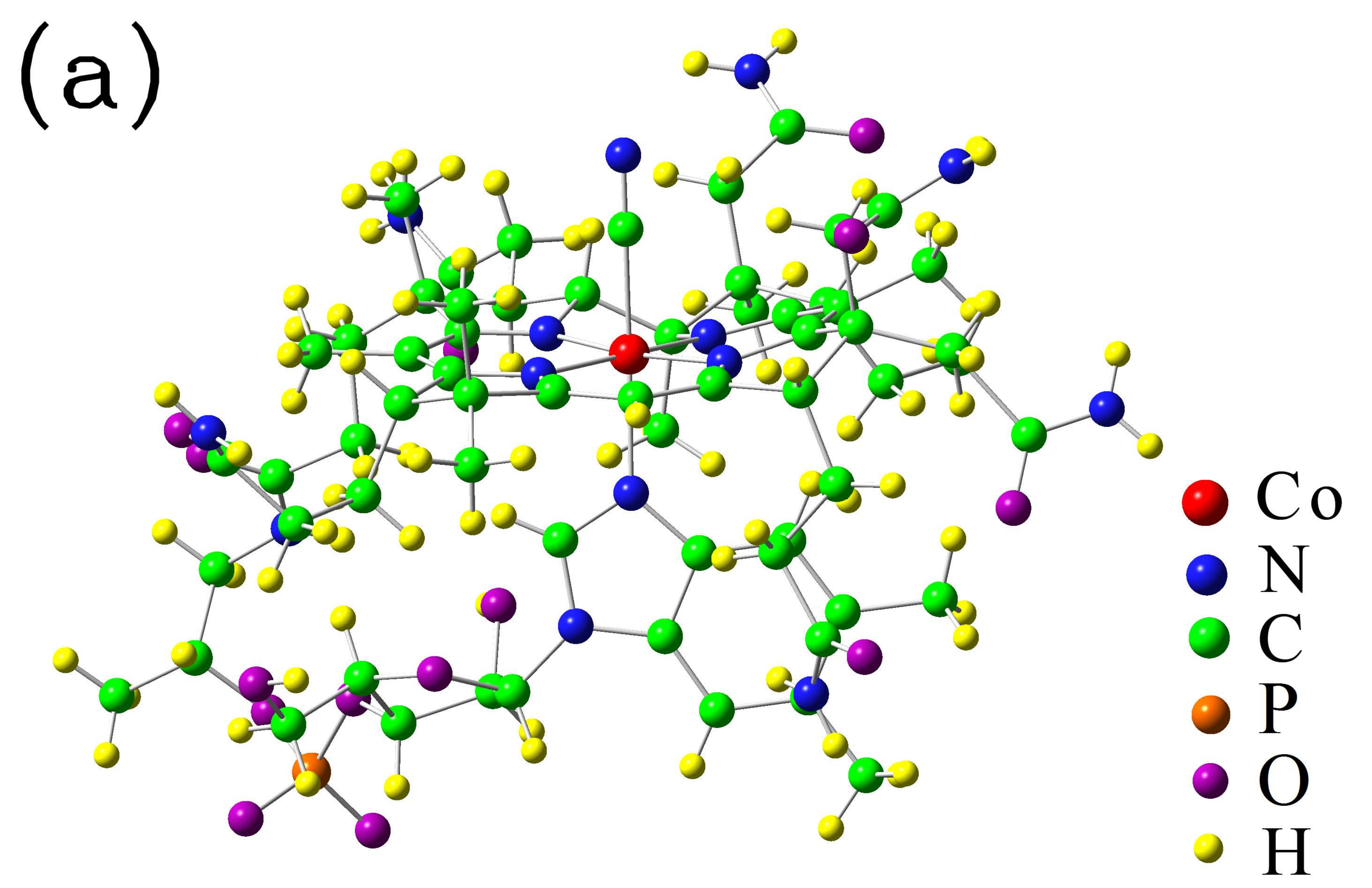}
\includegraphics[width=6.5cm]{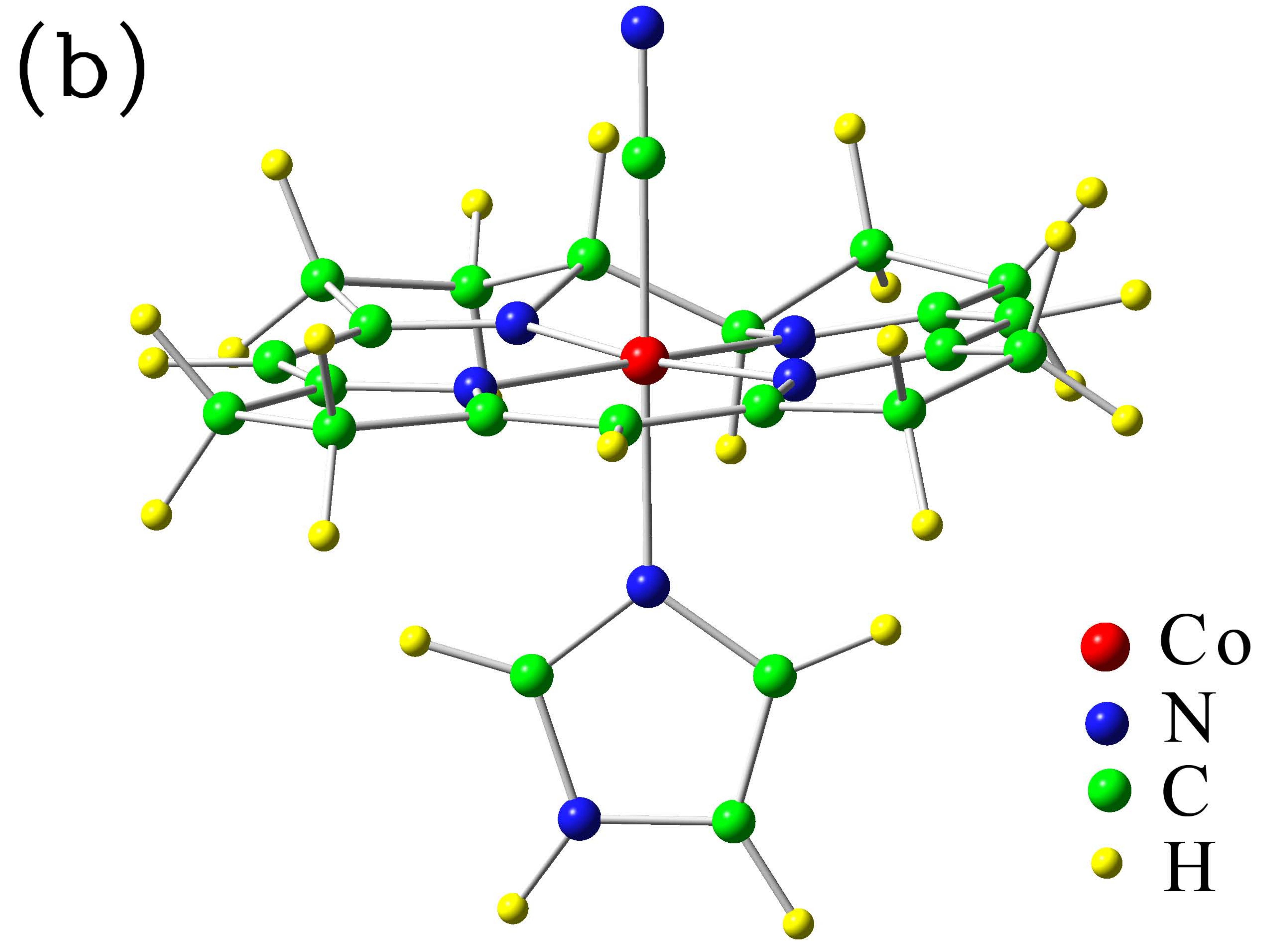}
\caption{(Color online) 
(a) Schematic plot of the molecular structure of cyanocobalamine
(C$_{63}$H$_{88}$CoN$_{14}$O$_{14}$P).
(b) Schematic plot of the simpler truncated structure for cyanocobalamine,
Im-[Co$^{\rm III}$(corrin)]-CN$^+$,
for which we perform numerical calculations. 
} 
\label{fig1}
\end{figure}

\subsection{Motivation}

The comparison with the DMS materials leads naturally to
the following questions:
Do similar many-body effects play a role in 
determining the electronic and magnetic properties of CNCbl
as well as the functioning of metalloenzymes 
and metalloproteins in general?
What is the special role of the transition metal atom
in the organometallic molecules?
In metalloenzymes and metalloproteins, 
are there electronic states which are analogous to the 
impurity bound state found in the DMS materials?
If so, 
what role to they play in the catalytic functioning of 
these molecules?
The research presented in this paper is motivated 
mainly by these questions.

\subsection{Haldane-Anderson model}

As for the DMS materials, 
we study the electronic structure and correlations 
of CNCbl by using the Haldane-Anderson model of a transition
metal impurity in a semiconducting host.
For this purpose, 
we first map the electronic structure of CNCbl onto the 
multi-orbital single-impurity Haldane-Anderson model 
by making use of the Hartree-Fock (HF) approximation.
Then we study this effective Anderson Hamiltonian \cite{Anderson} by
using QMC calculations with the Hirsch-Fye algorithm \cite{Hirsch}.
Hence, 
we study CNCbl by combining the HF and the QMC techniques,
which is called as HF+QMC.
There have been many previous studies of the electronic state of Cbl 
and other metalloproteins by using the density functional theory
\cite{Jensen2001,Andruinov,Ouyang2003,Kurmaev,Ouyang2004,Rovira,Mebs,Solheim,
Kornobis2011,Reiq,Kornobis2013,Chen}.
Here,
we study Cbl from the perspective of many-body physics 
within the framework of the Haldane-Anderson impurity model  
by using the HF+QMC approach.

The multi-orbital single-impurity Anderson Hamiltonian \cite{Anderson} 
is given by 
\begin{eqnarray}
H &=& \sum_{m,\sigma} (\varepsilon_m-\mu) c^{\dagger}_{m\sigma} c_{m\sigma} +
\sum_{\nu,\sigma} (\varepsilon_{d\nu}-\mu) 
d^{\dagger}_{\nu\sigma} d_{\nu\sigma}  \nonumber \\
&+& \sum_{m,\nu,\sigma} ( V_{m\nu} c^{\dagger}_{m\sigma} d_{\nu\sigma} + 
V^*_{m\nu} d^{\dagger}_{\nu\sigma} c_{m\sigma} ) \\
&+& \sum_{\nu} U_{\nu} 
n_{\nu\uparrow} n_{\nu\downarrow} \nonumber
\label{hamiltonian}
\end{eqnarray}
where $c^{\dagger}_{m\sigma}$ ($c_{m\sigma}$) creates (annihilates) an electron 
in host state $m$ with spin $\sigma$, 
$d^{\dagger}_{\nu}$ ($d_{\nu}$) is the creation
(annihilation) operator for a localized electron at 
the Co($3d_{\nu}$) orbital, and 
$n_{\nu\sigma}= d^{\dagger}_{\nu\sigma} d_{\nu\sigma}$.
Here, $\varepsilon_m$ and $\varepsilon_{d\nu}$ are the 
energies of the host and the $3d_{\nu}$ impurity states, respectively,
$V_{m\nu}$ is the hybridization matrix element 
between these states and $U_{\nu}$ 
is the intra-orbital Coulomb repulsion.  
We have introduced a chemical potential $\mu$ since the QMC 
calculations are performed in the grand canonical ensemble. 

This form of the Hamiltonian does not include the inter-orbital
Coulomb interaction along with the Hund's coupling. 
In cobaltates,
the Hund's coupling plays an important role \cite{Maekawa},
and we expect it to be important in Cbl also. 
In the present HF+QMC calculations the inter-orbital Coulomb 
correlations are taken into account only at the HF level,
and not in the QMC calculations.

We note that the charge and spin states of Fe in
hemoglobin were studied previously by using mean-field theory
within the extended Haldane-Anderson model
which contains the inter-orbital Coulomb interaction 
and the Hund's coupling \cite{Yamauchi}.
In addition,
the high-spin to low-spin transition due to oxygen
binding in myglobin 
was studied by using mean-field theory
for the extended Haldane-Anderson model \cite{Badaut}. 

\subsection{Truncated molecular structure for  cyanocobalamine}

We perform these calculations by using the simplified molecular structure
Im-[Co$^{\rm III}$(corrin)]-CN$^+$ shown in Fig.~1(b),
as has been done in previous DFT calculations \cite{Kornobis2011}.
Here, 
it is clearly seen that Co is coordinated by 
5 N and 1 C atoms,
which forms an approximately octahedral configuration.
In order to obtain this simplified structure,
the nucleotide group attached to the corrin ring is truncated 
and the side chains of the corrin ring are replaced by hydrogen atoms. 
The resulting
Im-[Co$^{\rm III}$(corrin)]-CN$^+$
has 56 atoms, instead of 181 in CNCbl,
and contains 238 electrons instead of 718.
Our use of this truncated structure for modelling CNCbl 
decreases the number of the host eigenstates
and the computational cost of the QMC calculations, 
because in the HF+QMC approach it is necessary 
to calculate all of the host Green's functions 
in order to determine the chemical potential.

\subsection{Outline of the paper}

The outline of this paper is as follows. 
In Section II, 
in order to obtain the one-electron parameters of 
the Anderson Hamiltonian, 
we perform Hartree-Fock calculations with the Gaussian program \cite{Gaussian}.
In particular, 
we first write the HF solution for the Fock matrix 
in the orthogonal basis set of the natural atomic orbitals (NAO) 
\cite{Weinhold,NBO}.
From this, 
we obtain the one-electron parameters
$\varepsilon_m$, $\varepsilon_{d\nu}$, and $V_{m\nu}$.
The HF calculation finds an energy gap of 8.5 eV 
for the host electronic states.
In addition, 
the hybridization matrix elements are
largest between the Co $e_g$-like states
and the atoms around Co, in particular, 
the CN axial ligand.
For the intra-orbital Coulomb repulsion $U_{\nu}$,
we use the value obtained from the two-electron integrals. 
However, we also perform calculations for different values
$U_{\nu}$
in order to see the dependence of the results. 
In the HF+QMC approach, 
the intra-orbital Coulomb repulsion is taken into account 
by both the HF and the QMC parts of the calculations.
In order to eliminate this double-counting, 
the bare Co($3d_{\nu}$) energy levels $\varepsilon_{d\nu}$ 
are shifted by a constant amount. 

In Section III,
we present QMC data on the single-particle and
magnetic correlation functions for 
the effective Anderson Hamiltonian. 
The Hartree-Fock calculations find that the host 
eigenstates have a semiconducting gap
in the energy interval 
$-10.2 \eV \leq \varepsilon \leq -1.7 \eV$.
The HF+QMC results presented in Section III
show that new single-particle states are induced 
in the intervals $-10 \eV \ltsim \varepsilon \ltsim -8.5 \eV$
and $-5.5 \eV \ltsim \varepsilon \ltsim -2 \eV$. 
We find that the states in the interval 
$-10 \eV \ltsim \varepsilon \ltsim -8.5 \eV$
arise mainly from the doubly-occupied Co $t_{2g}$-like 
states, 
while containing smaller amount of states from the corrin ring.
On the other hand, 
based on the QMC data on the single-particle and magnetic 
correlation functions,
we conclude that the states in the interval 
$-5.5 \eV \ltsim \varepsilon \ltsim -2 \eV$
represent impurity bound states arising from the 
strong hybridization of the Co $e_g$-like states with 
the host eigenstates from the valence band. 
These impurity bound states have sufficient spectral weight to 
accommodate 2 electrons including spin.
Interestingly, 
only 0.4 of these electrons originate from the Co $e_g$-like
orbitals, the remaining
1.6 electrons originate from the host states which 
consist of orbitals mainly from the CN axial ligand 
and to a lesser extent from the corrin ring. 
In Section III, 
we also present QMC data on the magnetic correlation functions. 
We study the magnetic moment formation as a function of the electron filling. 
In addition,
we show that magnetic moments develop in the valence-band host eigenstates 
which are strongly hybridized with the Co $e_g$-like orbitals. 
The size of these magnetic moments depend sensitively 
on the filling of the impurity bound states
located at $\approx -5.5 \eV$ and $\approx -4 \eV$.
In addition,
the magnetic moments of these host states are coupled antiferromagnetically
to the Co $e_g$-like magnetic moments,
and these correlations 
vanish rapidly with the electron filling of the impurity bound states. 
It is because of this filling dependence that we attribute the induced states
at $\approx -5.5 \eV$ and $\approx -4 \eV$
to be arising from impurity bound states. 
It would be useful to experimentally look for signatures of the
antiferromagnetic correlations which are predicted by the HF+QMC calculations.
It would also be useful to look for experimental evidence that the states
in the interval $-5.5 \eV \ltsim \varepsilon \ltsim \approx -2 \eV$ 
indeed arise from impurity bound states in CNCbl. 

In Section IV, 
we present a discussion in which we compare the HF+QMC 
data with the DFT calculations on the same molecule.
With respect to the HF results, 
the DFT calculations find that new single-particle spectral 
weight is induced in the intervals
$-10.5 \eV \leq \varepsilon \leq -8.0 \eV$
and $-5.5 \eV \leq \varepsilon \leq -1.7 \eV$.
This is similar to the HF+QMC data.
However,
the overall distributions of the 
spectral weight obtained by the DFT and the HF+QMC calculations
are very different. 
In the DFT calculations,
the states in the interval 
$-10.5 \eV \leq \varepsilon \leq -8.0 \eV$
do not arise from the upper-Hubbard states of the 
doubly-occupied Co $t_{2g}$ orbitals. 
In addition,
the states in the interval $-5.5 \eV \leq \varepsilon \leq -1.7 \eV$
do not correspond to an impurity bound state. 
Furthermore,
the magnetic correlations are absent in the DFT calculations.
In Section IV, 
we also compare the HF+QMC data with the 
experimental data on the photoabsorption spectrum of CNCbl,
which is characterized by several distinctive peaks. 
The HF+QMC data suggest that the lowest-energy excitations are dominated 
by the electron transfer from the Co $t_{2g}$-like states to the  
impurity bound states. 
In Section IV, 
we also discuss that the combined DFT+QMC technique
may also be applied to the same problem. 
In addition,
we note that the inter-orbital Coulomb interaction 
with the Hund's coupling needs to be included 
in the QMC calculations before carrying out
quantitative comparisons with the experimental data 
on CNCbl.

Section V gives the summary and conclusions of the paper. 

\section{Construction of the effective Anderson Hamiltonian}

Various methods for the {\it ab initio} calculation of the 
parameters of the Anderson Hamiltonian had been developed
for rare-earth, actinide and $3d$ compounds \cite{Gunnarsson}.
Here, we use the Hartree-Fock approximation 
to determine the parameters of the effective Haldane-Anderson
model for the truncated molecule Im-[Co$^{\rm III}$(corrin)]-CN$^+$.

In Section II.A, 
we discuss how the self-consistent
Hartree-Fock solution for the Fock matrix is obtained.
In Section II.B, we introduce the natural atomic orbitals (NAO's) 
which form an orthogonal basis\cite{Weinhold}. 
In Section II.C, 
we use the Fock matrix written in the NAO basis to determine the bare
energy levels and the hybridization parameters of the effective 
Haldane-Anderson model. 
In Section II.D, we present results on the density of states
obtained from the Hartree-Fock approximation.
By mapping to the effective Haldane-Anderson model, 
the energies of the host states
$\varepsilon_m$ and the impurity levels $\varepsilon_{d\nu}$ are obtained. 
Here, we show the density of states of the host part of the model
along with the positions of the bare Co($3d$) levels. 
The hybridization matrix elements obtained from the mapping to the 
Anderson Hamiltonian are presented in Section II.E.
In Section II.F, 
we discuss the values of the onsite Coulomb repulsion
used in the QMC simulations. 
In order to eliminate the effects of the double counting 
of the Coulomb interaction in HF+QMC,
the Co($3d_{\nu}$) energy levels are shifted by $\mu_{\rm DC}$,
which is discussed in Section II.G.
The results presented in Section II are based on Ref. [\onlinecite{Kandemir}].

At this point, 
it is necessary to discuss how the atomic coordinates 
of the molecule Im-[Co$^{\rm III}$(corrin)]-CN$^+$,
which are needed as input
parameters for the Hartree-Fock calculations,
are determined.
Here,
we use the approach taken by Kornobis {\it et al.} \cite{Kornobis2011},
which may be summarized as follows.
X-ray measurements had been used to obtain with high accuracy 
the geometrical parameters for the molecule MeCbl \cite{Randaccio}.
In order to have an initial estimate of the atomic coordinates of CNCbl,
the axial ligand CH$_3$ in MeCbl was replaced by CN.
The resulting molecular structure for CNCbl was then truncated 
obtaining the simplified structure
Im-[Co$^{\rm III}$(corrin)]-CN$^+$.
The final atomic coordinates 
were determined after performing
an optimization on the coordinates 
of the simplified structure by using the DFT technique
with the 6-31G(d) Gaussian basis set \cite{Gaussian} 
and the BP86 exchange functionals {\cite{Becke,Perdew}.
This way we obtain a reasonable set of values for 
the geometrical parameters, 
which are used as input 
in the Hartree-Fock calculations.
The atomic coordinates used in this paper for
Im-[Co$^{\rm III}$(corrin)]-CN$^+$
are the same as in Ref. [\onlinecite{Kornobis2011}].

\subsection{Hartree-Fock solution and the Fock matrix}

According to the electronic structure theory, 
the many-body wave function 
$\Psi$ 
of the molecule is given by the 
Slater determinant of the one-electron molecular
orbitals $\psi_n$. 
Within the Hartree-Fock mean-field approximation \cite{Slater},
the one-electron wave function $\psi_n({\bf r})$
is determined from the self-consistent solution of 
\begin{eqnarray}
\bigg(
- {\hbar^2 \over 2m_e}  &\nabla&_{\bf r}^2 
+ V_{\rm ion}({\bf r}) 
+ \sum_{m}  
\int d^3r' { |\psi_m({\bf r'})|^2 \over |{\bf r} - {\bf r'}| } \bigg) \psi_n({\bf r}) \\ 
&-& \sum_{m} 
\int d^3r' { \psi_m^*({\bf r'})\psi_n({\bf r'}) \over |{\bf r} - {\bf r'}|}  
\psi_m({\bf r}) = E_n \psi_n({\bf r}) \nonumber
\label{fockeqn}
\end{eqnarray} 
with $E_n$ the molecular-orbital energy.
Here, the left-hand side includes the kinetic-energy and 
the electron-ion potential $V_{\rm ion}({\bf r})$ terms
in addition to the Hartree and the Fock contributions. 

The molecular orbitals $|\psi_n\rangle$ can be expressed in terms of the 
$N$ atomic orbitals $|\phi_i\rangle$,
\begin{eqnarray}
|\psi_n\rangle = \sum_i^N C_{ni}|\phi_i\rangle,
\label{psin}
\end{eqnarray}
where $C_{ni}$ are the elements of the coefficient matrix ${\bf C}$. 
By substituting Eq. (3) into Eq. (2), the Roothan equation is obtained 
\cite{FJensen},
\begin{equation}
{\bf C \, F} = {\bf E \, C \, S}.
\end{equation}
Here, the overlap matrix ${\bf S}$ has the matrix elements 
$S_{ij} = \langle \phi_i |\phi_j \rangle$.
Since the atomic orbitals $|\phi_i\rangle$ do not form an orthogonal basis,
the overlap matrix is not diagonal. 
The matrix ${\bf E}$ is diagonal with the diagonal elements giving 
the molecular orbital energies, $E_{nm}= \delta_{nm} E_m$.
The elements of the Fock matrix ${\bf F}$ are given by 
\begin{eqnarray}
F_{ij} = \int d^3r \, \phi^*_i({\bf r}) 
\bigg( &-& {\hbar^2 \over 2m_e} \nabla_{\bf r}^2 
+ V_{\rm ion}({\bf r}) \bigg)
\phi_j({\bf r}) \nonumber \\
+
\sum_{i',j'}^N D_{i'j'}
\int &d^3r & \int d^3r' \,
{ \phi^*_i({\bf r}) \phi^*_{i'}({\bf r'})  \over  
| {\bf r} - {\bf r'} | } \\
&\times& 
\bigg( \phi_j({\bf r}) \phi_{j'}({\bf r'}) -
\phi_{j'}({\bf r}) \phi_j({\bf r'}) \bigg) \nonumber
\label{Fij}
\end{eqnarray} 
where 
\begin{equation}
D_{ij} = \sum_n^{N_{\rm el}} C_{ni} C_{nj}
\end{equation}
forms the density matrix. 
Here, $N_{\rm el}$ denotes the total number of the electrons 
and also twice the 
number of the occupied molecular orbitals.  
For  Im-[Co$^{\rm III}$(corrin)]-CN$^+$,
we have $N_{\rm el}=238$.
Using as input the one- and two-electron integrals,
the solution of the Roothan equation for $E_n$ and the density matrix 
are obtained self-consistently through iteration.
In order to obtain the HF solution for 
Im-[Co$^{\rm III}$(corrin)]-CN$^+$,
we have used 
the Gaussian program \cite{Gaussian} with the 6-31G basis set and
$N=347$ basis functions. 
We perform the HF calculations with the Gaussian 6-31G basis set 
instead of 6-31G(d), which was used in the initial optimization,
in order to increase the speed of the QMC calculations reported in Section III. 
After the HF solution has been obtained, 
we have formed the Fock matrix. 
From the Fock matrix we have obtained the one-electron parameters of the 
effective Anderson Hamiltonian 
through a procedure outlined in the following two subsections. 

\subsection{Basis of the natural atomic orbitals}

The overlap matrix ${\bf S}$ defined above
is not diagonal, since
the atomic orbitals do not form an orthogonal basis set.
A convenient orthogonal basis for expressing $|\psi_n\rangle$ 
is provided by the basis of the natural atomic orbitals 
$|\tilde{\phi}_i\rangle$ \cite{Weinhold}.
The NAO's form a maximally localized basis set; 
they are similar to the atomic orbitals but they are also orthogonal.
The NAO's are obtained by minimizing the function
\begin{equation}
\sum_i w_i \int d^3\,r \, | \tilde{\phi}_i({\bf r}) - \phi_i({\bf r}) |^2,
\end{equation}
where $w_i$ denotes the electron occupation number 
of the atomic orbital $|\phi_i\rangle$
obtained by the Hartree-Fock calculation.
Here, we will first express the Fock matrix in the NAO basis,
and from this deduce the parameters of the Anderson Hamiltonian
except for the Coulomb repulsion.

Using the basis of the NAO's, the molecular orbitals $|\psi_n\rangle$ 
can be expressed as 
\begin{equation}
|\psi_n\rangle = \sum_{\nu =1}^5 \beta_{n\nu} |\tilde{\phi}_{d\nu}\rangle + 
\sum_{i=1}^{N-5} \beta_{ni}|\tilde{\phi}_i\rangle,
\end{equation}
where, $|\tilde{\phi}_{d\nu}\rangle$'s represent 
the five NAO's for the Co($3d_{\nu}$) orbitals,
and $|\tilde{\phi}_i\rangle$'s represent the remaining NAO's. 
We have obtained the elements of the Fock matrix within the NAO basis
by using the Gaussian program \cite{Gaussian}
along with the NBO 6.0 software package \cite{NBO}.

\subsection{One-electron parameters of the Anderson Hamiltonian from the Fock matrix}

Within the HF+QMC approach, 
we determine 
the one-electron parameters of the Anderson Hamiltonian from 
the matrix elements of the Fock matrix 
expressed in the orthogonal NAO basis. 
This procedure is sketched in Fig. 2(a) and (b). 
Here, the $N\times N$ Fock matrix  ${\bf F}$
consists of the $3d$ part ${\bf F}_{3d}$, 
the host part ${\bf F}_0$, and
the hybridization parts ${\bf M}$ and its transpose ${\bf M}^T$
as illustrated in Fig. 2(a). 
The host part of the Fock matrix ${\bf F}_0$ 
is not diagonal in the NAO basis.
By diagonalizing ${\bf F}_0$, 
we obtain the host energy levels 
$\varepsilon_m$ of the effective Haldane-Anderson model,
and the corresponding host eigenstates $|u_m\rangle$.
In turn, the hybridization matrix elements $V_{m\nu}$ 
between the host eigenstates $|u_m\rangle$ and 
the Co($3d_{\nu}$) NAO's $|\tilde{\phi}_{d\nu}\rangle$ are obtained.
We denote by ${\bf F'}$ the Fock matrix expressed in the basis 
formed from 
the Co($3d_{\nu}$) NAO's and 
the orthogonal host states,
which is illustrated in Fig. 2(b). 
We determine the one-electron parameters of the Anderson Hamiltonian
from ${\bf F'}$.
The impurity energies $\varepsilon_{d\nu}$ of the effective model
are given by the diagonal matrix elements of ${\bf F}_{3d}$.
We neglect the off-diagonal elements of ${\bf F}_{3d}$,
which are denoted by $t_{\nu,\nu'}$ in Fig. 2(b),
since they are negligible in comparison to the diagonal terms. 

\begin{figure}[t]
{\includegraphics[height=6.5cm]{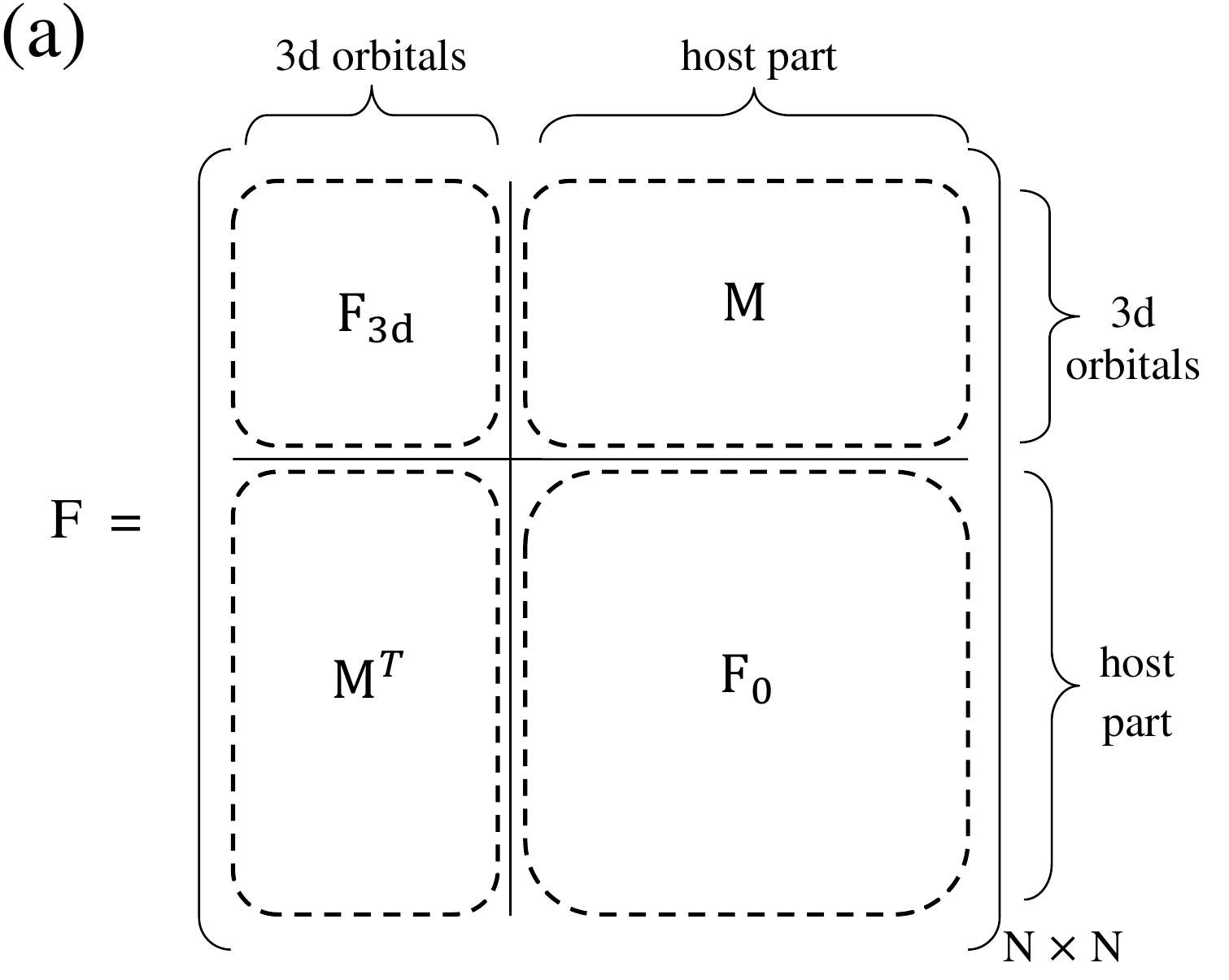}} 
{\includegraphics[height=6.5cm]{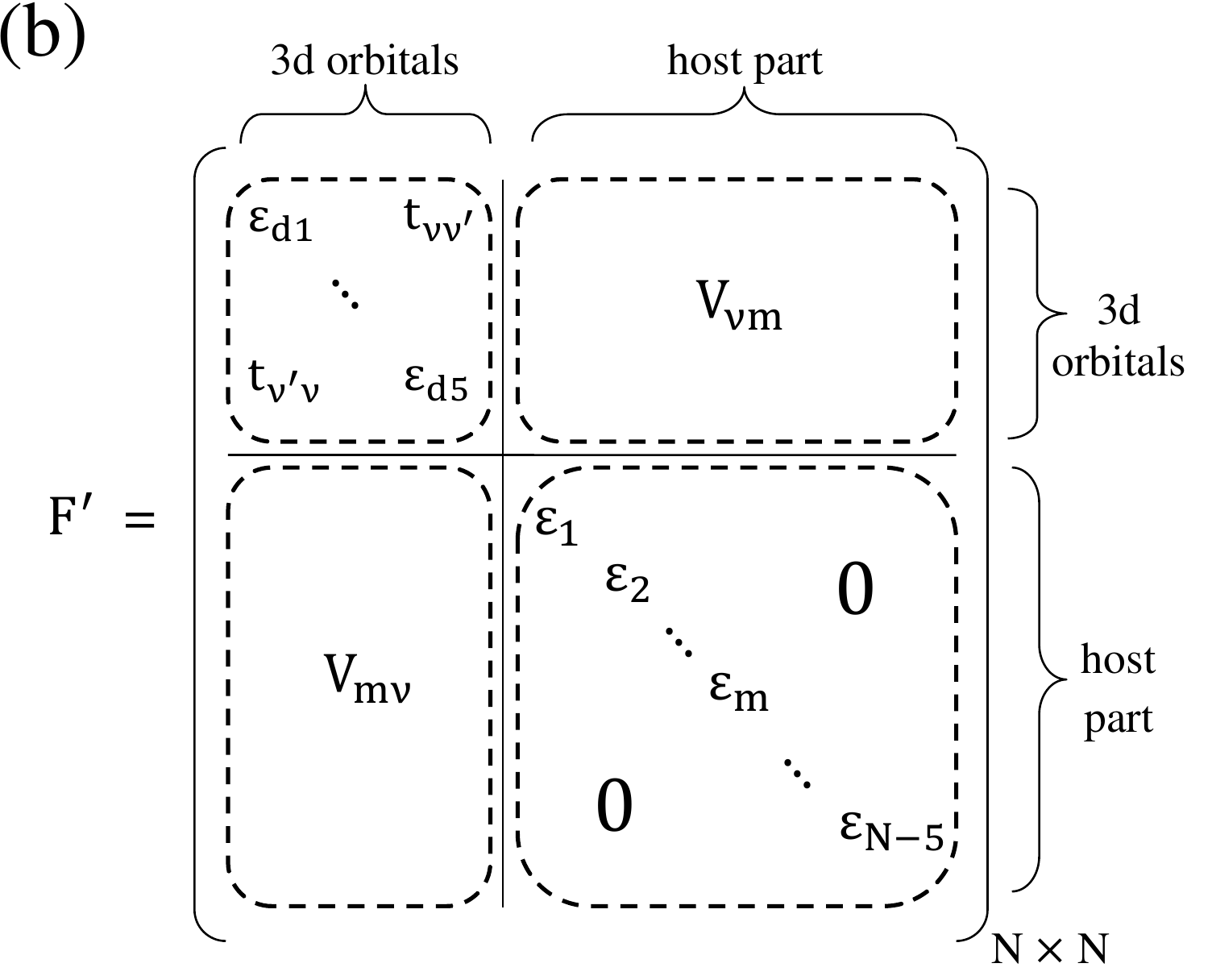}}
\caption{ 
(a) Sketch of the Fock matrix ${\bf F}$ 
in the natural atomic orbital basis.
Here, the $3d$ and the host parts of the Fock matrix are separated
as ${\bf F}_{3d}$ and ${\bf F}_0$, respectively. 
The hybridization parts are ${\bf M}$ and ${\bf M}^T$. 
The host part ${\bf F}_0$ is not diagonal in the NAO basis.
(b) Sketch of the Fock matrix ${\bf F'}$
which is obtained after diagonalizing the host part.
We obtain the one-electron energies and 
the hybridization matrix elements of the effective Haldane-Anderson model
from ${\bf F'}$.
Eigenvalues of the host part of the Fock matrix ${\bf F'}$ yields $\varepsilon_m$, 
the host energy levels of the Anderson model. 
The off-diagonal matrix elements yield the hybridization parameters $V_{m\nu}$
between the host states and the Co($3d_{\nu}$) NAO's.
Since $t_{\nu\nu'}$ are negligible, 
$\varepsilon_{d\nu}$'s 
become the energy levels of the impurity states. 
}
\label{fig2}
\end{figure}

Using the procedure outlined above, 
we obtain the model parameters
$\varepsilon_m$, $\varepsilon_{d\nu}$, and $V_{m\nu}$ 
from the transformed Fock matrix ${\bf F'}$.
Considering the effective Hamiltonian Eq.~(1),
we note that now $d_{\nu\sigma}^{\dagger}$ 
($d_{\nu\sigma}$) creates (annihilates)
a fermion in the Co($3d_{\nu}$) NAO state $|\tilde{\phi}_{d\nu}\rangle$,
while $c_{m\sigma}^{\dagger}$ ($c_{m\sigma}$) creates (annihilates)
a fermion in the orthogonal host eigenstate $|u_m\rangle$.
Hence, the $V_{m\nu}$'s represent the hybridization matrix elements
between the host eigenstates $|u_m\rangle$ and 
the Co($3d_{\nu}$) NAO's.

There have been previous studies where the electronic structure 
calculations are combined 
with the many-body techniques
\cite{Held}. 
The use of NAO's here for the case of a molecule is similar to the use of 
maximally-localized Wannier orbitals \cite{Marzari-Vanderbilt}
to obtain the model parameters for various compounds
\cite{Schnell1,Schnell2}.

\begin{figure} 
{\includegraphics[width=7cm]{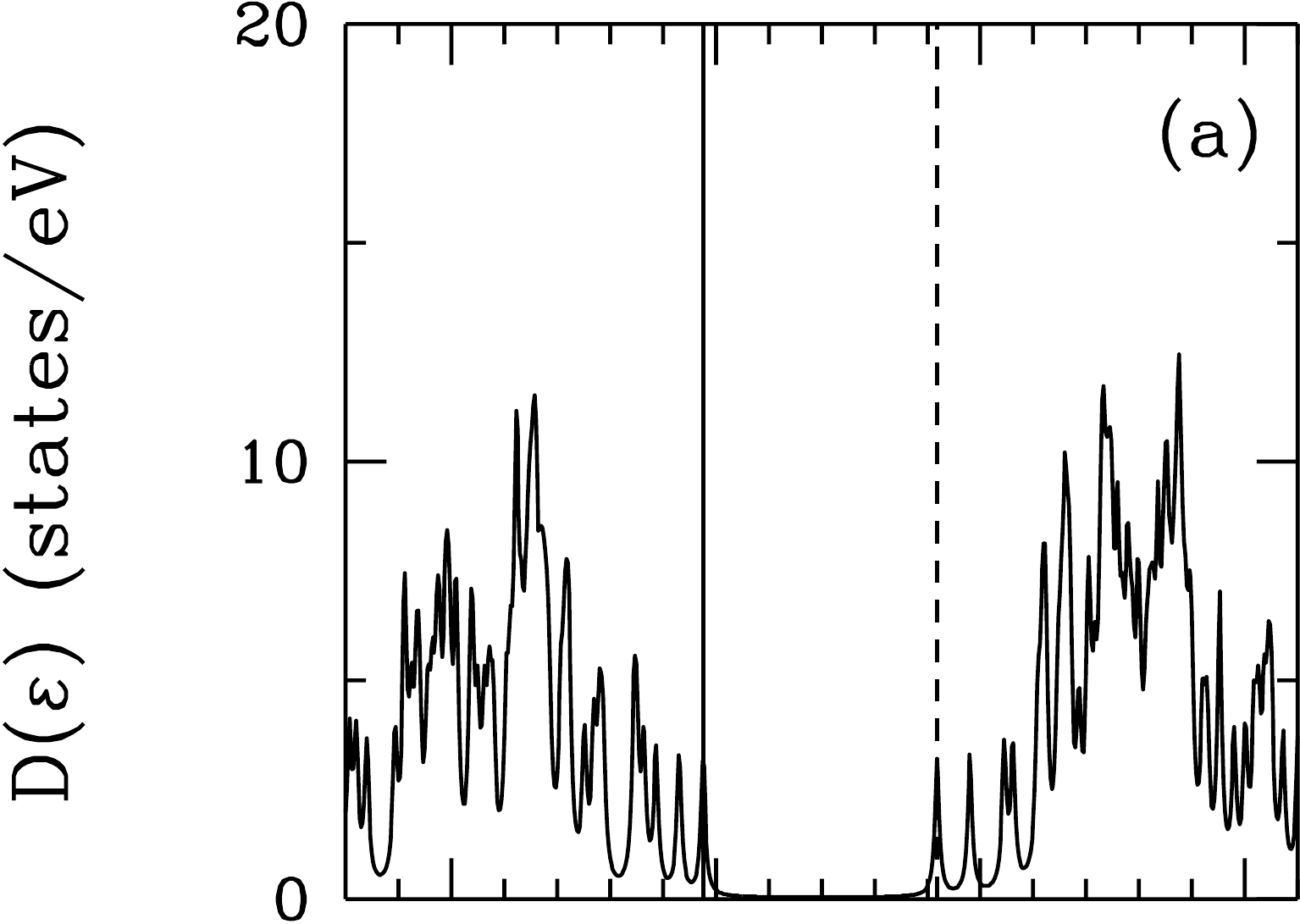}}
{\includegraphics[width=7cm]{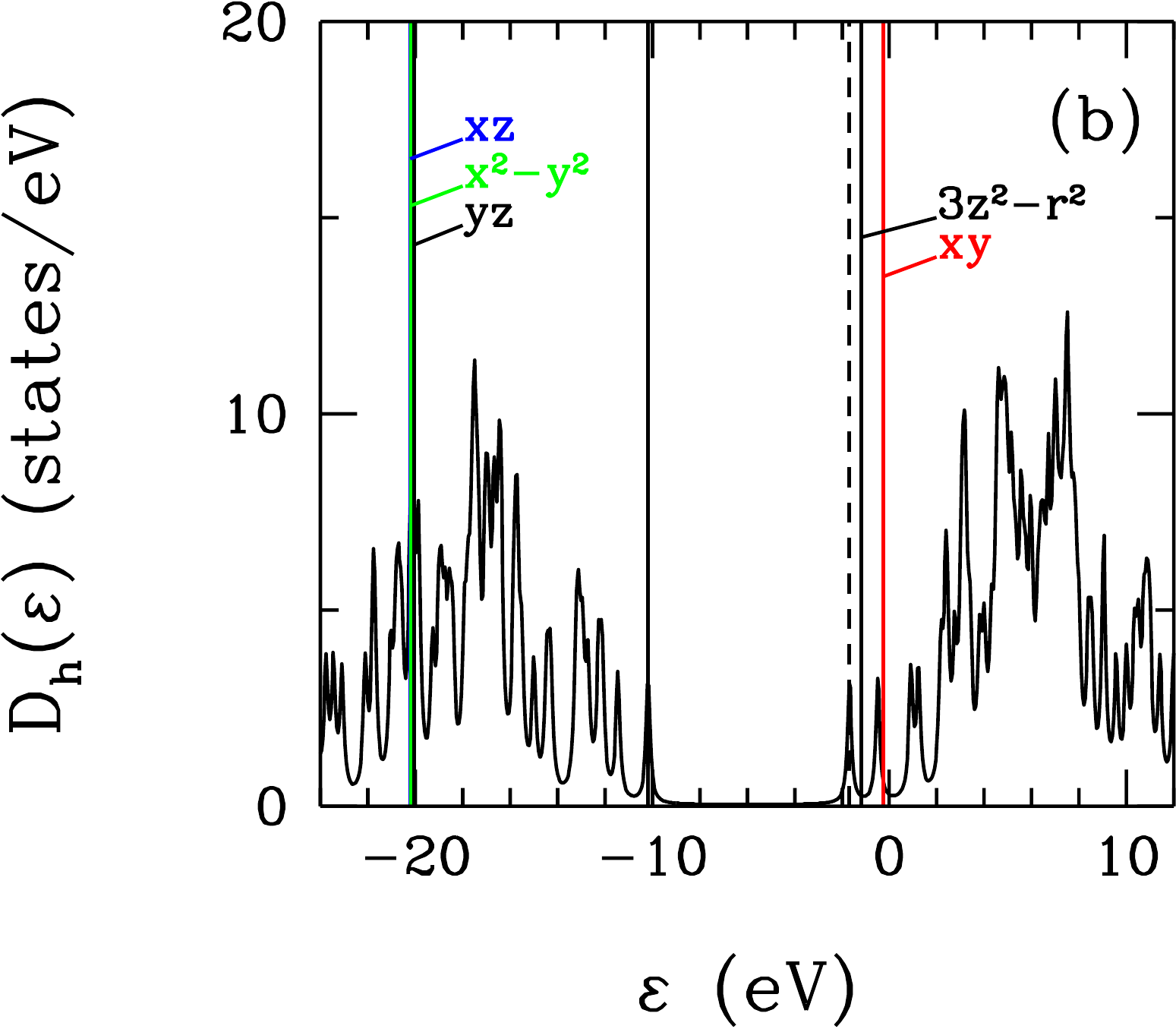}}
\caption{(Color online) 
(a) Total density of states of Im-[Co$^{\rm III}$(corrin)]-CN$^+$,
$D(\varepsilon)$,
obtained within the Hartree-Fock approximation by using the Gaussian program. 
The vertical solid and dashed lines denote the HOMO and LUMO levels, respectively.
(b) Density of states of the host eigenstates of the effective Haldane-Anderson model, 
$D_h(\varepsilon)$.
Here, 
the HOMO and LUMO levels are slightly different than in (a).
In addition,
the energy levels of the Co($3d_{\nu}$) NAO's 
are indicated by the vertical lines. 
}
\label{fig3}
\end{figure}

\subsection{Host density of states and the Co($\bm{3d_{\nu}}$) energy levels}

Figure 3(a) shows the total density of states $D(\varepsilon)$ defined by 
\begin{equation}
D(\varepsilon) = \sum_{n=1}^N \delta(\varepsilon-E_n)
\end{equation}
where $E_n$ are the eigenvalues of the Fock matrix. 
Here, the highest occupied molecular orbital (HOMO) is located at -10.2 eV,
and the lowest unoccupied molecular orbital (LUMO) is at -1.7 eV.
Hence, the occupied and the unoccupied states are separated by an energy gap 
of 8.5 eV. 
In Fig. 3(b), the density of the host eigenstates defined by 
\begin{equation}
D_h(\varepsilon) = \sum_{m=1}^{N-5} \delta(\varepsilon-\varepsilon_m),
\end{equation}
where $\varepsilon_m$ are the energy 
levels of the host eigenstates,
is shown.
We observe that in Fig. 3(b) 
the HOMO and the LUMO levels are slightly shifted
in comparison to Fig. 3(a). 
The energy levels of the Co($3d_{\nu}$) NAO states, 
$\varepsilon_{d\nu}$, 
are also shown as vertical lines in Fig. 3(b).
We observe that the $xz$, $x^2-y^2$, and $yz$ orbitals are degenerate
and are much lower in energy with respect to the $3z^2-r^2$ and $xy$ orbitals,
which are located above the energy gap and near the LUMO level.

In CNCbl, 
the 5 nitrogen atoms and the C atom surrounding Co form roughly 
a local octahedral environment for Co.
Hence, the Co($3d_{\nu}$) levels separate into two groups, 
one containing the $3z^2-r^2$ and $xy$ orbitals 
and the other containing the $xz$, $x^2-y^2$, and $yz$ orbitals.
This is similar to the separation of the $3d$ orbitals into the groups of $e_g$
and $t_{2g}$ orbitals in a crystal field with cubic symmetry. 
However, here we have taken a coordinate system such that 
the $x$ and $y$ axis are located at 45 degrees to the
Co-N bond direction in the corrin layer
instead of being parallel to the Co-N bond direction.
Hence, 
the $x^2-y^2$ and $xy$ orbitals have been exchanged.
In the following, 
because of our choice of the coordinate system
which is the same as that of the Gaussian program,
we will denote 
the $3z^2-r^2$ and $xy$ orbitals as the $e_g$-like states,
and the $xz$, $x^2-y^2$, and $yz$ orbitals 
as the $t_{2g}$-like states.

\subsection{Host-Co($\bm{3d_{\nu}}$) hybridization}

In this section, 
we present Hartree-Fock 
data on the hybridization matrix elements 
of the Haldane-Anderson model, 
which were obtained as discussed in Subsection II.C.
The square of the hybridization matrix elements $|V_{m\nu}|^2$ between 
the $m$'th host eigenstate $|u_m\rangle$ and 
the Co($3d_{\nu}$) NAO 
are shown as a function of the host energy $\varepsilon_m$ in Fig. 4.
Here, 
we observe that the magnitude of $|V_{m\nu}|^2$ 
for the cobalt $3z^2-r^2$ and $xy$ orbitals can be an order of magnitude larger 
than those for $xz$, $x^2-y^2$ and $yz$.
This is not surprising since the $3z^2-r^2$ and $xy$ orbitals
lie along the Co-N and Co-C bond directions
according to the coordinate system used here.
Figure 4(a) shows that, for the $\nu=3z^2-r^2$ orbital, 
$|V_{m\nu}|^2$ has the largest values at $m=114$ and 116. 
For $\nu=xy$, the largest values occur at $m=111$ and 112.
In Section III, we will see that the host eigenstates $m=114$, 111, 112 and 116
are the states which are most affected by the presence of the cobalt atom.

We note that $|V_{m\nu}|^2$'s have large values for a few host states,
while being negligibly small for the remainder.
This feature may be used in the future to increase the speed
of the numerical calculations without losing accuracy.

\begin{figure}
\includegraphics[width=7cm]{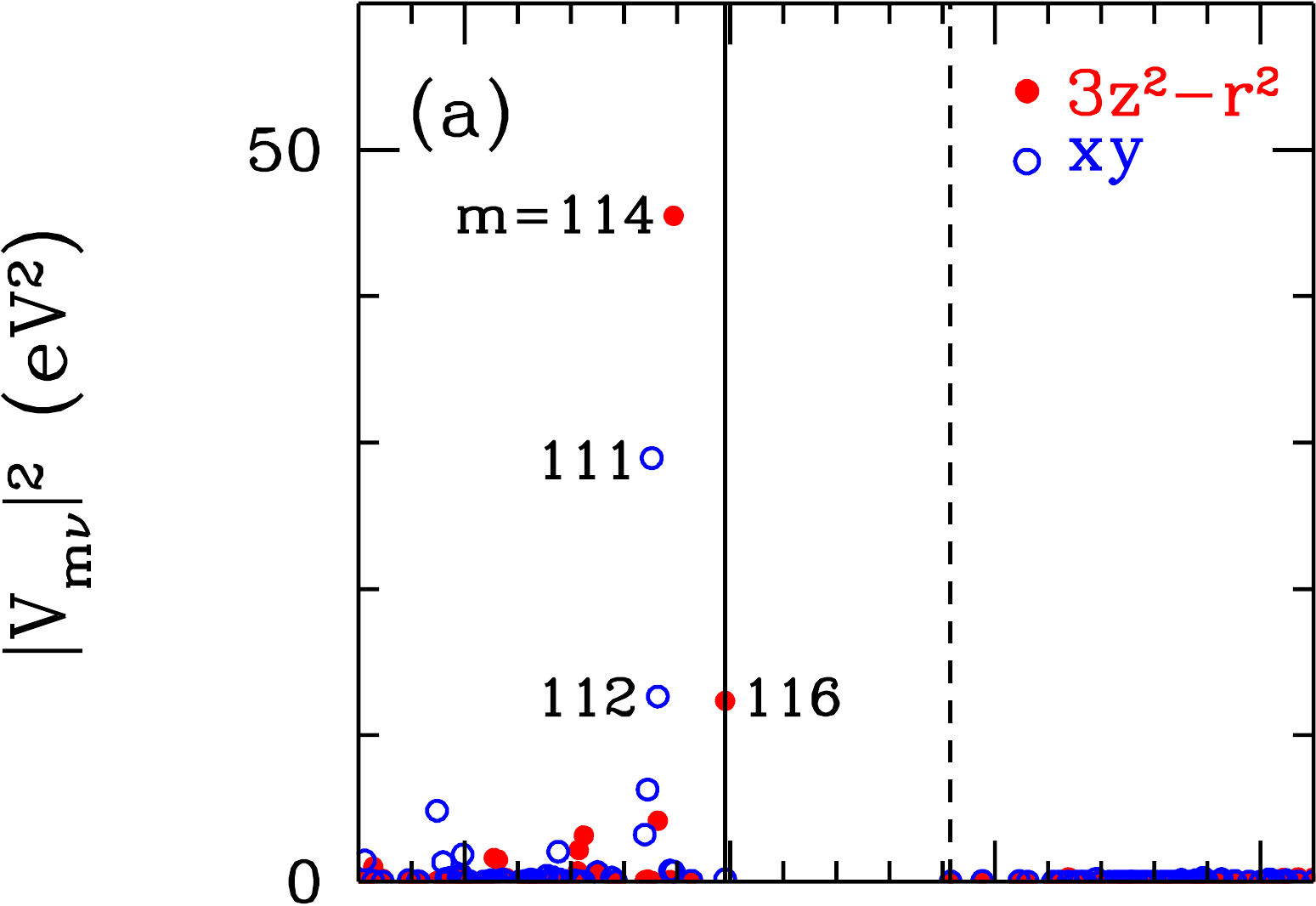} 
\includegraphics[width=7cm]{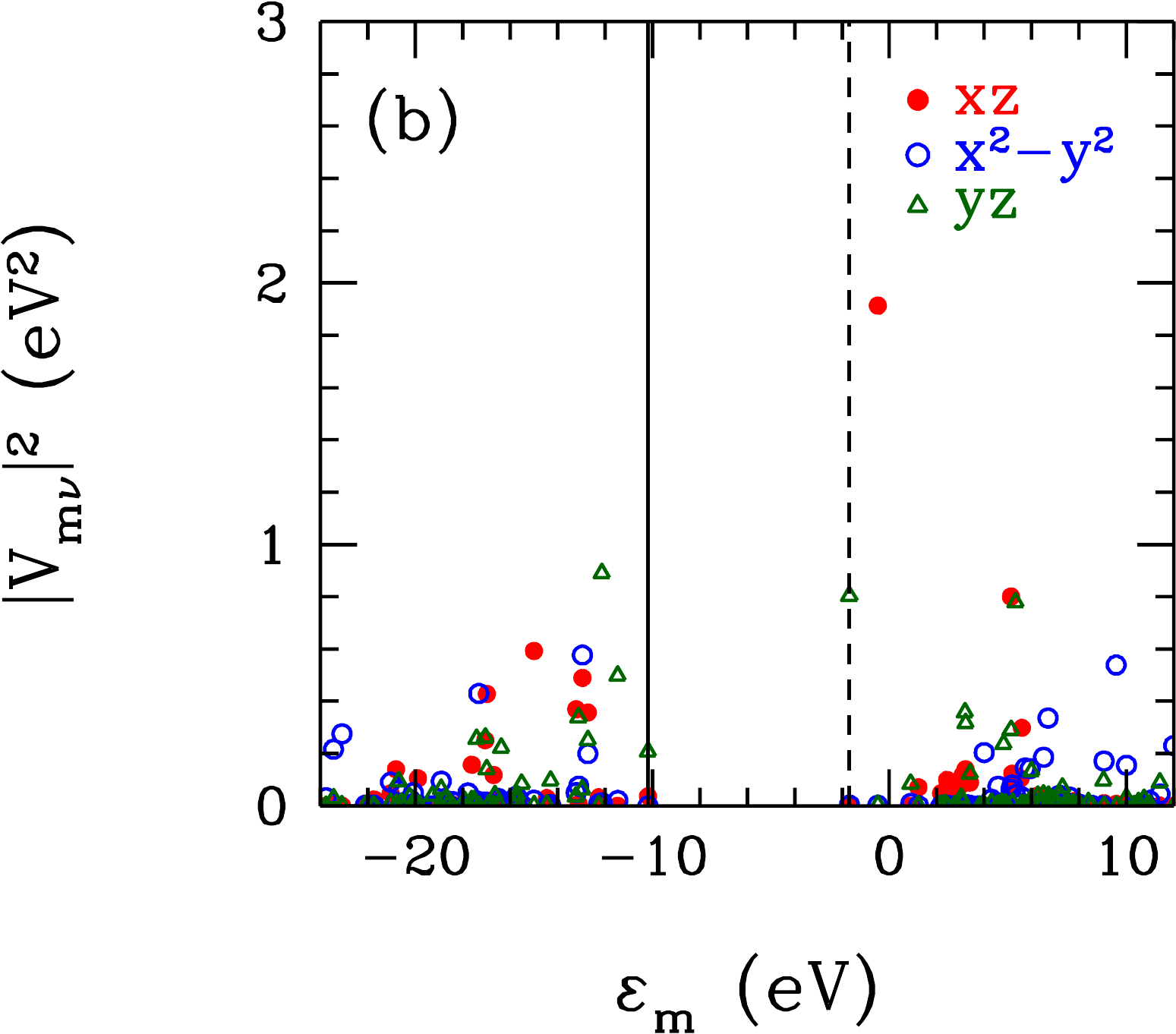}
\caption{(Color online) 
Hartree-Fock results on the 
square of the hybridization matrix elements $|V_{m\nu}|^2$ between 
the host eigenstates $|u_m\rangle$ and 
the Co($3d_{\nu}$) NAO states 
$|\tilde{\phi}_{d\nu}\rangle$ plotted as a function
of the energy of the $m$'th host eigenstate $\varepsilon_m$.
In (a) results are shown for the $e_g$-like $\nu = 3z^2-r^2$ and $xy$ orbitals, 
and in (b) for the $t_{2g}$-like $\nu = xz$, $x^2-y^2$, and $yz$.
Here,
the vertical solid and dashed lines denote the HOMO and LUMO levels, 
respectively.
We observe that 
the host states $m=114$, 111, 112 and 116 have the strongest 
hybridization matrix elements. 
The matrix elements with the $t_{2g}$-like states are much smaller
than those with the $e_g$-like states. 
}
\label{fig4}
\end{figure}

\subsection{Coulomb repulsion at the Co($\bm{3d_{\nu}}$) orbitals}

The intra-orbital Coulomb repulsion at the Co($3d_{\nu}$) 
atomic orbitals is given by 
the two-electron integral
\begin{equation}
U_{\nu} = \int d^3r \int d^3r'
|\phi_{d\nu}({\bf r})|^2 {1 \over |{\bf r} - {\bf r'}| }
|\phi_{d\nu}({\bf r'})|^2.
\end{equation}
Using the 6-31G Gaussian basis set \cite{Gaussian},
we obtain $U_{\nu}=36.8 \eV$ for the Co($3d_{\nu}$) orbitals.
However, we note that the value of $U_{\nu}$ depends on 
which Gaussian basis set is used.
While for the 3-21 basis set
$U_{\nu} = 42.8 \eV$, 
for the 6-31G and the 6-31G(d) sets, 
we obtain $U_{\nu} = 36.8 \eV$.

Since the effective Anderson Hamiltonian introduced 
above uses the basis of the Co($3d_{\nu}$) NAO's,
the Coulomb two-electron integrals also need to be evaluated in this basis
instead of the basis of the Co($3d_{\nu}$) atomic orbitals.
When we evaluate the inter-orbital Coulomb repulsion given by Eq. (11) using the 
Co($3d_{\nu}$) NAO's with the Gaussian 6-31G basis set,
we obtain $\approx 28$ eV.

In the remaining of this paper, 
we will use a constant $U$ for $U_{\nu}$,
and present QMC results particularly for $U$ varying between 
28 eV and 36 eV.
We are neglecting the inter-orbital Coulomb
interactions in the QMC calculations, 
which are actually
close in value to the intra-orbital terms.
Hence,
in the QMC part of the calculations,
we are taking into account the effects of the Coulomb repulsion at a
simple level,
treating $U$ more as a variable than a true {\it ab initio} parameter.
Inclusion of the inter-orbital Coulomb interaction 
will clearly provide more accurate results
as discussed in Subsection IV.D.

\subsection{Double counting in HF+QMC}

At this point, 
it is necessary to note that  
in the HF+QMC approach the onsite Coulomb repulsion $U$
is taken into account twice, once in the Hartree-Fock calculation, 
and a second time in the QMC calculations. 
In order to prevent the double-counting of the effects of $U$ 
in the many-body approaches
which combine the electronic-structure calculations with the QMC
simulations, 
a constant $\mu_{\rm DC}$ is usually subtracted 
from the Co($3d_{\nu}$) levels \cite{Anisimov,Czyzyk,Kunes,Karolak}, 
$\varepsilon_{d\nu} 
\rightarrow 
\tilde{\varepsilon}_{d\nu} = \varepsilon_{d\nu}  - \mu_{\rm DC}$.
Here, 
$\mu_{\rm DC}$ is given by
\begin{equation}
\mu_{\rm DC} = U { \langle n^{\rm HF}_d \rangle \over 10},
\end{equation}
where $\langle n^{\rm HF}_d \rangle$ is the total number of the electrons 
(including spin) in the Co($3d$) levels obtained within Hartree-Fock. 
Since the Hartree-Fock calculations yield $\langle n^{\rm HF}_d \rangle=6.8$, 
we use $\mu_{\rm DC}=0.68 U$ in this paper.

It is possible to argue that, 
because we are using only the intra-orbital Coulomb repulsion in QMC, 
the double-counting shift should be orbital dependent
and given by
$\mu^{\nu}_{\rm DC} = U \langle n^{\rm HF}_{d\nu} \rangle /2$,
where $\langle n^{\rm HF}_{d\nu} \rangle$ is the orbital dependent 
Co($3d_{\nu}$) occupation number including spin from the HF calculation. 
However,
if we include the inter-orbital Coulomb interaction in QMC,
then the shift would be orbital independent.
Since we try to model the actual system as much as possible,
we use a constant $\mu_{\rm DC}$. 

In Fig. 5, 
the shifted Co($3d_{\nu}$) NAO energy levels 
$\tilde{\varepsilon}_{d\nu}$
are shown for $U=36$ eV and $\mu_{\rm DC}=0.68 U$.
Also, shown here is the host density of states
$D_h(\varepsilon)$.
For these parameters, 
the $xz$, $x^2-y^2$ and $yz$ states are located below 
the bottom of the valence band,
while the $3z^2-r^2$ and $xy$ states are located in the middle. 
 
\begin{figure} 
{\includegraphics[width=7cm]{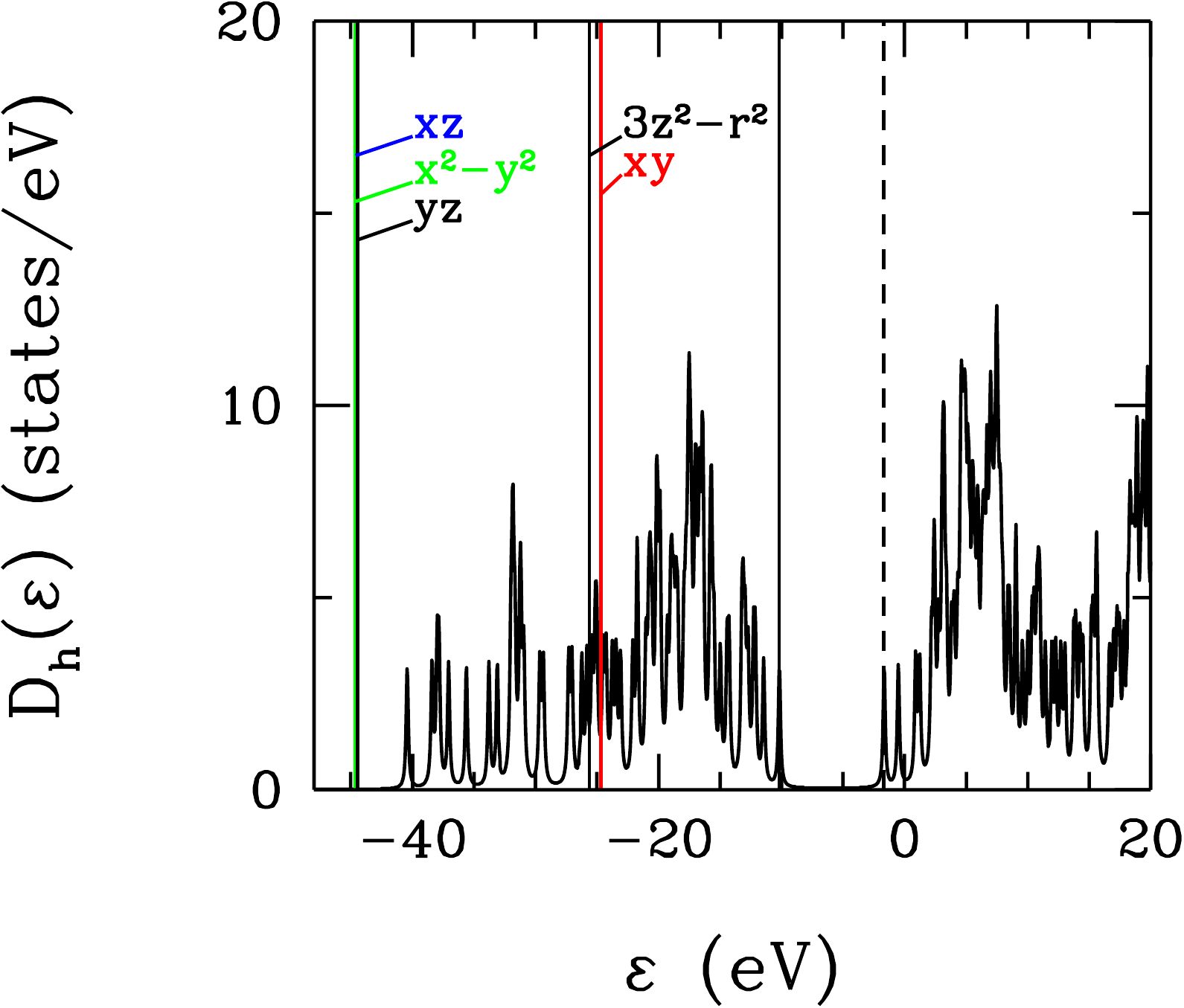}}
\caption{(Color online) 
Density of states of the host states of the effective Haldane-Anderson model
$D_h(\varepsilon)$ from the Hartree-Fock calculation. 
Here, the Co($3d_{\nu}$) levels have been shifted by $\mu_{DC}$
in order to compensate for the double-counting of the Coulomb 
repulsion by the HF+QMC approach.
The shift $\mu_{\rm DC}$ has been evaluated for $U=36$ eV. 
}
\label{fig5}
\end{figure} 

\section{Quantum Monte Carlo results}

In this section, we present QMC data on the effective 
Haldane-Anderson model for Im-[Co$^{\rm III}$(corrin)]-CN$^+$,
of which parameters were obtained in the previous section.
For this model, we have performed QMC calculations 
by using the Hirsch-Fye QMC algorithm \cite{Hirsch}. 
In particular, we have calculated the
Co($3d_{\nu}$) and the host single-particle 
Green's functions and the magnetic correlations functions.

In Subsection III.A, 
we introduce the correlation functions which we evaluate with QMC.
In Subsection III.B, 
we present data on the electron occupation number 
and the magnetic moment formation
at the Co($3d_{\nu}$) NAO's.
Here, we observe that in-gap states develop arising from the 
Co($3d_{\nu}$) states in the semiconductor energy gap. 
The induced in-gap states reduce the value of the 
semiconductor gap found by the Hartree-Fock calculation.
We find that the Co($3d_{\nu}$) states induced above the top of the valence band
correspond to the upper Hubbard states of the Co $t_{2g}$-like orbitals.
On the other hand, 
the Co($3d_{\nu}$) states induced in the middle of the semiconductor gap
correspond to the impurity bound states originating from 
the Co $e_g$-like states. 
We reach this assignment based on the QMC results 
on the chemical potential dependence of the 
Co($3d_{\nu}$) electron occupation number and magnetic moment formation, 
host electron occupation numbers and magnetic moment formation
as well as the antiferromagnetic correlations between the Co $e_g$
states and the host eigenstates.

The QMC results on the host electrons are shown in Subsection III.C.
Here, we observe that the host electrons form new states inside the 
semiconductor gap at the same energies where the Co($3d_{\nu}$) NAO's
develop in-gap states. 
These host states are the ones which have the strongest
hybridization matrix elements with the Co $e_g$-like states. 
These host states also develop magnetic moments
which depend strongly on the electron filling 
of the impurity bound states. 

In Section III.D, we present data on the antiferromagnetic correlations 
which develop around the Co atom. 
We find that the antiferromagnetic correlations 
develop mainly between the Co $e_g$-like states
and the electron spins at the CN axial ligand, 
and they depend strongly on the filling 
of the impurity bound states. 
The results presented in Section III are based on Ref. [\onlinecite{Mayda}].

The QMC calculations are performed within the grand canonical ensemble,
and the total electron number $\langle n_T\rangle$
is obtained by evaluating with QMC,
\begin{equation}
\langle n_T \rangle = \sum_{\nu=1}^{5} \sum_{\sigma} 
\langle d_{\nu\sigma}^{\dagger} d_{\nu\sigma} \rangle +
\sum_{m=1}^{N-5} \sum_{\sigma} 
\langle c_{m\sigma}^{\dagger} c_{m\sigma} \rangle.
\end{equation}
In presenting the QMC data in the next section, 
we will obtain results as a function of $\mu$
and evaluate the corresponding $\langle n_T\rangle$
also with QMC. 
For the truncated structure Im-[Co$^{\rm III}$(corrin)]-CN$^+$,
the total electron number is
$N_{\rm el}=238$.
By comparing $\langle n_T\rangle$ with $N_{\rm el}$, 
we obtain the corresponding value for $\mu$.

\subsection{QMC measurements}

The Matsubara-time dependent single-particle Green's function 
for the Co($3d_{\nu}$) NAO states is defined by
\begin{equation}
G_{\nu\sigma}(\tau) = -
\langle T_{\tau} d_{\nu\sigma}(\tau) d_{\nu\sigma}^{\dagger}(0) \rangle,
\end{equation}
where $T_{\tau}$ is the usual Matsubara $\tau$-ordering operator and 
$d_{\nu\sigma}(\tau) = e^{H\tau} d_{\nu\sigma} e^{-H\tau}$.
Similarly, the host Green's function is defined by
\begin{equation}
G_{m\sigma}(\tau) = -
\langle T_{\tau} c_{m\sigma}(\tau) c_{m\sigma}^{\dagger}(0) \rangle.
\end{equation}
In addition, 
we calculate the square of the magnetic moment at
each Co($3d_{\nu}$) orbital, 
$\langle (M_{\nu}^z)^2 \rangle$, 
where 
\begin{equation}
M_{\nu}^z = d_{\nu\uparrow}^{\dagger} d_{\nu\uparrow} -
d_{\nu\downarrow}^{\dagger} d_{\nu\downarrow}.
\end{equation}
In order to probe the magnetic correlations around the Co atom,
we also calculate the equal-time Co($3d_{\nu}$)-host 
magnetic correlation function
$\langle M_{\nu}^z M_m^z \rangle$,
where 
\begin{equation}
M_{m}^z = c_{m\uparrow}^{\dagger} c_{m\uparrow} -
c_{m\downarrow}^{\dagger} c_{m\downarrow}.
\end{equation}
We present data on $\langle (M_m^z)^2\rangle$ as well. 

In the following subsections, we show QMC data on the 
Co($3d_{\nu}$) electron number $\langle n_{\nu}\rangle$
and the square of the Co($3d_{\nu}$) moments
$\langle (M_{\nu}^z)^2 \rangle$ as a function of the 
chemical potential $\mu$. 
We also present data on the total host electron number 
$\langle n_h\rangle$
as well as the total electron number 
$\langle n_T\rangle$
as a function of $\mu$.
In addition, we study the local moment formation at the host 
eigenstates and their magnetic coupling to the 
Co($3d_{\nu}$) NAO's. 
We find that antiferromagnetic correlations develop between the 
Co $e_g$ states and the electronic spins 
at the CN axial ligand
depending on the value of $\mu$. 

In obtaining the QMC data shown below, 
Matsubara-time discrete steps between
$\Delta\tau = 0.04 \eV^{-1}$ and $0.07 \eV^{-1}$ were
used. 
The results are presented for temperature $T=2000 K$
in the grand canonical ensemble. 
Hence, a temperature broadening of order 0.17 eV is 
expected for the energy resolution of the QMC data.

\subsection{QMC results on the Co($\bm{3d_{\nu}}$) electrons}

We begin presenting QMC data with 
Figure 6(a) which 
shows the electron number $\langle n_{\nu}\rangle$ 
of the Co($3d_{\nu}$) NAO states 
as a function of the chemical potential $\mu$ 
for $U=36$ eV.
Similarly,
Figure 6(b) shows the square of the magnetic moment 
$\langle (M^z_{\nu})^2\rangle$ at the Co($3d_{\nu}$) orbitals
as a function of $\mu$. 
We observe that the results for the Co($3d$) $xz$, $x^2-y^2$, and $yz$ 
states are similar to each other. 
The results for the $3z^2-r^2$ and $xy$ states are also similar. 
From the $\mu$ dependence shown in these figures 
it is possible to extract information
on the single-particle spectral weight distribution and 
the local moment formation. 

\begin{figure}
\includegraphics[width=7cm]{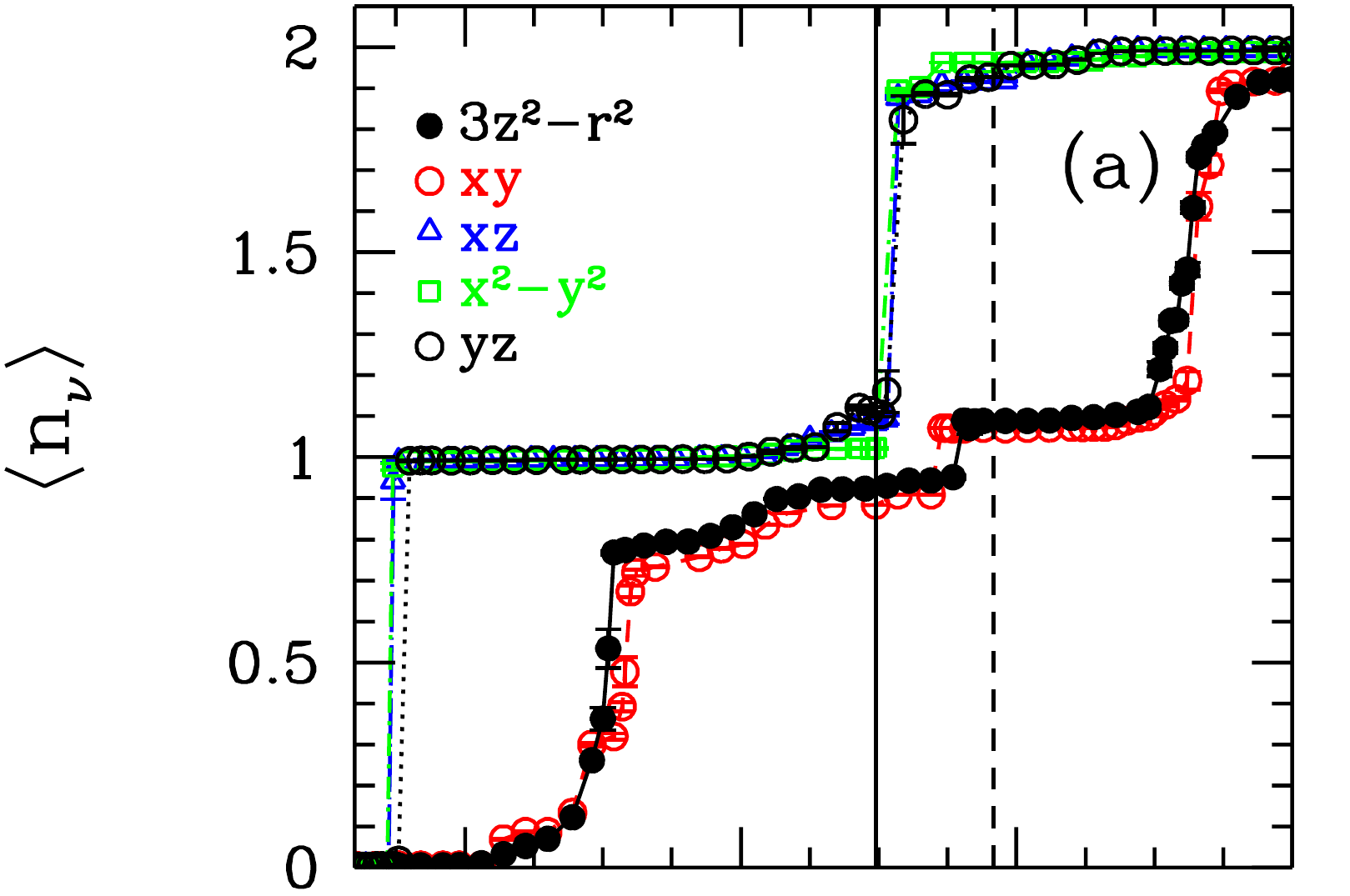} 
\includegraphics[width=7cm]{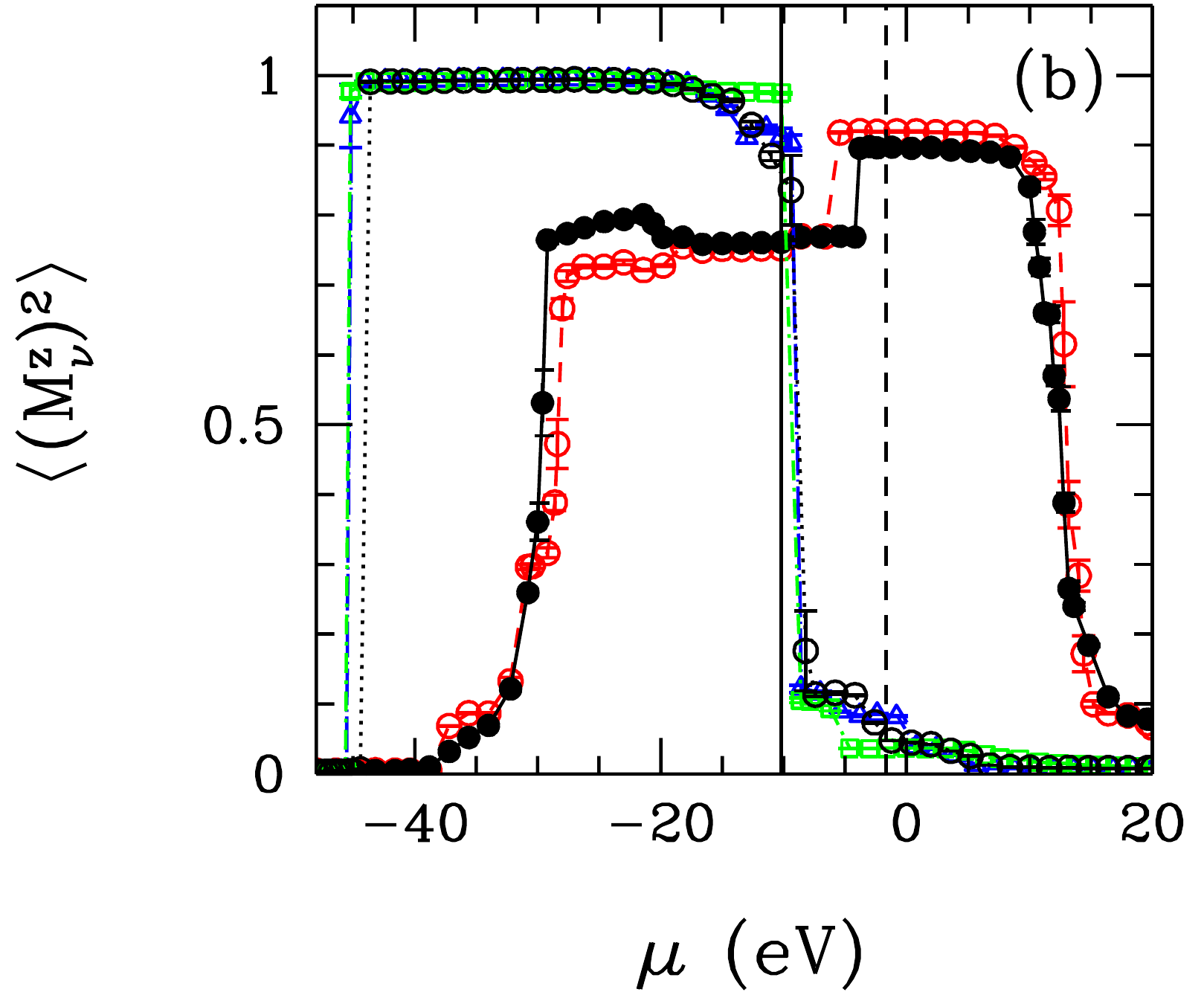} 
\caption{(Color online) 
(a) QMC results on the electron
occupation number $\langle n_{\nu}\rangle$ 
of the Co($3d_{\nu}$) natural atomic orbitals
plotted as a function of the chemical potential $\mu$.
(b) Square of the magnetic moment $\langle (M_{\nu}^z)^2 \rangle$
at the Co($3d_{\nu}$) natural atomic orbitals
versus the chemical potential $\mu$.
Here, the vertical solid and dashed lines denote the HOMO and the LUMO levels, 
respectively.
These results are for $U=36 \eV$.
}
\label{fig6}
\end{figure}

The bare energy levels $\tilde{\varepsilon}_{d\nu}$ of the
Co($3d$) $xz$, $x^2-y^2$, and $yz$ NAO states
are almost degenerate and are located 
at $\varepsilon\approx -45$ eV below the lower edge of the valence band,
as seen in Fig. 5. 
In Fig. 6(a) we see that for these orbitals  
$\langle n_{\nu}\rangle$ exhibits a jump of unit magnitude 
at $\mu\approx -45$ eV.
These orbitals remain singly occupied as $\mu$ is increased upto
the interval $-10 \eV \ltsim \mu \ltsim -8 \eV$, 
at which point these orbitals become 
nearly doubly occupied.
In Fig. 6(b), 
we see that the magnetic moments at these orbitals 
have their maximum value in the interval 
$-45 \eV \ltsim \mu \ltsim -10 \eV$.
Above $\approx -10$ eV, these moments decrease rapidly due to double occupancy.
The in-gap states 
in the energy interval $-10 \eV \ltsim \varepsilon \ltsim -8 \eV$
have sufficient spectral weight to accommodate 
a total of nearly three electrons, and
their location corresponds to 
approximately 
$\tilde{\varepsilon}_{d\nu} + U$.
Hence, 
we deduce that
these in-gap states correspond to the upper-Hubbard states
of the Co $t_{2g}$-like orbitals.
We note that they are located inside the gap,
and hence new impurity states are formed inside the gap found by HF.

Figure 5 shows that the bare energy levels $\tilde{\varepsilon}_{d\nu}$ 
of the $3z^2-r^2$ and $xy$ orbitals are near $-25$ eV.
In Fig. 6(a) we observe that for these two orbitals,
$\langle n_{\nu}\rangle$ becomes finite at $\mu\approx -38$ eV,
increases rapidly around $\mu\approx -30 \eV$,
and remains less than one until $\mu$
reaches $\approx -5.5 \eV$. 
This type of $\mu$ dependence suggests that 
the $U=0$ single-particle spectral weight,
which is a $\delta$-function 
near $-25$ eV, has been shifted down in energy and has been broadened.
These orbitals remain less than singly occupied 
until $\mu$ reaches $\approx -5.5$ eV,
where $\langle n_{xy} \rangle$ exhibits a sudden increase of about 0.2.
We think that this sudden increase 
corresponds to an impurity bound state as 
found in the mean-field solution of the Haldane-Anderson model\cite{Haldane}. 
For the $3z^2-r^2$ orbital, 
a similar increase occurs at
a slightly higher energy of $\approx -4 \eV$.
We note that these jumps take place in the energy gap
found by HF.
Hence, new impurity states are formed in the energy gap.
As seen in Fig. 6(b), 
the magnetic moments at the $xy$ and $3z^2-r^2$ orbitals
also exhibit sudden increases at these energies.
The occupation numbers of these states remain constant 
at unit magnitude as $\mu$ is increased upto $\approx 10 \eV$.
This value of $\mu$ corresponds approximately to 
$\tilde{\varepsilon}_{d\nu} + U$ for the $xy$ and $3z^2-r^2$ 
orbitals.
Above this energy,
the $xy$ and $3z^2-r^2$ orbitals rapidly become doubly occupied,
and the magnetic moments decrease. 

In summary, 
from Fig. 6(a)
we deduce that new in-gap states originating 
from the Co($3d$) $xz$, $x^2-y^2$ and $yz$
orbitals form in the energy interval 
$-10 \eV \ltsim \varepsilon \ltsim -8 \eV$,
while new states from the $xy$ and $3z^2-r^2$ orbitals 
form at energies $\approx -5.5 \eV$ and $\approx -4 \eV$,
respectively.
These newly formed states reduce the value of the semiconductor gap
found by the Hartree-Fock calculation,
which is $8.5 \eV$, 
down to $\approx 3 \eV$.

\begin{figure}[t]
\includegraphics[width=7cm]{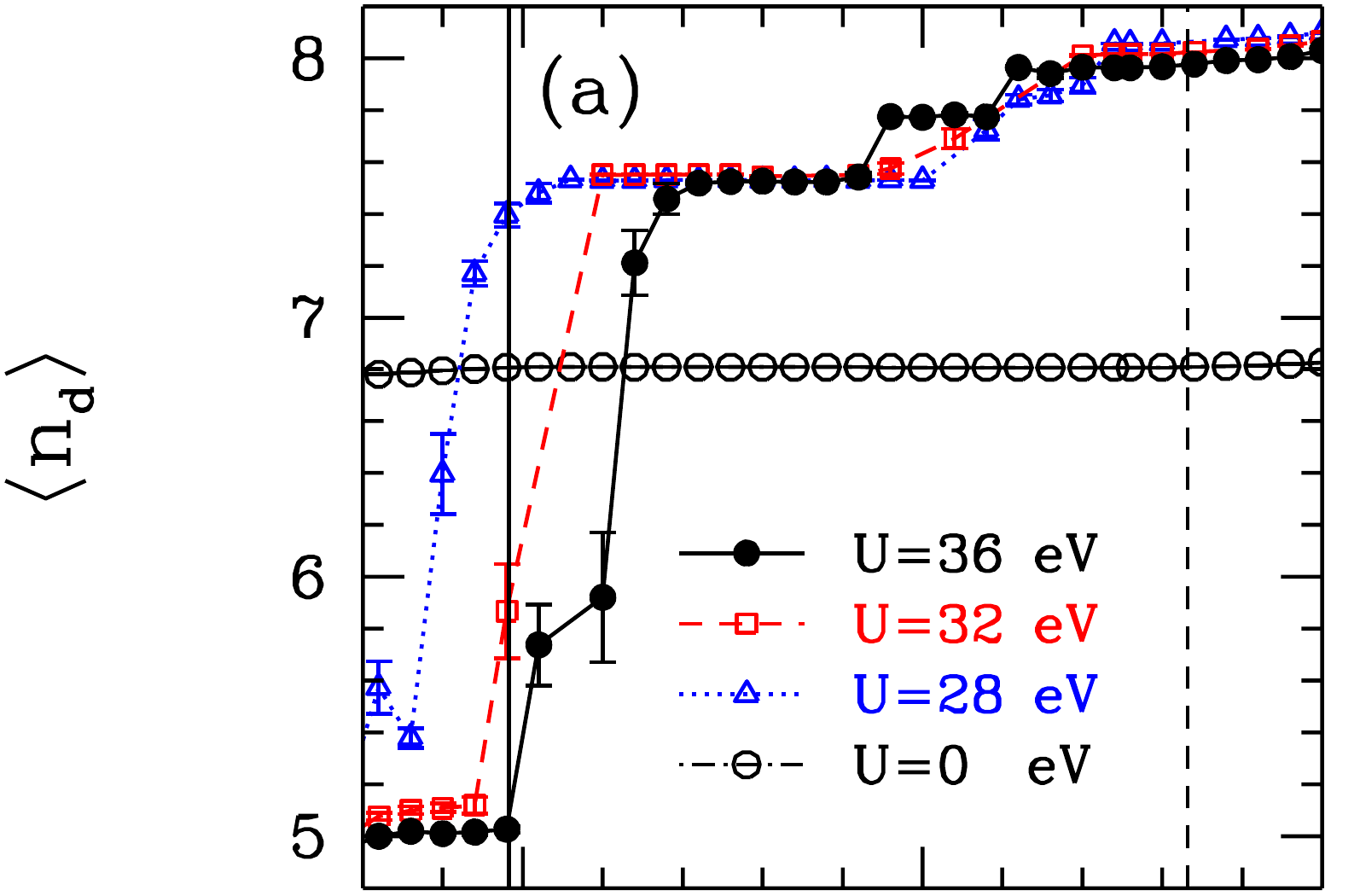} 
\includegraphics[width=7cm]{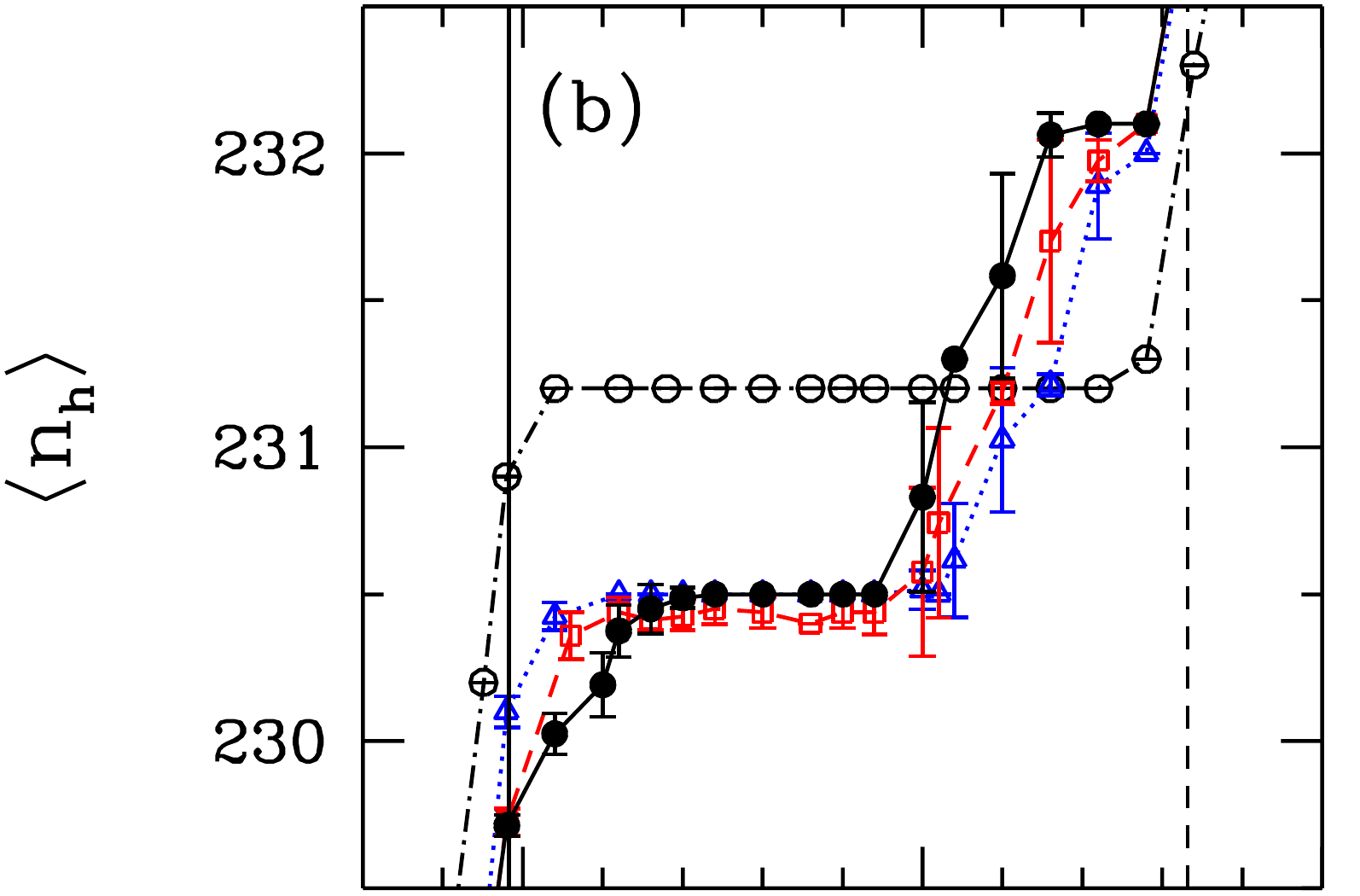} 
\includegraphics[width=7cm]{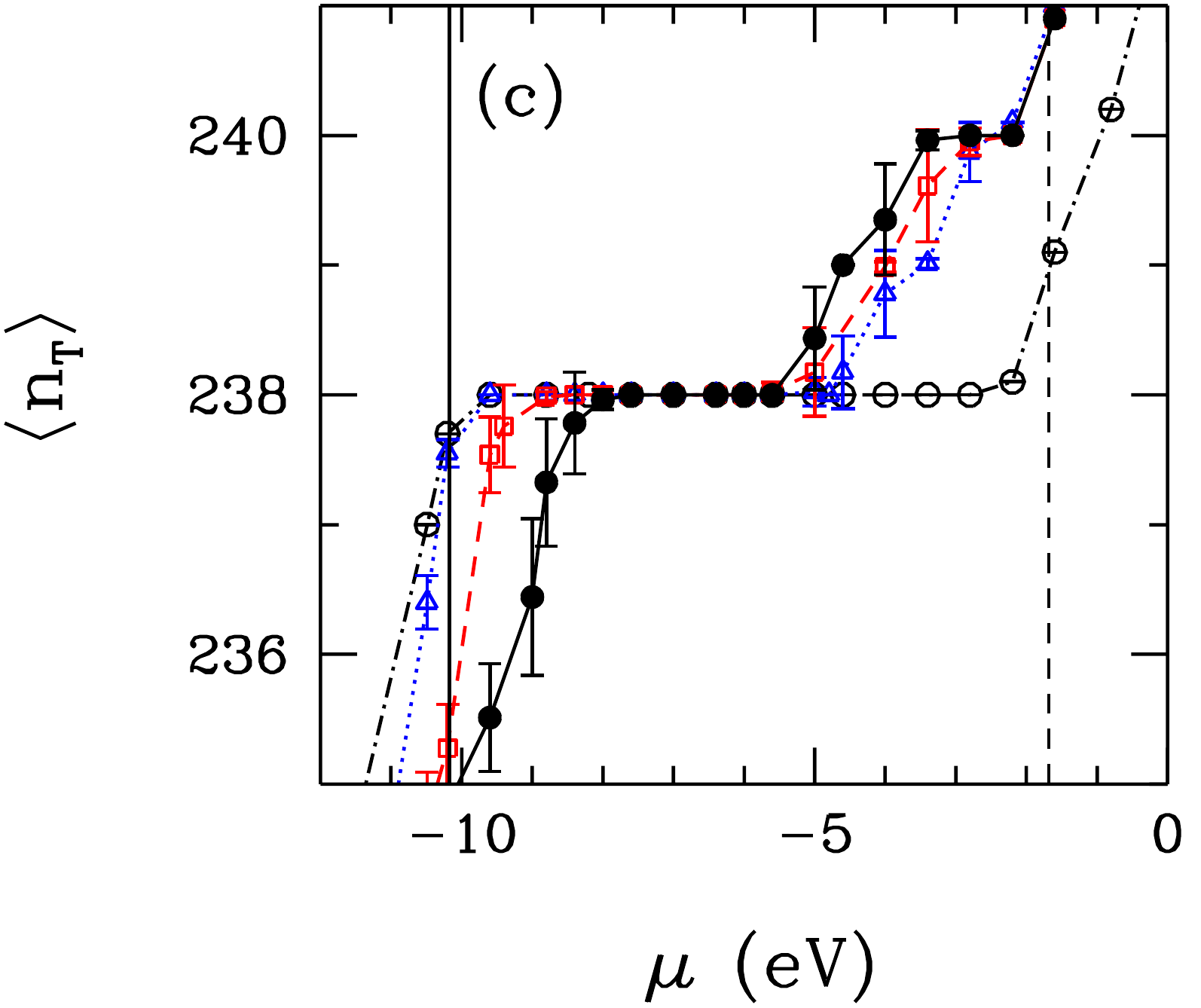} 
\caption{(Color online) 
(a) Total electron occupation number $\langle n_d\rangle$ 
of the Co($3d$) natural atomic orbitals
as a function of the chemical potential $\mu$.
(b) Total number of the host electrons $\langle n_h\rangle$ versus 
$\mu$.
(c) Total number of electrons, $\langle n_T\rangle = \langle n_d\rangle + \langle n_h\rangle$, 
versus $\mu$.
Here, results are shown for various values of the intra-orbital Coulomb repulsion $U$.
In addition, 
the vertical solid and dashed lines denote 
the Hartree-Fock results for the HOMO and LUMO levels, 
respectively.
}
\label{fig7}
\end{figure}

Figure 7(a) shows the QMC data on the total electron occupation 
of the Co($3d_{\nu}$) states
\begin{equation}
\langle n_d\rangle = 
\sum_{\nu=1}^{5} \sum_{\sigma} 
\langle d_{\nu\sigma}^{\dagger} d_{\nu\sigma} \rangle
\end{equation}
versus the chemical potential $\mu$.
Here, results are shown for  various values of $U$
in order to see the dependence on $U$.
We note that in these calculations, as $U$ is varied, 
$\tilde{\varepsilon}_{d\nu}$ also varies according to Eq. (12). 

For $U=36 \eV$,
we observe that $\langle n_d\rangle$
increases as $\mu$ is varied from -10 eV to -8 eV,
because the $xz$, $x^2-y^2$ and $yz$ orbitals 
become nearly doubly occupied in this interval. 
In the energy gap, 
for $-8 \eV \ltsim \mu \ltsim -5.5 \eV$, we get
$\langle n_d\rangle \approx 7.55$.
In comparison,
the X-ray scattering experiments find 
that the value of $\langle n_d\rangle$ in CNCbl is 
$7.7 \pm 0.1$ \cite{Mebs}.
Above $\mu\approx -5.5 \eV$, we observe two 
step-like increases in $\langle n_d\rangle$
coming from the $xy$ and $3z^2-r^2$ states.
For $U=36 \eV$, 
the energy gap between the highest occupied and the lowest 
unoccupied Co($3d$) levels is $\approx 3 \eV$. 
In comparison,
the experimental value for the semiconductor gap
of CNCbl is about 2.2 eV \cite{Firth}.
As $U$ is decreased from 36 eV to 28 eV, 
we observe that the in-gap states move closer to 
the semiconductor gap edges, and 
the magnitude of the energy gap increases.

Also shown in Fig. 7(a) are results for $U=0$,
in which case
we obtain $\langle n_d\rangle = 6.8$
in the interval $-12 \eV \leq \mu \leq 0 \eV$.
As seen in Fig. 3(b), for $U=0$, 
the bare energy level for the $xz$, $x^2-y^2$ and $yz$ orbitals
is $\approx -20 \eV$,
while for the $3z^2-r^2$ and $xy$ orbitals one has 
$\approx -1 \eV$ and $\approx 0 \eV$, respectively.
If the hybridization is turned on while keeping $U=0$,
then spectral weight of the $3z^2-r^2$ and $xy$ orbitals 
are broadened and shifted to higher energies. 
For this reason, 
$\langle n_d\rangle$ remains constant in the interval
$-12 \eV \leq \mu \leq 0 \eV$
for $U=0$. 

Finally,
we note that 
the  hybridization matrix elements $V_{m\nu}$'s 
are dependent on the atomic 
coordinates used in the HF calculation.
Because of uncertainties in the atomic coordinates,
it is necessary to study how
the general features of the single-particle spectrum
depend on $V_{m\nu}$.
With this in mind,
we have repeated the above QMC calculations
by reducing the hybridization matrix elements by 10\%. 
In this case, we found that, for $U=36 \eV$,
the lowest energy impurity bound state occurs
at $\approx -6.5 \eV$ instead of $-5.5 \eV$,
and the energy gap is reduced to $\approx 2 \eV$. 
Hence,
we see that the impurity bound state is robust 
with respect to small variations in $V_{m\nu}$. 
This is important 
because the values of the $V_{m\nu}$'s depend on the Co-C bond length,
and we are using simple estimates for 
the geometrical parameters 
as described in the introduction of Section II.

\subsection{QMC results on the host electrons}

Figure 7(b) shows the total number of the host electrons
\begin{equation}
\langle n_h \rangle = 
\sum_{m=1}^{N-5} \sum_{\sigma} 
\langle c_{m\sigma}^{\dagger} c_{m\sigma} \rangle,
\end{equation}
where $N=347$,
plotted as a function of the chemical potential $\mu$.
Comparing Figs. 7(a) and (b), we observe that 
$\langle n_h\rangle$ and $\langle n_d\rangle$ exhibit jumps 
at the same values of $\mu$.
Hence,
the host states also develop spectral weight 
inside the gap at the same energies as the Co($3d_{\nu}$) states. 
We note that the error bars in the data on $\langle n_h\rangle$ 
are larger than in $\langle n_d\rangle$,
since the QMC calculation of $\langle n_h\rangle$ is computationally
more costly. 

For the interval
$-10 \eV \ltsim \varepsilon \ltsim -8 \eV$,
the jump in $\langle n_d\rangle$ is $\approx 2.5$, 
while the jump in $\langle n_h\rangle$ is $\approx 0.5$. 
On the other hand,
we note that, in the interval 
$-5.5 \eV \ltsim \varepsilon \ltsim -2 \eV$,
the jump in $\langle n_h\rangle$ is about $\approx 1.6$,
while that in $\langle n_d\rangle$ is $\approx 0.4$. 
Hence, 
actually there are more host states induced in the gap
for $\approx -5.5 \eV \ltsim \varepsilon \ltsim -2 \eV$
than from the Co($3d_{\nu}$) orbitals. 

The total number of the electrons
$\langle n_T \rangle = \langle n_d \rangle + \langle n_h \rangle$
versus $\mu$ is shown in Fig. 7(c). 
For Im-[Co$^{\rm III}$(corrin)]-CN$^+$, 
the total electron number is 238. 
From this figure 
we identify $\mu\approx -8.5 \eV$ as the new value of the HOMO level, 
and $\mu \approx -5.5 \eV$ as the new LUMO level. 
Here, we also observe that the in-gap states located 
in the interval  $-5.5 \eV \ltsim \mu \ltsim -3.5 \eV$
have sufficient spectral weight to accommodate a total of two electrons.
In comparison,
the in-gap states in 
$-10 \eV \ltsim \mu \ltsim -8 \eV$
accommodate 3 electrons.

We also note that $\tilde{\varepsilon}_{d\nu}$ 
for $3z^2-r^2$ and $xy$ states are near
$-25$ eV.
Hence, for $U=36$ eV, the doubly-occupied states occur at 
$\mu \approx \tilde{\varepsilon}_{d\nu}+U = 11$ eV. 
Hence, the in-gap states 
which form from the $3z^2-r^2$ and $xy$ orbitals do not originate from the 
upper Hubbard states.
Instead, they are similar in origin to the impurity 
bound states found in the mean-field solution of the Haldane-Anderson model 
of a transition-metal impurity in a semiconductor host\cite{Haldane}.

Since for Im-[Co$^{\rm III}$(corrin)]-CN$^+$
the total electron number is 238, 
in Fig. 7(c)
we see that the effective energy gap decreases 
as $U$ is increased.
For $U=36$ eV, the effective energy gap becomes $\approx 3$ eV.
Hence, the energy gap can be significantly reduced with respect 
to the Hartree-Fock value of $8.5 \eV$.

\begin{figure}[t]
\includegraphics[width=7cm]{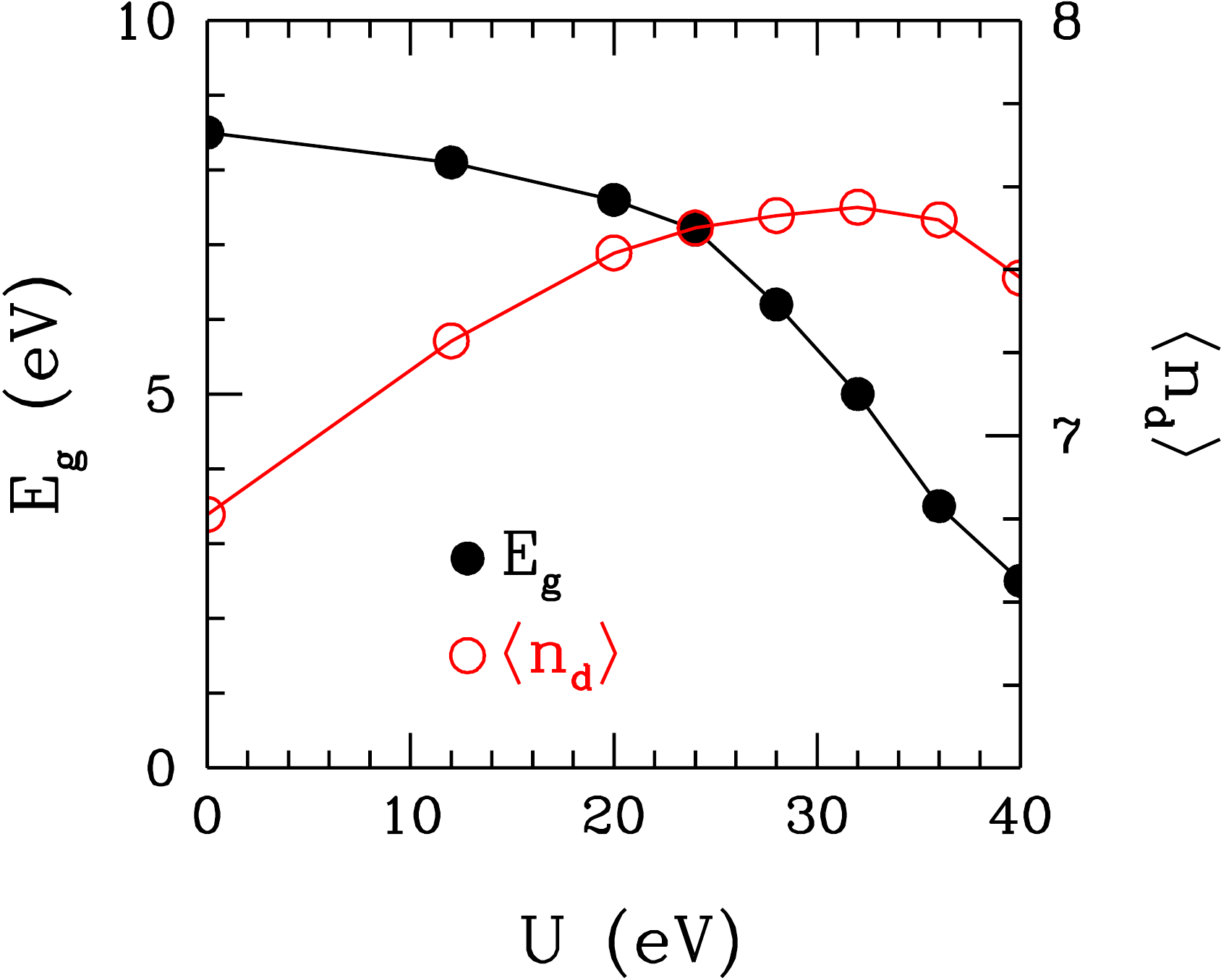}
\caption{(Color Online) 
HF+QMC results on the 
energy gap $E_g$ between the HOMO and LUMO states
versus the intra-orbital Coulomb repulsion $U$ (left axis).
Total electron number at the Co($3d$) orbitals,
$\langle n_d\rangle$
when the chemical potential is in the energy gap, versus $U$ (right axis).
Here, $\mu_{\rm DC}$ which shifts the bare Co($3d_{\nu}$) energy levels 
is varied along with $U$ according to Eq. (12). 
}
\label{fig8}
\end{figure}

The values of the effective energy gap are plotted 
in Fig. 8 as a function of $U$.
We note that in this calculation, as $U$ is varied, 
the shift $\mu_{\rm DC}$ 
is also varied according to Eq. (12). 
Also plotted in Fig. 8 is $\langle n_d\rangle$ versus $U$. 
For $\langle n_d\rangle$,
the experimental estimate \cite{Mebs} 
is $7.7\pm 0.1$.
Hence, for $U\approx 36 \eV$, 
HF+QMC yields values for the energy gap and $\langle n_d\rangle$
which are comparable to the experimental values. 
For this reason,
in the remainder of this paper
we will present QMC data for $U=36 \eV$.

\begin{figure}[t]
\includegraphics[width=7cm]{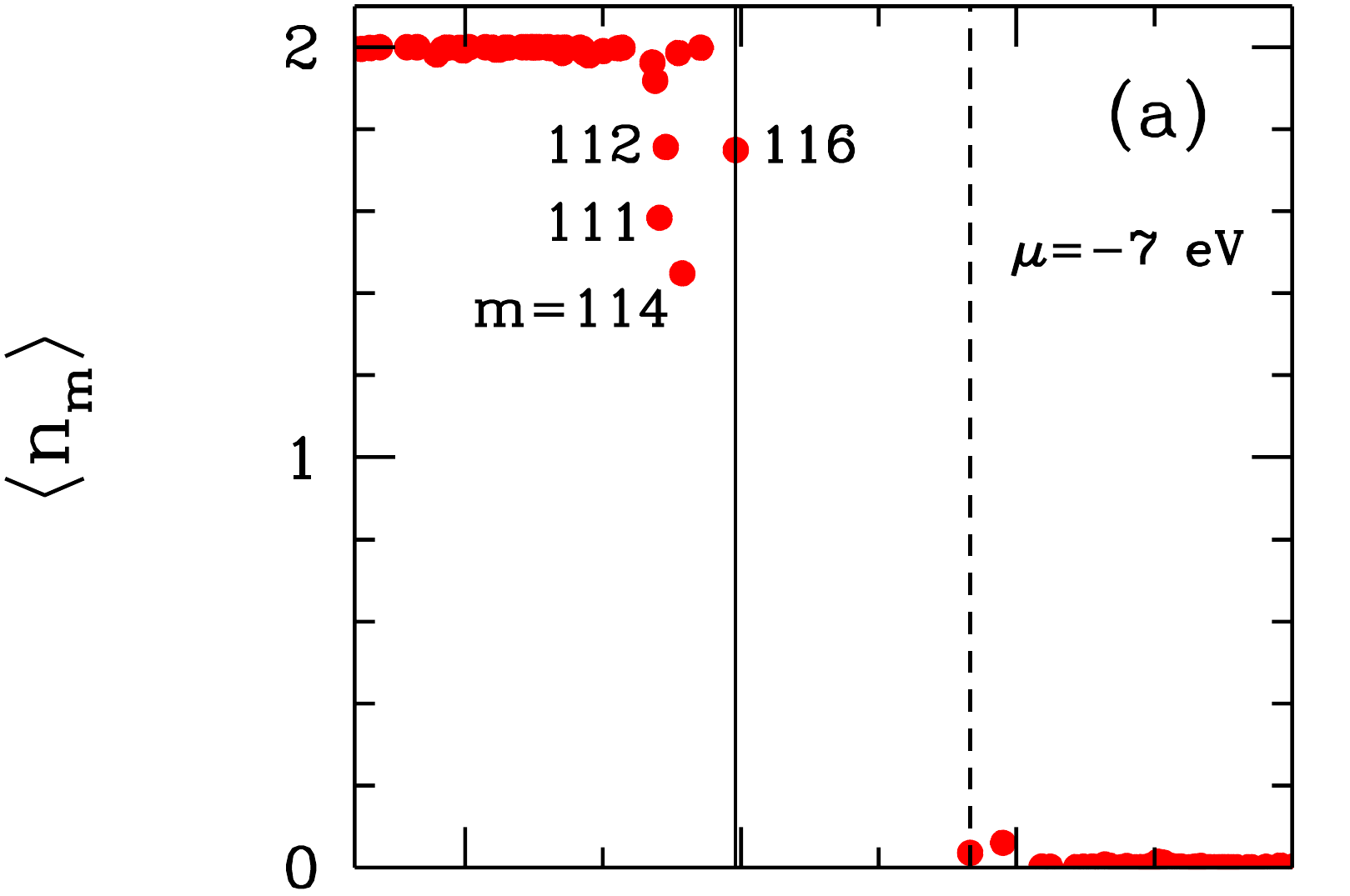}
\includegraphics[width=7cm]{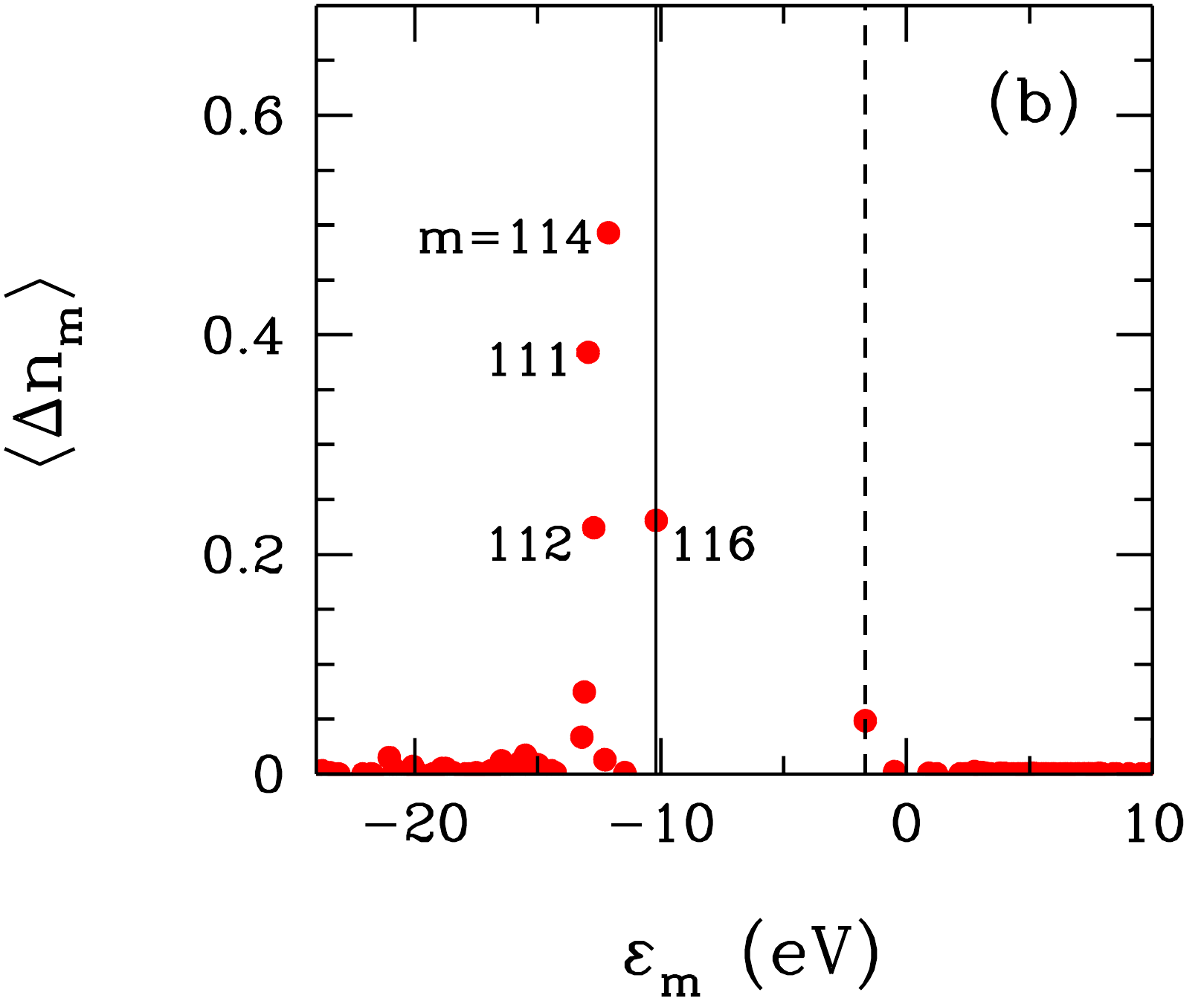}
\caption{(Color Online)
(a) Occupation number of the host eigenstates $\langle n_m\rangle$ 
versus the host energy $\varepsilon_m$ for 
chemical potential 
$\mu=-7.0$ eV.
Here, 
we see that the host eigenstates 
$m=111$, 112, 114 and 116 are not doubly occupied
even though the corresponding 
$\varepsilon_m$ are located deep below $\mu$.
(b) Difference in the occupation number of the $m$'th host eigenstate 
as $\mu$ is changed from $-7.0$ eV to $-2.2$ eV, 
$\langle \Delta n_m\rangle = \langle n_m\rangle|_{\mu=-2.2 \eV} - 
\langle n_m\rangle|_{\mu=-7.0 \eV}$,
plotted as a function of $\varepsilon_m$.
These results are for $U=36$ eV.
In addition, here,
the vertical solid and dashed lines denote the HOMO and the LUMO levels, 
respectively.
}
\label{fig9}
\end{figure}

\begin{figure}[t]
\includegraphics[width=7cm]{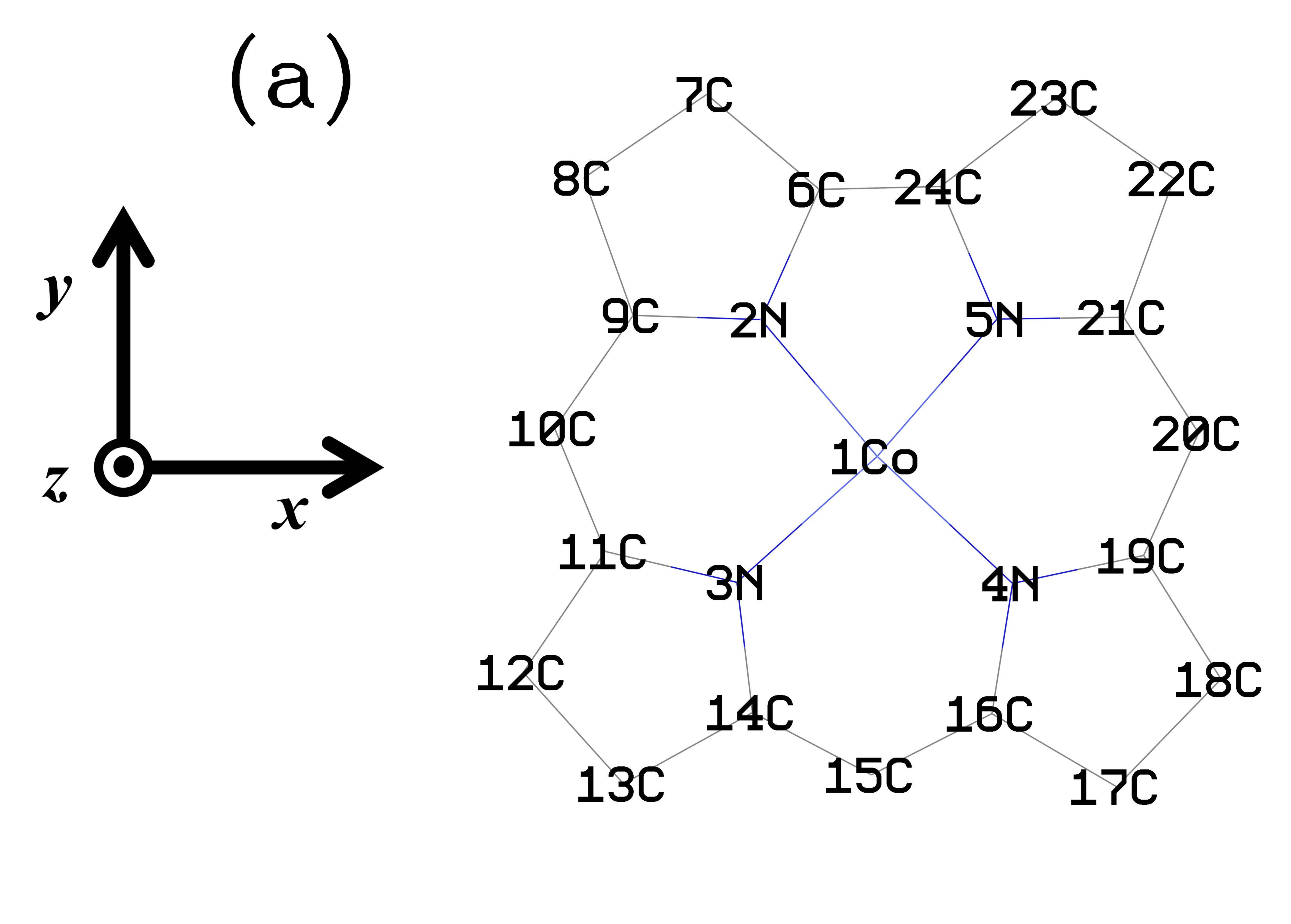} 
\includegraphics[width=7cm]{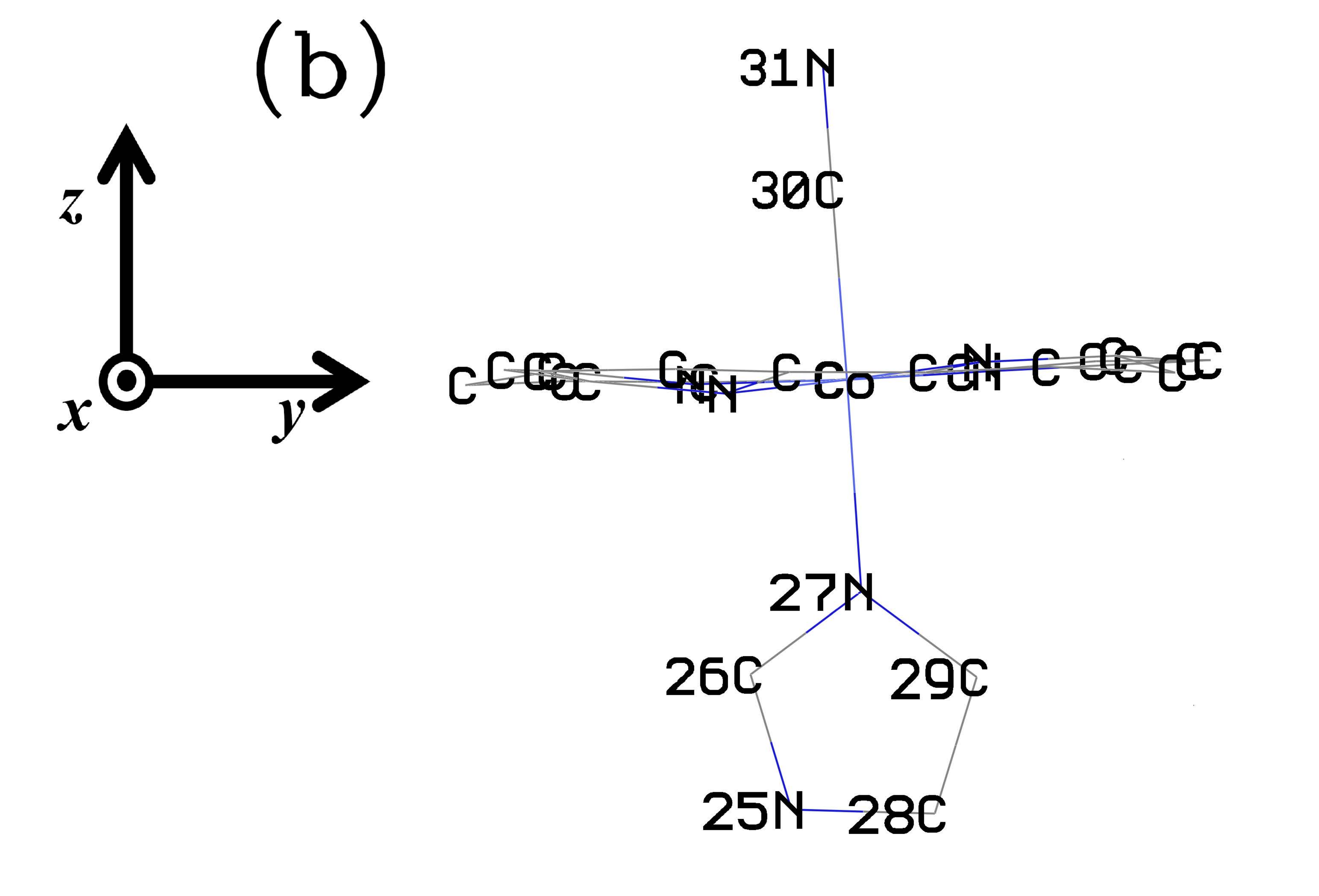} 
\caption{(Color online)  
(a) Top view of 
Im-[Co$^{\rm III}$(corrin)]-CN$^+$.
We note that the $x$ and $y$ axes are located
at 45 degrees to the Co-N bonds in the corrin plane,
and the $z$ axis is perpendicular to the corrin plane.
(b) Side view of Im-[Co$^{\rm III}$(corrin)]-CN$^+$,
where the imidazole and CN ligands attached to Co
below and above the corrin ring, respectively,
are seen. 
Here,
the carbon and nitrogen atoms have been labelled
for convenience,
while the hydrogen atoms are not shown. 
}
\label{fig10}
\end{figure} 

\begin{figure}[t]
\includegraphics[width=5.7cm]{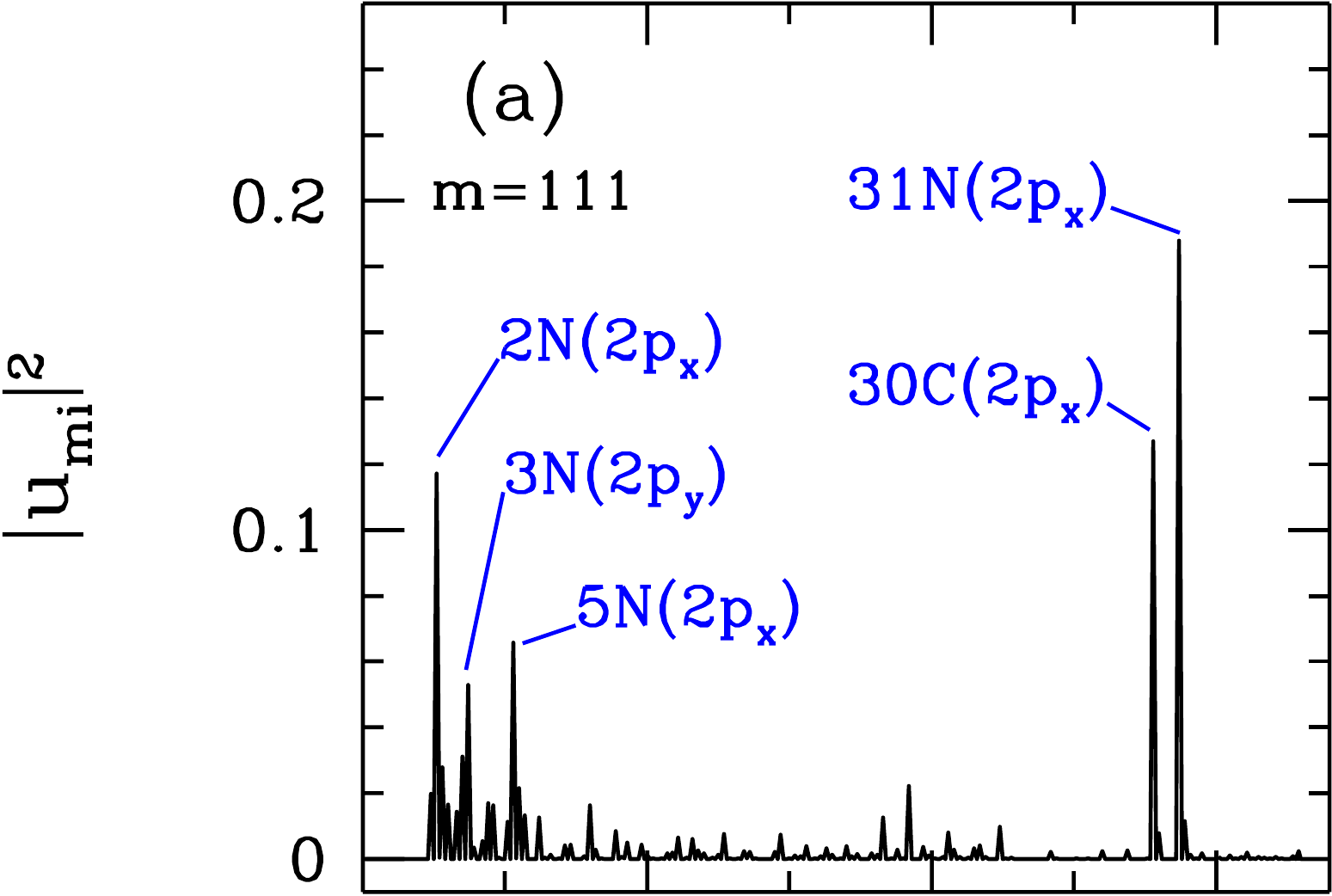} 
\includegraphics[width=5.7cm]{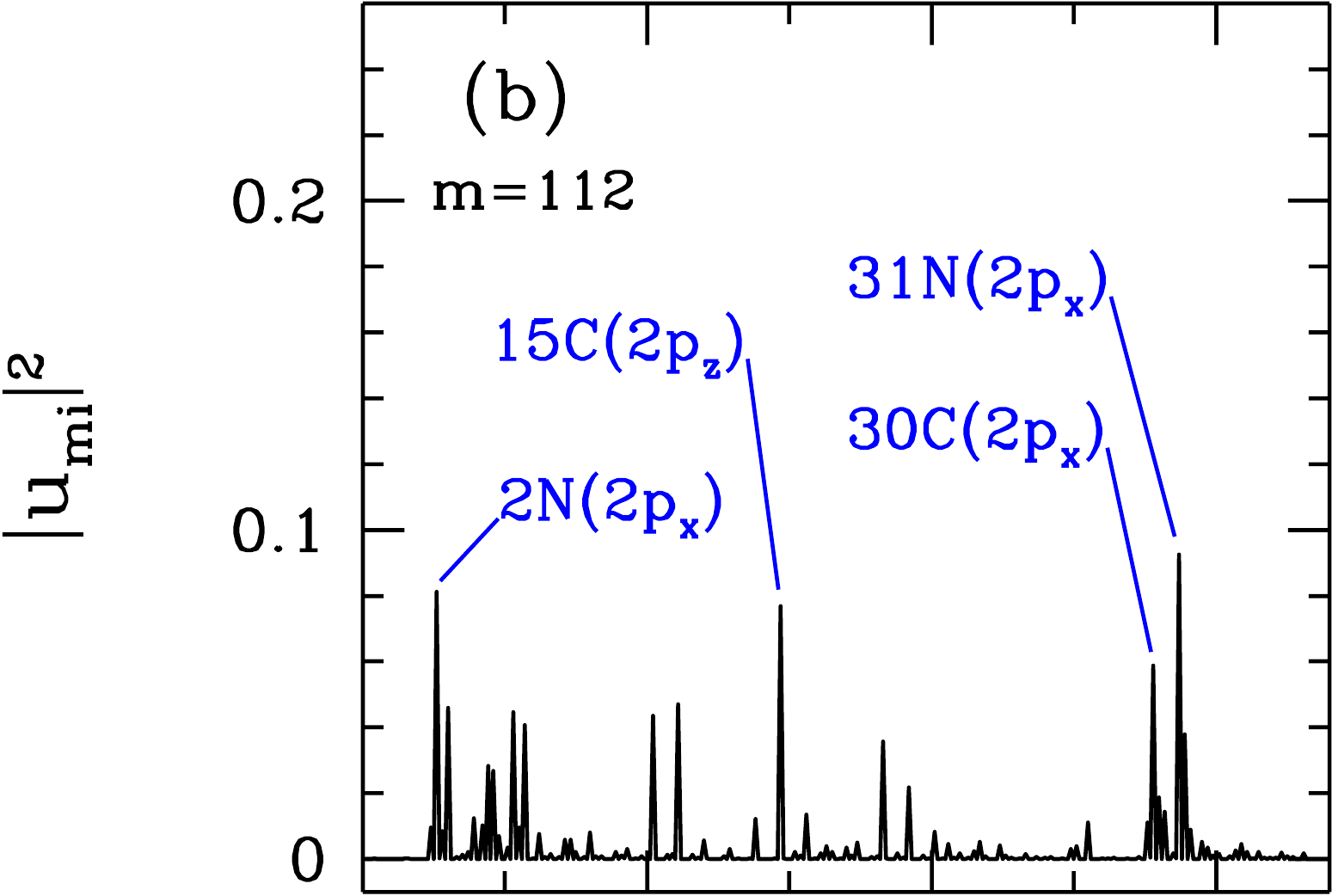} 
\includegraphics[width=5.7cm]{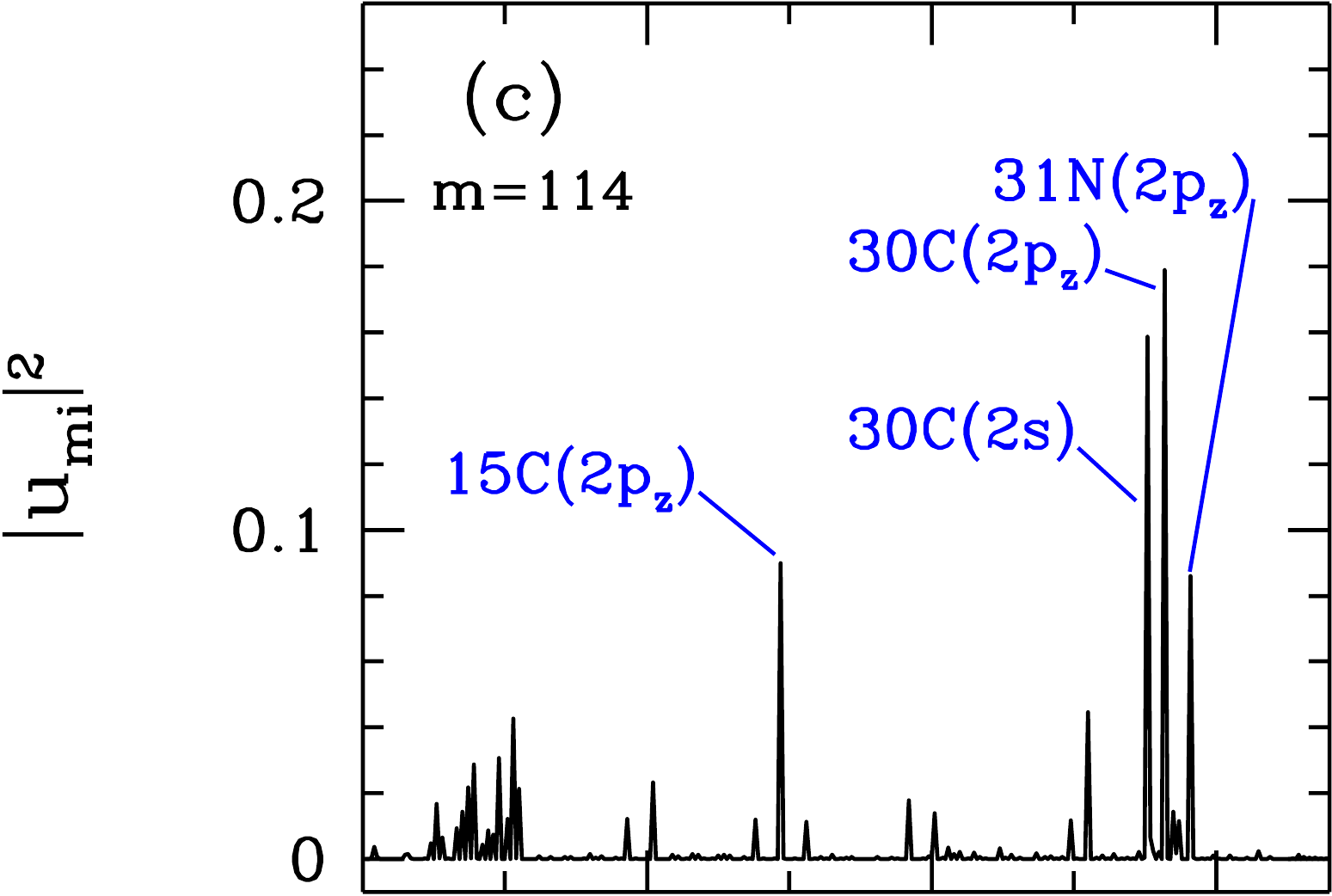} 
\includegraphics[width=5.7cm]{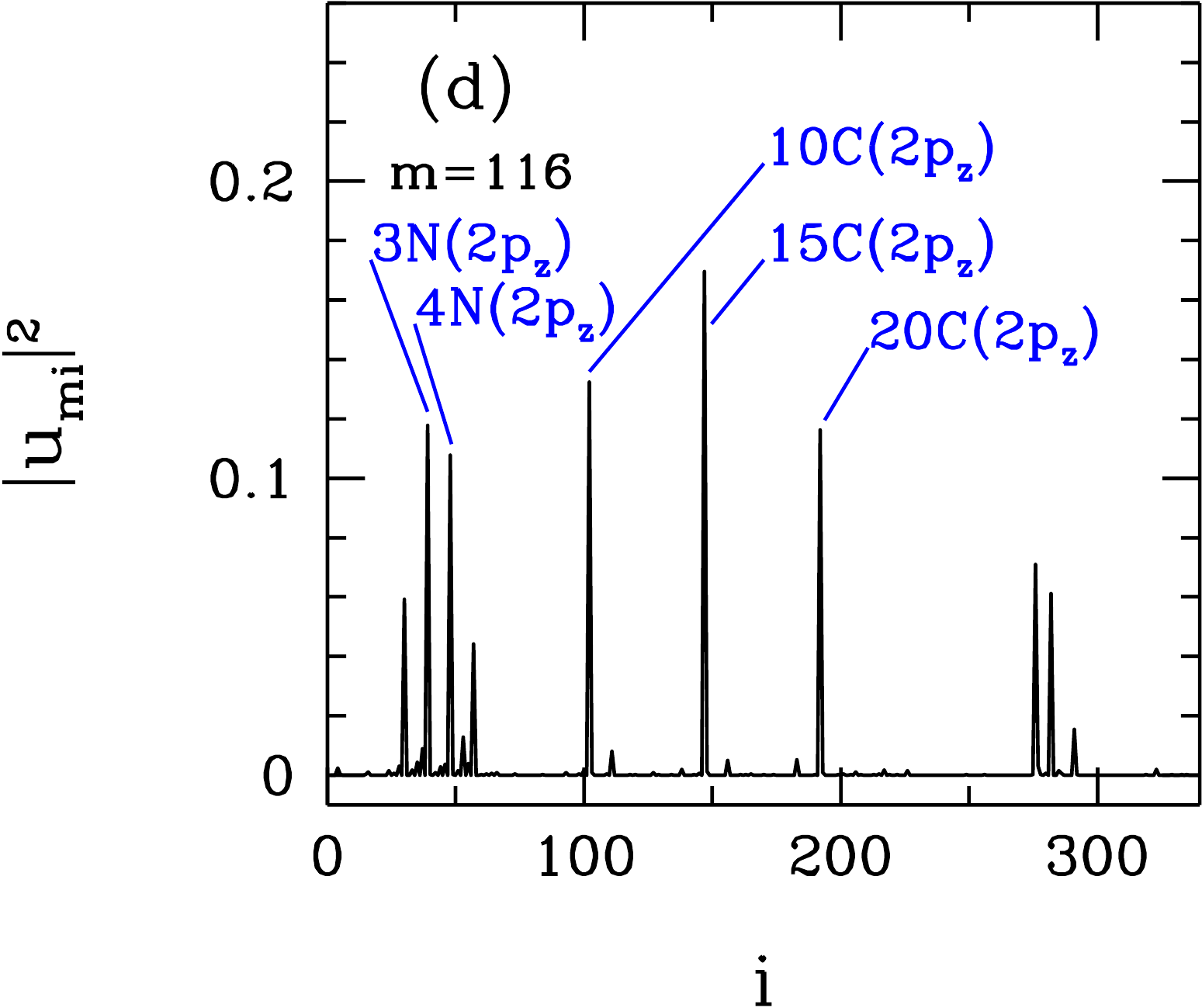} 
\caption{(Color online)  
NAO weight of the $m$'th host eigenstate 
$|u_{mi}|^2=|\langle \tilde{\phi}_i|u_m\rangle|^2$ 
versus the label $i$ of the host NAO
for (a) $m=111$, (b) 112, (c) 114, and (d) 116.
The maximum value for index $i$ is 347, 
which is the total number of the host NAO's. 
}
\label{fig11}
\end{figure} 

We next explore the nature of the in-gap states
originating from the host eigenstates. 
For this purpose,
in Figure 9(a) 
we show the occupation of the host eigenstates $\langle n_m\rangle$
as a function of the host energy $\varepsilon_m$ for 
$\mu=-7.0$ eV and $U=36$ eV.
Here, $\mu$ is located inside the energy gap
found by QMC.
In this figure, 
we notice that there are valence-band host eigenstates 
which are not doubly occupied even though they are located deep below
the chemical potential.
In particular, the host eigenstates with labels $m=114$, 
111, 112 and 116 exhibit the most reduced values of $\langle n_m\rangle$.   
We note that these host eigenstates have the strongest hybridization 
matrix elements with the $3z^2-r^2$ and $xy$ orbitals,
as it was seen in Fig. 4(a). 

Next, 
we increase $\mu$ from $-7.0$ eV 
to $-2.2$ eV, 
and plot in Fig. 9(b)
the increase in $\langle n_m\rangle$, 
$\langle \Delta n_m\rangle = \langle n_m\rangle|_{\mu=-2.2 \eV} - 
\langle n_m\rangle|_{\mu=-7.0 \eV}$,
as a function of the host energy $\varepsilon_m$ 
in Fig. 9(b).
We already know from Fig. 7(b) that,
for $U=36$ eV, the in-gap states become fully
occupied when $\mu=-2.2$ eV.
Here, we see that the largest increase in $\langle n_m\rangle$ takes place
for $m=114$, 111, 116 and 112.
The rest of the host eigenstates exhibit negligible change as $\mu$ 
is increased from $-7.0$ eV to $-2.2$ eV.
Hence, 
the impurity bound states derive mainly from the 
$m=114$, 111, 116 and 112 host eigenstates. 
In turn, within the rigid band picture, 
it is possible to state that the host eigenstates 
$m=114$, 111, 116 and 112 have significant amount of 
single-particle spectral weight between the 
energies $-7.0 \eV$ and $-2.2 \eV$.

In order to gain insight into the real-space structure
of the in-gap host eigenstates located in the interval 
between $-7.0 \eV$ and $-2.2 \eV$,
we study the NAO composition of these host eigenstates. 
However, 
first,
in Fig. 10 we discuss the geometrical structure of 
Im-[Co$^{\rm III}$(corrin)]-CN$^+$.
Fig. 10(a) shows a top view of the corrin ring around the Co atom
in Im-[Co$^{\rm III}$(corrin)]-CN$^+$.
Here, we have only shown the C and N atoms around Co,
and have not included the H atoms. 
According to our coordinate system, 
which is the same as that of the Gaussian program, 
the corrin ring lies in the 
$xy$ plane, and the $x$ and $y$ axes make $45$ degrees with the  
Co-N bond directions. 
Figure 10(b) shows the side view of 
Im-[Co$^{\rm III}$(corrin)]-CN$^+$.
Here, 
we see the imidazole ring attached to Co below the corrin plane 
and the CN axial ligand attached to Co above the corrin plane. 
Here, 
the C and N atoms in the axial ligand are labelled as 30C and 31N
as in the notation of the Gaussian program. 

Figure 11(a) shows the NAO composition of the $m=111$'th host eigenstate.
In particular, here we plot the NAO weight of the 
$m=111$'th host eigenstate defined as 
\begin{equation}
|u_{mi}|^2=|\langle \tilde{\phi}_i|u_m\rangle|^2
\end{equation}
versus the label $i$ of the host NAO state.
Figure 11(b)-(d) show similar results for the $m=112$, 114 and 116'th host states. 
In these figures,
we see that the $m=111$ and 112'th states have large weight from the 
$2p\pi$ NAO's of the 
CN axial ligand. 
Similarly, 
the $m=114$ host state mainly consists of the 
$2p\sigma$ NAO's of the CN axial ligand.
On the other hand,
the $m=116$ host state consists of the 
$2p\sigma$ NAO's on the 
10'th, 15'th and 20'th C sites as well as N sites 
located around Co in the corrin plane
as sketched in Fig. 10(a).
Thus in Figs. 10 and 11,
we see the real-space composition around the Co site 
of the in-gap host states induced in the interval 
$-7.0 \eV \ltsim \mu \ltsim -2.2 \eV$ by the presence of Co.

In Fig. 12, 
we show results on the occupation number 
and the magnetic moment for the
same host eigenstates $m=111$, 112, 114 and 116.
Figure 12(a) shows the occupation number $\langle n_m\rangle$ versus $\mu$
for these host eigenstates.
We observe that 
a sharp increase in $\langle n_m\rangle$ occurs at 
$\mu \approx \tilde{\varepsilon}_m$
in the valance band.
We see that
the single-particle spectral weight of the
$m=111$'th host state is broadened in the valence band. 
In addition, 
a step-like increase exists inside the energy gap
at $\approx -5.5$ eV,
above which this host state reaches double occupancy.
Similar $\mu$ dependencies are observed for the others. 
The host state $m=112$ also exhibits a jump at $\approx -5.5 \eV$,
while for $m=114$ and 116 the jump occurs at $\approx -4 \eV$.
The jumps in $\langle n_m\rangle$ are largest for 
$m=114$ and 111.

\begin{figure}[tb]
\includegraphics[width=7cm]{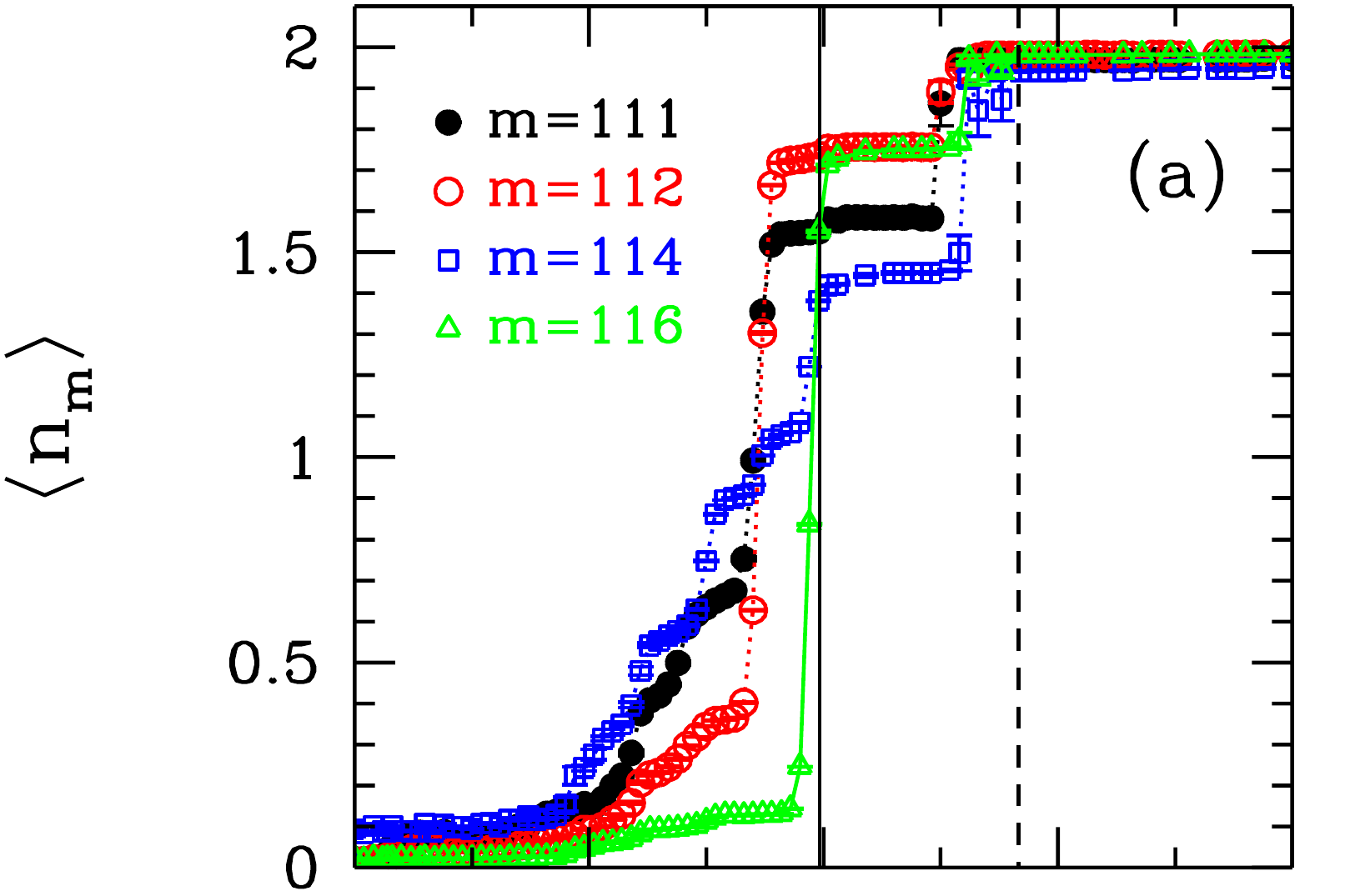}
\includegraphics[width=7cm]{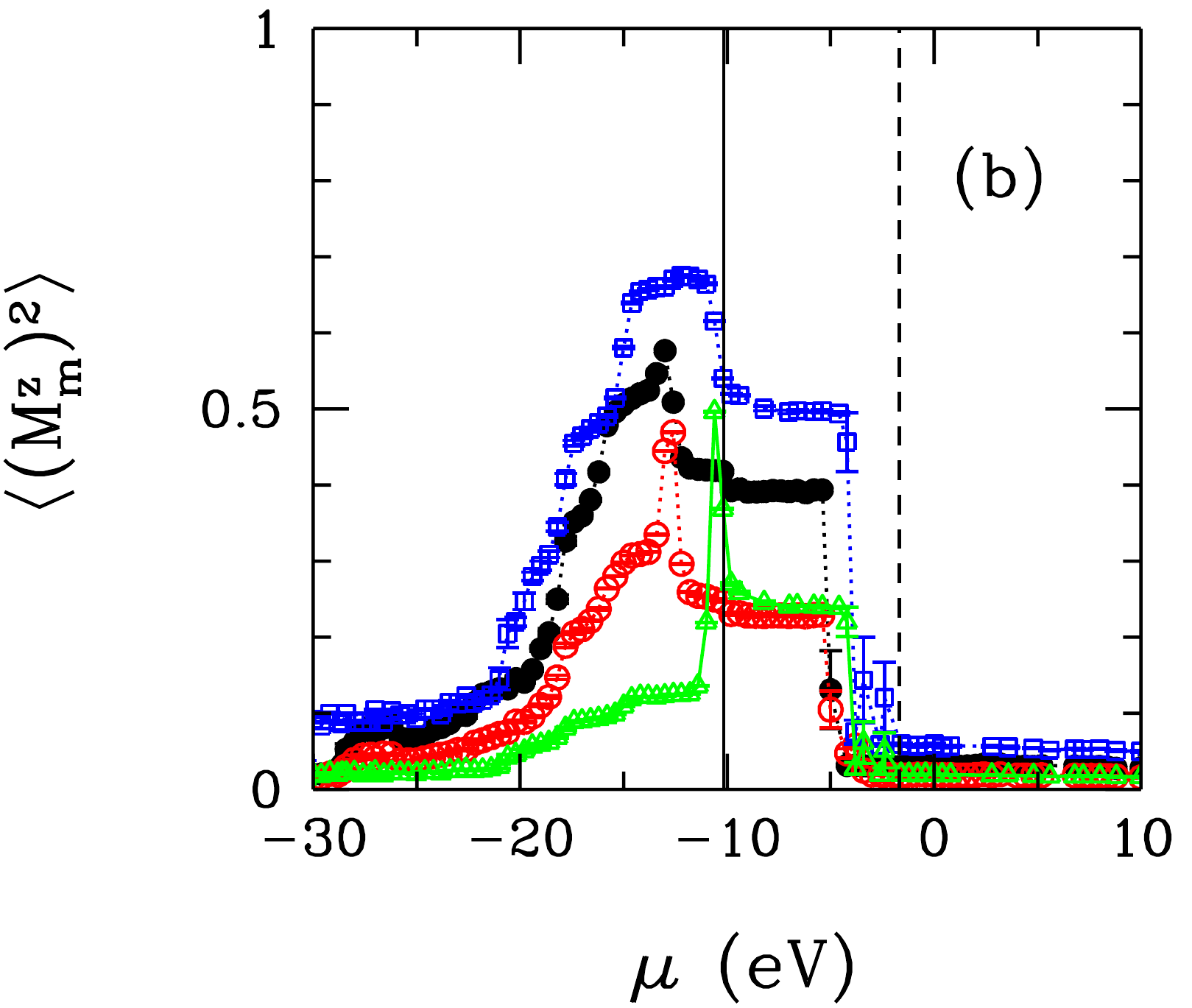} 
\caption{(Color online)  
(a) Occupation of the $m$'th host eigenstate $\langle n_m\rangle$ versus $\mu$.
(b) Square of the magnetic moment of the $m$'th host eigenstate 
$\langle (M_m^z)^2 \rangle$ versus $\mu$.
Here, 
the vertical solid and dashed lines denote the HOMO and the LUMO levels, 
respectively.
In addition,
these results are for $U=36$ eV.
}
\label{fig12}
\end{figure} 

Figure 12(b) shows the square of the magnetic moment 
$\langle (M^z_m)^2\rangle$ for the same host eigenstates.
We observe that significant size moments develop for these host states until the 
in-gap states become occupied.
As the in-gap host states are filled and
they approach double-occupancy, 
the magnetic moments decrease rapidly.

Finally, 
we note that the states induced in the interval 
$-10 \eV \ltsim \varepsilon \ltsim -8 \eV$
originate mainly from the Co $t_{2g}$-like states.
In Figs. 7(a)-(c), 
we observe that in this interval 
there is sufficient spectral weight to accommodate 3 electrons,
of which about 2.5 electrons have the Co $t_{2g}$-like
character. 
The remaining 0.5 electrons originate from 
the $m=115$ host state, which mainly consists of 
the $2p\pi$ orbitals of the C and N atoms in the corrin ring.
We do not show the composition of the $m=115$'th host state
in order not to increase the length of the manuscript. 

\subsection{QMC results on the Co($\bm{3d_{\nu}}$)-host magnetic correlations}

In Fig. 13, we discuss the magnetic 
correlations between the Co($3d_{\nu}$) electrons 
and the host eigenstates $m=111$, 112, 114 and 116.
In particular, these figures show 
the magnetic correlation function 
$\langle M^z_{\nu} M^z_m \rangle$
between the Co($3d_{\nu}$) state 
and the $m$'th host eigenstate
plotted as a function of $\mu$.

\begin{figure}[tb]
\includegraphics[width=5.5cm]{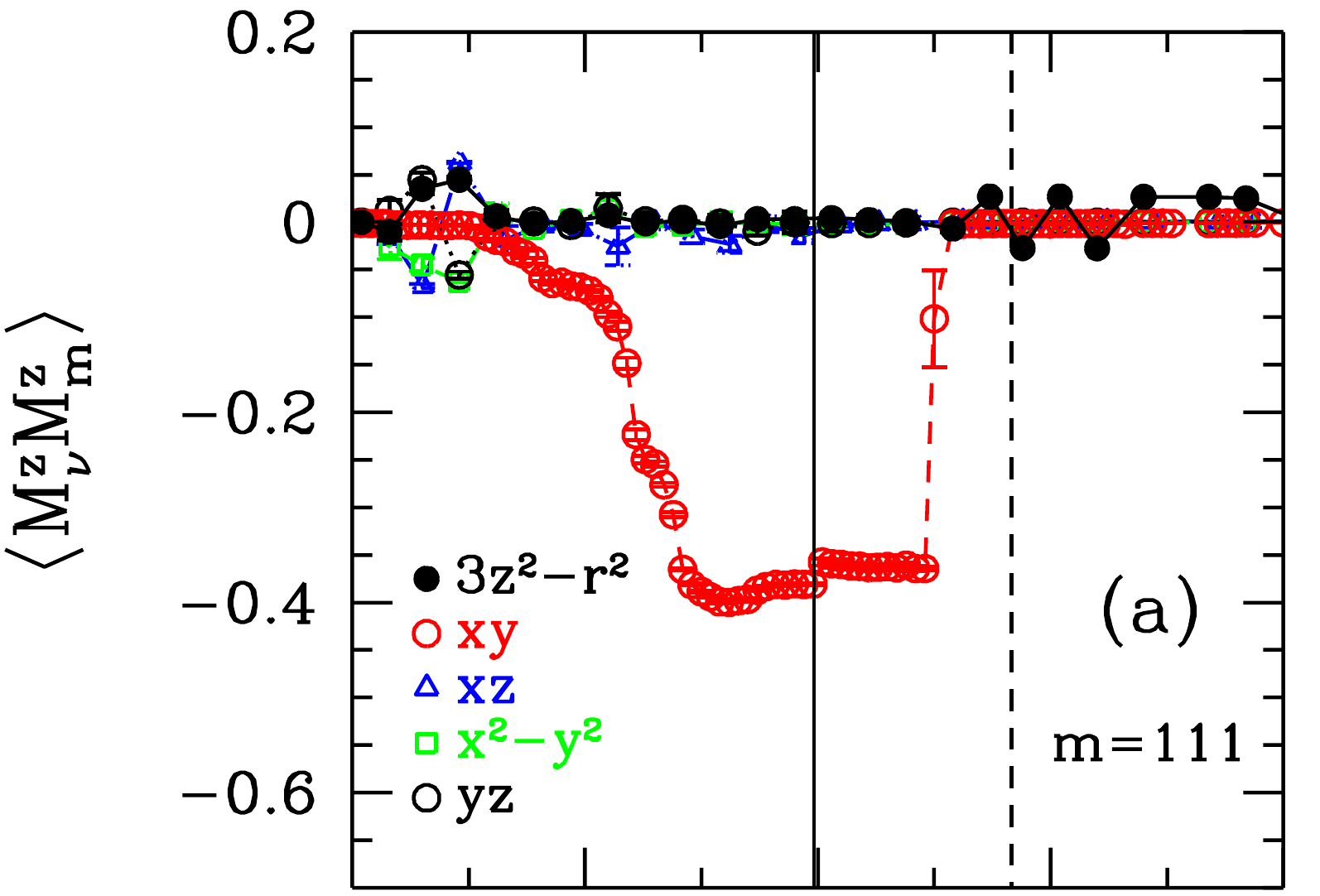}
\includegraphics[width=5.5cm]{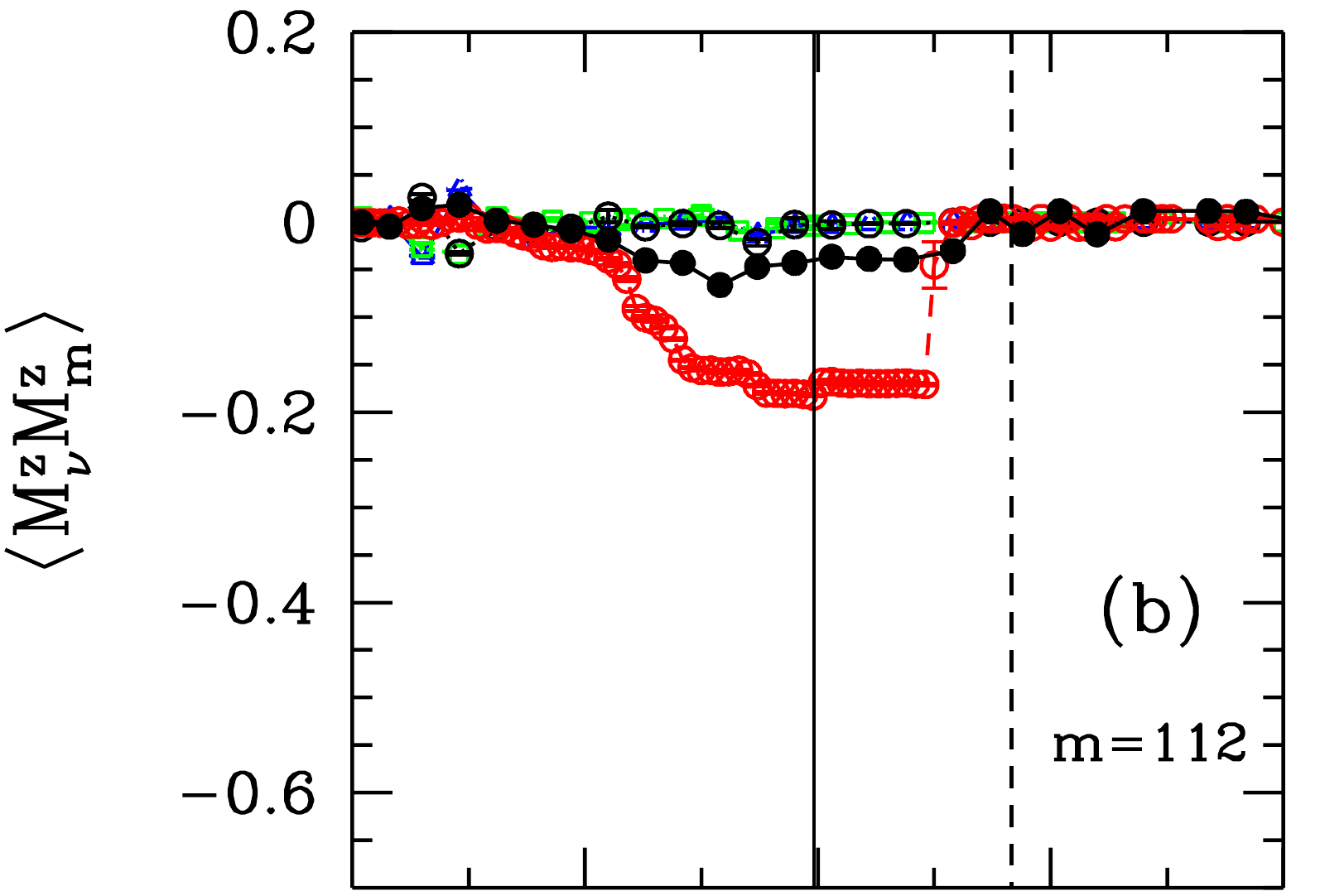}  
\includegraphics[width=5.5cm]{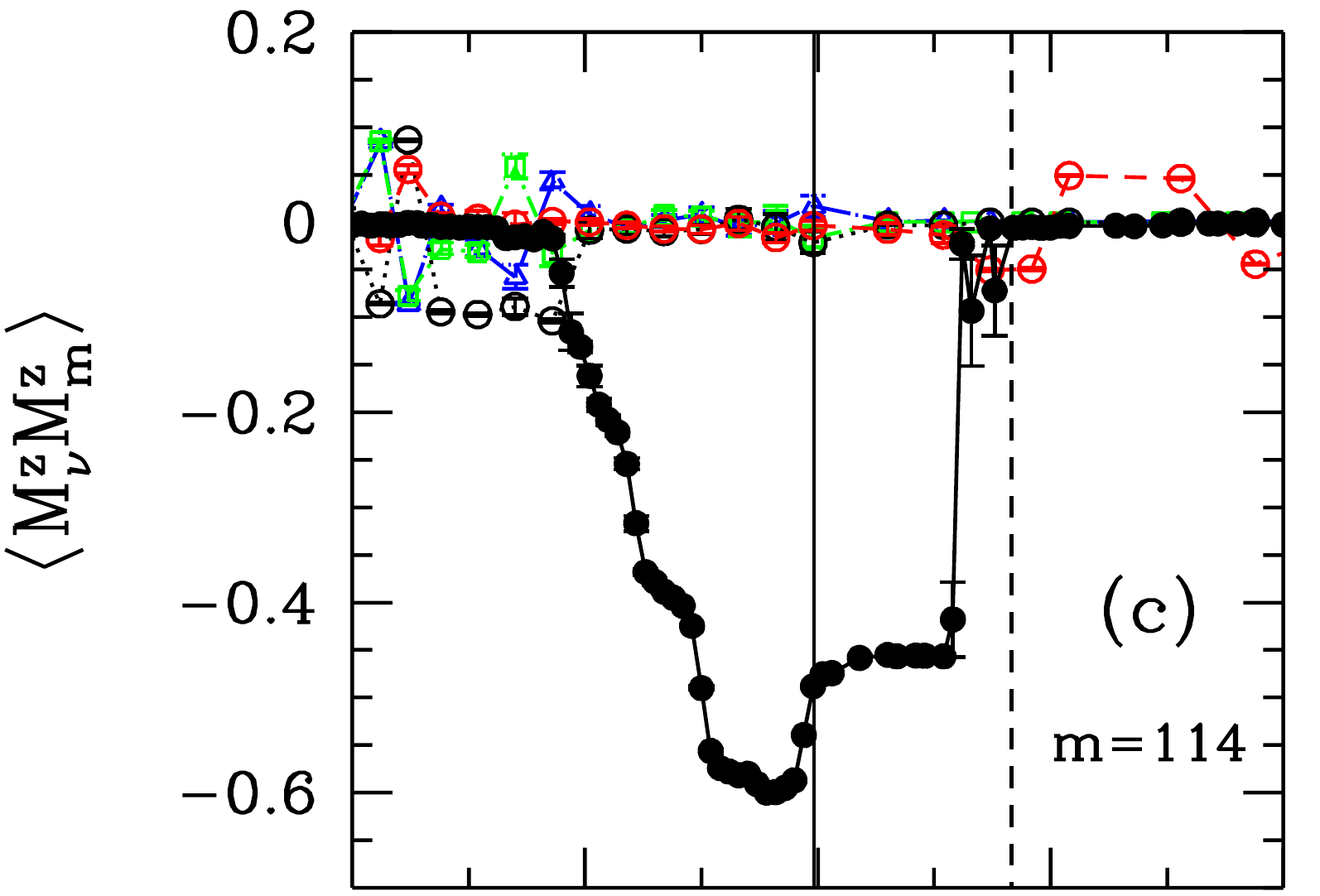}
\includegraphics[width=5.5cm]{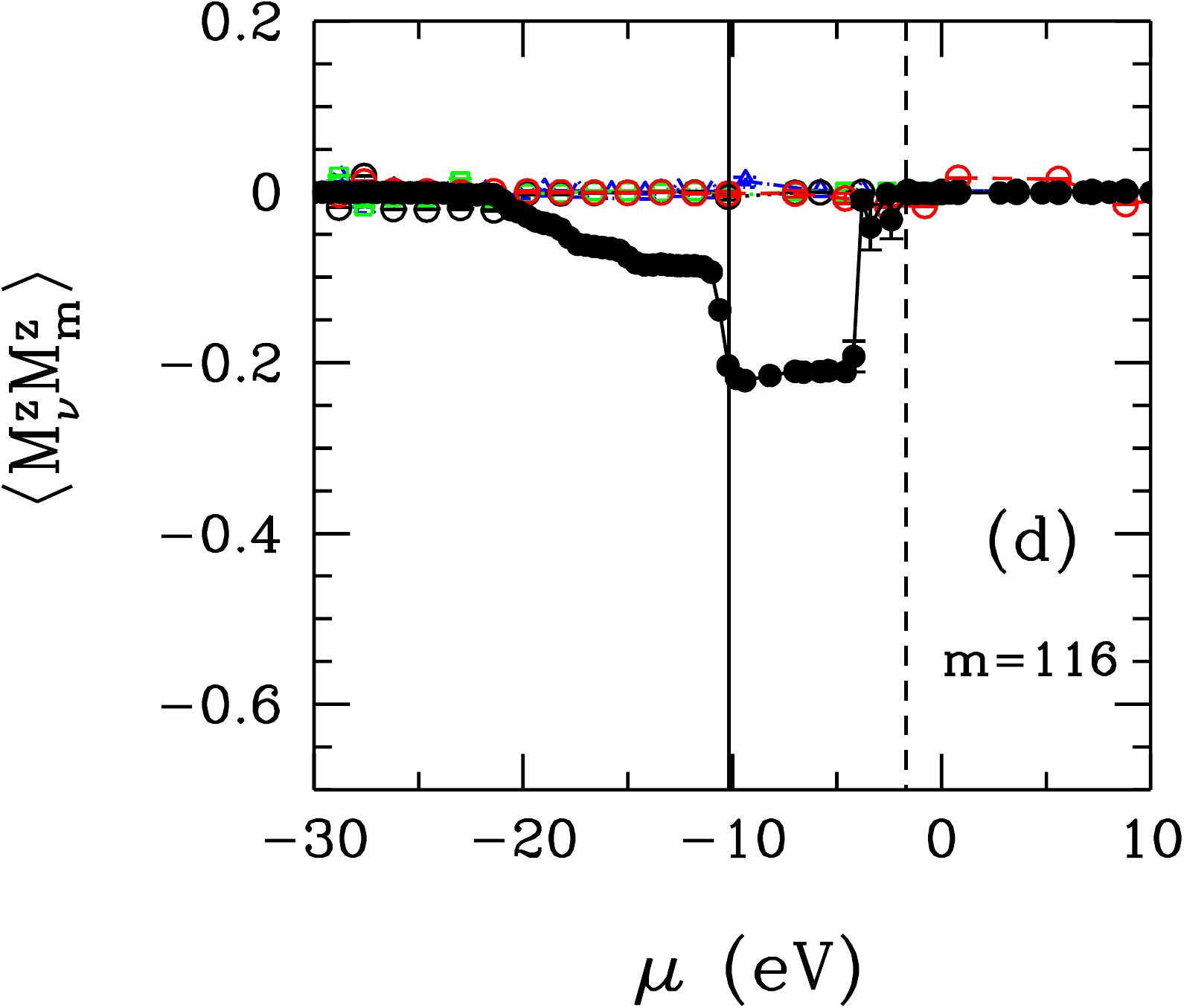}
\caption{(Color online)  
Magnetic correlation function
$\langle M_{\nu}^z M_m^z \rangle$
between the $m$'th host eigenstate and 
the various the Co($3d_{\nu}$) NAO's.
Here, results are shown for host eigenstates 
(a) $m=111$, (b) 112, (c) 114, and (d) 116.
In addition, 
the vertical solid and dashed lines denote the HOMO and the LUMO levels, 
respectively.
These results are for $U=36$ eV.
}
\label{fig13}
\end{figure} 

In Figs. 13(a) and (b)
we observe that 
the Co($3d_{xy}$) NAO 
develops antiferromagnetic correlations with the 
$m=111$ and 112'th host states.
These antiferromagnetic correlations diminish rapidly as the in-gap
state located at $\approx -5.5 \eV$ becomes occupied.
The $m=111$ and 112'th states do not exhibit magnetic correlations 
with the other Co($3d_{\nu}$) orbitals.
In Figs. 13(c) and (d) we observe that 
the Co($3d_{3z^2-r^2}$) NAO 
develops antiferromagnetic correlations with the 
$m=114$ and 116'th host states.
In Figs. 11(a)-(c), we have seen that the 
$m=111$, 112 and 114'th states contain large weight of 
the NAO's localized at the CN axial ligand.
Thus we deduce that the electrons at the CN axial ligand
develop antiferromagnetic correlations with the 
Co $e_g$ orbitals. 
These antiferromagnetic correlations diminish rapidly 
when the impurity bound states become filled 
with electrons. 
Hence, as 
Im-[Co$^{\rm III}$(corrin)]-CN$^+$ 
goes from the $N_{\rm el}=238$ state to $N_{\rm el}=239$ 
and 240 states, the antiferromagnetic correlations between the CN ligand 
and the Co $e_g$ electrons disappear rapidly.
We note that the $m=116$ host state contains spectral weight from the 
$2p\sigma$ NAO's at the C and N sites in the corrin ring as seen in 
Fig. 11(d). 
Figure 13(d) shows that these NAO's also develop antiferromagnetic correlations
with the Co($3d_{3z^2-r^2}$) orbital. 

It is important to note that we identify the in-gap induced states
at $\approx -5.5 \eV$ and $\approx -4 \eV$as impurity bound states
because of the $\mu$ dependencies discussed in Figs. 6, 7, 12 and 13. 
Similar behaviour is obtained in studying the impurity bound states found 
in the DMS materials \cite{Ichimura,Bulut,Tomoda}. 

We have also studied the magnetic correlations between 
the magnetic moments forming at the 
Co $3d_{3z^2-r^2}$ and $3d_{xy}$ orbitals.
We find that these orbitals exhibit only weak magnetic correlations.
This is different than the case for the DMS materials,
where an indirect ferromagnetic coupling is generated between 
two impurity moments due to their antiferromagnetic coupling to the 
same continuum of host electrons \cite{Ichimura,Bulut,Tomoda}. 
Here, 
the Co $3d_{3z^2-r^2}$ and $3d_{xy}$ orbitals 
do not have strong hybridization   
with the same host eigenstates 
as seen in Subsection II.E.
In modelling the DMS materials, 
a continuum of host states 
with constant hybridization is used,
while for CNCbl the host eigenstates have discrete energy levels
with varying hybridization matrix elements. 
Hence,
an indirect ferromagnetic coupling is not generated.
However,
the inclusion of the Hund's coupling can 
generate a ferromagnetic coupling
between the magnetic moments at 
the Co $3d_{3z^2-r^2}$ and $3d_{xy}$ NAO's. 

It would be interesting to probe experimentally 
the antiferromagnetic correlations 
found in the HF+QMC calculations.
It is known that Cbl exhibits weak diamagnetism 
\cite{Grun,Diehl},
which has been intepreted as Co in Cbl having a low spin state. 
This is also partly responsible for the notion that 
the interaction effects are not important in Cbl.
In the HF+QMC results, 
we see that magnetic moments can develop 
in the Co $e_g$ orbitals and in the host eigenstates coupled to these orbitals.
It is possible to calculate the magnetic susceptibility 
within HF+QMC in order to extract the effective moment size
for Co and the whole molecule.
But such a calculation would be more meaningful after 
including the Hund's coupling. 

\section{Discussion}

In this section, 
we discuss the meaning and implications of the HF+QMC data
shown in Section III. 
For this purpose,
in Subsection IV.A, we compare the HF+QMC data with 
the DFT calculations on the same truncated molecule 
Im-[Co$^{\rm III}$(corrin)]-CN$^+$.
We find that there are important differences between
the results of these two approaches.
In particular,
while the HF+QMC calculations for the Haldane-Anderson model
find that the LUMO state
corresponds to an impurity bound state, 
this is not the case for the DFT results.
In Subsection IV.B, 
we compare the HF+QMC data with the photoabsorption 
experiments on CNCbl.
Here,
we discuss the nature of the 
lowest excited states in the photoabsorption spectrum.
The HF+QMC data suggest that the lowest-energy excitations
are dominated by electron transfer from the 
Co $t_{2g}$-like orbitals to the impurity bound states which 
contain spectral weight mainly from the 
CN axial ligand and the corrin ring. 
In Subsection IV.C, 
we note that the DFT+QMC method may also be applied to the same problem. 
In Subsection IV.D,
we emphasize the importance of including 
the inter-orbital Coulomb repulsion and 
the Hund's coupling.

\subsection{Comparison of the HF+QMC data with the DFT results}

Various DFT based calculations have been performed for Cbl
\cite{Jensen2001,Andruinov,Ouyang2003,Kurmaev,Ouyang2004,Rovira,Mebs,Solheim,
Kornobis2011,Reiq,Kornobis2013}.
In this subsection, we compare the HF+QMC data obtained 
for Im-[Co$^{\rm III}$(corrin)]-CN$^+$
by using the Haldane-Anderson model with the results  
obtained from the density functional theory (DFT) calculations
on the same truncated molecule.
We note that 
these DFT results are similar to the results of the 
time-dependent DFT calculations 
on the same system \cite{Kornobis2011}.

\begin{figure}[t]
\includegraphics[width=7cm]{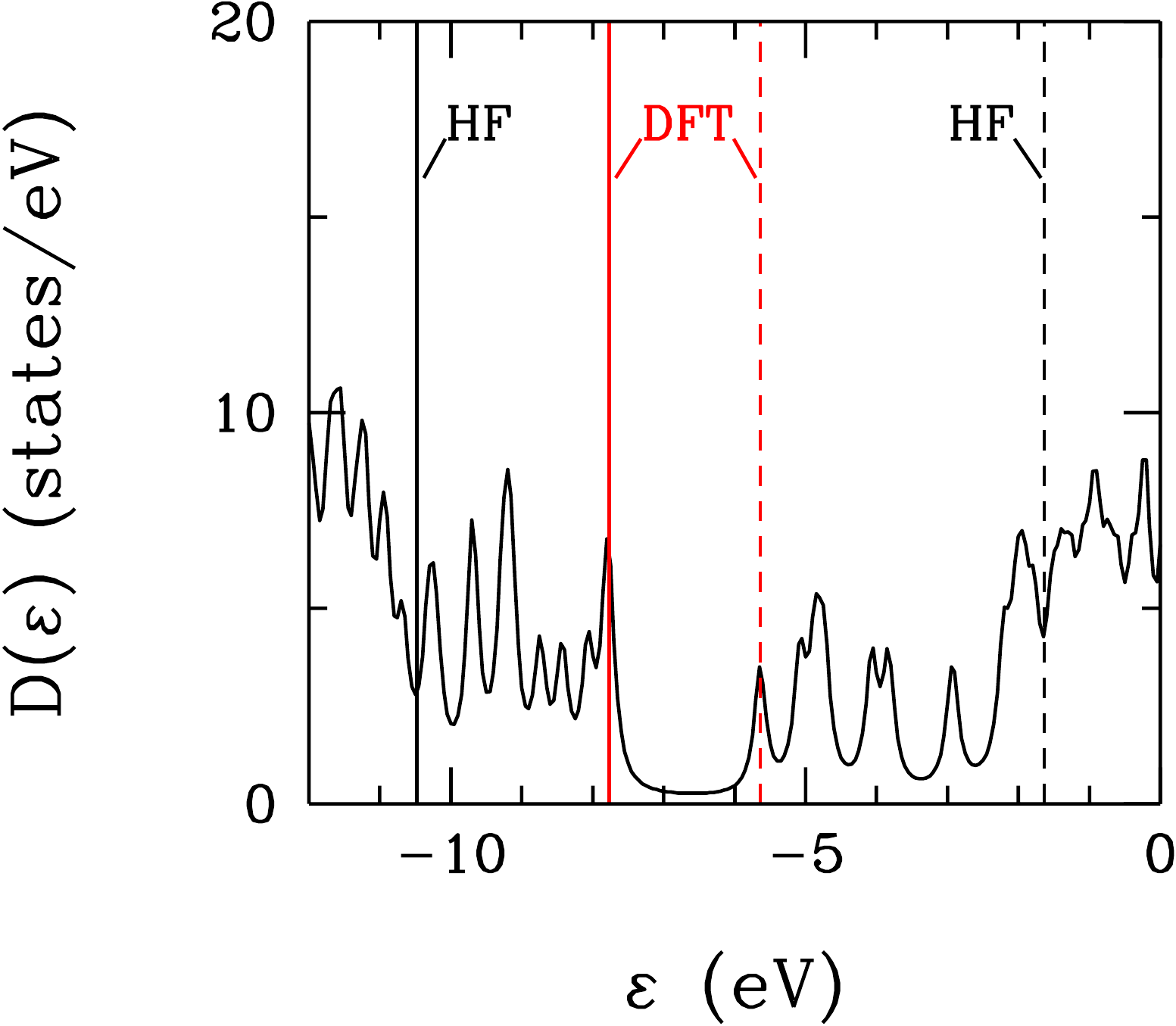} 
\caption{(Color online) 
DFT results on the 
total density of states $D(\varepsilon)$ 
of Im-[Co$^{\rm III}$(corrin)]-CN$^+$
obtained by using the Gaussian program 
with the BP86 exchange potentials and the 6-31G basis functions.
The solid and dashed vertical red lines denote the 
HOMO and LUMO levels obtained by the DFT calculation.
Similarly, the solid and dashed vertical black lines denote the 
HOMO and LUMO levels obtained by the Hartree-Fock calculation
described in Section II. 
Here, 
we see that in the DFT calculations 
additional spectral weight is induced in the intervals
$-10.5 \eV \leq \varepsilon \leq -7.8 \eV$ and 
$-5.7 \eV \leq \varepsilon \leq -1.8 \eV$ 
when compared with the HF results. 
}
\label{fig14}
\end{figure} 

Figure 14 shows the total density of states 
$D(\varepsilon)$ versus $\varepsilon$ 
for Im-[Co$^{\rm III}$(corrin)]-CN$^+$ 
obtained by using the Gaussian program \cite{Gaussian} 
with the 6-31G basis functions
and the BP86 exchange potentials \cite{BP86}.
Here, the solid and dashed vertical red lines denote the 
HOMO and LUMO levels obtained by DFT.
According to the DFT results
the energy gap is $\approx 2.2 \eV$,
which is in agreement with the experimental results.
The vertical black lines, on the other hand, denote the 
HOMO and LUMO levels obtained by the HF calculations
shown in Fig. 3(a). 
Comparing the HF results on $D(\varepsilon)$ shown in Fig. 3(a) 
with the DFT results of Fig. 14, 
we observe that the HF HOMO 
and LUMO levels located at $-10.5 \eV$ and $-1.7 \eV$ move to
$-7.8 \eV$ and $-5.6 \eV$, respectively. 
Hence,
in comparison to the HF results,
the DFT calculation yields additional spectral weight 
in the energy intervals 
$-10.5 \eV \ltsim \varepsilon \ltsim -7.8 \eV$ and 
$-5.6 \eV \ltsim \varepsilon \ltsim -1.7 \eV$. 

Next, we compare the DFT results on $D(\varepsilon)$, 
Fig. 14, with the single-particle spectral weight deduced from the 
HF+QMC data on the total electron number 
$\langle n_T\rangle$ versus $\mu$ shown in Fig. 7(c)
for $U=36 \eV$. 
We observe that the locations of the HOMO 
and LUMO levels obtained by DFT and 
HF+QMC are comparable. 
In addition, 
the DFT result for the total number of the electrons in the 
Co($3d_{\nu}$) NAO states is 7.4,
which is similar to the HF+QMC result seen in Fig. 8.
So, the DFT and the HF+QMC results for the energy gap and 
$\langle n_d\rangle$
are comparable to each other as well as to the 
the experimental results.
However, we will see in the following that 
there are important differences between the DFT and 
HF+QMC results when the overall distribution 
of the single-particle spectral weight near the semiconductor gap edges 
is considered.

In the HF+QMC data presented in Section III for $U=36 eV$, 
we have seen that
new states are induced in the energy interval 
$-10 \eV \ltsim \varepsilon \ltsim -8 \eV$ which originate mainly
from the doubly-occupied Co $3d_{xz}$, $3d_{x^2-y^2}$ and $3d_{yz}$ 
orbitals. 
This is because 
$\tilde{\varepsilon}_{d\nu}+U$ 
for the Co $t_{2g}$-like orbitals corresponds to this energy interval. 
In the same interval, there also exist 
smaller amount of host states due to hybridization
with the Co $t_{2g}$-like states.
On the other hand, 
the states induced in the interval 
$-5.5 \eV \ltsim \varepsilon \ltsim -2 \eV$ 
in the HF+QMC data originate from 
the Co $3d_{3z^2-r^2}$ and $3d_{xy}$ states.
These states are similar in nature to the 
impurity bound states found in the mean-field and QMC studies of the 
Haldane-Anderson model for the DMS materials 
\cite{Ichimura,Bulut,Tomoda}. 
We have also seen that host states, 
which are mainly originating from the CN axial ligand,
are induced in the energy interval 
$-5.5 \eV \ltsim \varepsilon \ltsim -2 \eV$.
In fact, here,
the amount of the single-particle spectral weight 
originating from the host states is larger than that from 
the $e_g$-like states. 
From Figs. 7(a) and (b),
we observe that in the interval $-5.5 \eV \ltsim \varepsilon \ltsim -2 \eV$
the host states accommodate about 1.6 electrons while 
the $e_g$-like states accommodate about 0.4 electrons. 

\begin{table*}[t]
\addtolength{\tabcolsep}{0.6pt} 
\renewcommand{\arraystretch}{1.2} 
\begin{tabular}{|cc|c|ccccc|cc|}
\hline
\multirow{2}{*}{\bf Molecular orbital}&
\multirow{2}{*}{$\bm n$}&
\multirow{2}{*}{${\bm{E_{n}}}$  \bf (eV)}&
\multicolumn{1}{|c}{}&
\multicolumn{1}{c} {}&
\multicolumn{1}{c} {\bf Co(${\bm {3d_{\nu}}}$)} &
\multicolumn{1}{c} {}&
\multicolumn{1}{c|}{}&
\multicolumn{1}{c}{${\bm {30}}$\bf C}&
\multicolumn{1}{c|}{${\bm {31}}$\bf N} \\
& & & {$\bm {3z^2-r^2}$} & {$\bm {xy}$} & {$\bm {xz}$} & {$\bm {x^{2}-y^{2}}$} & {$\bm {yz}$} & 
{\footnotesize {\bf (summed over $\bm {2p}$'s)}} &
{\footnotesize {\bf (summed over $\bm {2p}$'s)}}  \\\hline
\bf HOMO-10 & \bf 109 & -10.22 & 0.0  & 0.0  & 0.0      & 0.0     & 0.02     & 0.01    & 0.01     \\\hline
\bf HOMO-9  & \bf 110 & -9.70  & 0.0  & 0.0  &\bf 0.10  & 0.0     & 0.01     &\bf 0.14 & \bf 0.14 \\\hline
\bf HOMO-8  & \bf 111 & -9.69  & 0.0  & 0.0  & 0.01     & 0.0     &\bf 0.16  &\bf 0.25 & \bf 0.24 \\\hline
\bf HOMO-7  & \bf 112 & -9.27  & 0.04 & 0.0  & 0.0      & 0.0     & 0.0      &\bf 0.09 & \bf 0.49 \\\hline
\bf HOMO-6  & \bf 113 & -9.19  & 0.01 & 0.0  & 0.0      & 0.0     & 0.0      &\bf 0.14 & \bf 0.28 \\\hline
\bf HOMO-5  & \bf 114 & -9.15  & 0.0  & 0.0  & 0.0      & 0.0     & 0.0      & 0.0     & 0.01     \\\hline
\bf HOMO-4  & \bf 115 & -8.74  & 0.0  & 0.0  & 0.0      &\bf 0.08 &\bf 0.15  &\bf 0.09 & \bf 0.21 \\\hline
\bf HOMO-3  & \bf 116 & -8.43  & 0.0  & 0.02 & 0.0      &\bf 0.83 & 0.02     & 0.01    & 0.03     \\\hline
\bf HOMO-2  & \bf 117 & -8.07  & 0.0  & 0.01 & \bf 0.66 & 0.0     & 0.0      & 0.02    &\bf 0.12  \\\hline
\bf HOMO-1  & \bf 118 & -7.80  & 0.02 & 0.0  & 0.0      & 0.0     & \bf 0.38 & 0.01    & 0.03     \\\hline
\bf HOMO & \bf 119  & -7.77 & 0.01  & 0.0  & 0.01 & 0.01 &\bf 0.10 & 0.01 & 0.02 \\\hline\hline
\bf LUMO & \bf 120  & -5.64 & 0.0   & 0.0  & 0.0  & 0.0  &\bf 0.05 & 0.0  & 0.0  \\\hline
\bf LUMO+1 &  \bf 121 & -5.08  & 0.03     & \bf 0.49 & 0.02 & 0.01 & 0.0  & 0.0     & 0.0  \\\hline
\bf LUMO+2 &  \bf 122 & -4.87  & \bf 0.38 & \bf 0.07 & 0.01 & 0.0  & 0.01 &\bf 0.06 & 0.02 \\\hline
\bf LUMO+3 &  \bf 123 & -4.74  & \bf 0.09 & 0.01     & 0.04 & 0.0  & 0.0  & 0.01    & 0.01 \\\hline
\bf LUMO+4 &  \bf 124 & -4.07  & 0.0      & 0.01     & 0.0  & 0.0  & 0.0  & 0.0     & 0.0  \\\hline
\bf LUMO+5 &  \bf 125 & -3.83  & 0.0      & 0.0      & 0.0  & 0.0  & 0.0  & 0.0     & 0.0  \\\hline
\bf LUMO+6 &  \bf 126 & -2.93  & 0.0      & 0.0      & 0.0  & 0.0  & 0.0  & 0.0     & 0.0  \\\hline
\bf LUMO+7 &  \bf 127 & -2.20  & 0.0      & 0.0      & 0.0  & 0.0  & 0.0  & 0.01    & 0.01 \\\hline
\bf LUMO+8 &  \bf 128 & -2.03  & 0.0      & 0.0      & 0.0  & 0.0  & 0.0  & 0.01    & 0.01 \\\hline
\bf LUMO+9 &  \bf 129 & -1.93  & 0.0      & 0.0      & 0.0  & 0.0  & 0.01 & 0.01    & 0.01 \\\hline
\bf LUMO+10 & \bf 130 & -1.78  & 0.0      & 0.0      & 0.0  & 0.0  & 0.0  & 0.0     & 0.0  \\\hline
\end{tabular}
\caption{
DFT results on the NAO spectral weights $|\tilde{C}_{ni}|^2$
of the molecular orbitals 
near the semiconductor gap edges 
for Im-[Co$^{\rm III}$(corrin)]-CN$^+$.
Here, $n$ is the index for the molecular orbital, and $i$ is for the NAO. 
The first two columns denote the molecular orbital and the corresponding index.
The following five columns denote how much Co($3d$)
NAO spectral weight exists 
in the $n$'th molecular orbital.
Similarly,
the last two columns denote the amount of the total $2p$ spectral weight 
at the C and N atoms in the CN axial ligand.
For convenience,
when $|\tilde{C}_{ni}|^2 \geq 0.05$,
the numbers are printed in bold face.
}
\label{tab1}
\end{table*}

Now, we discuss the nature and origin of the states forming 
in the energy intervals 
$-10 \eV \ltsim \varepsilon \ltsim -8 \eV$
and
$-5 \eV \ltsim \varepsilon \ltsim -2 \eV$ 
in the DFT calculations. 
For this purpose, 
we present DFT results in Table I
on the NAO spectral weights of the molecular orbitals 
located in these intervals. 
Using the DFT data, 
we expand the one-electron molecular orbitals
$|\psi_n\rangle$ in terms of the NAO's 
of the whole molecule including
both the host and the Co($3d$) states, 
\begin{equation}
|\psi_n\rangle = \sum_i^N \tilde{C}_{ni} | \tilde{\phi}_i\rangle.
\end{equation}
Here, the sum over the NAO index $i$ goes upto N=347.
Table I shows DFT results on $|\tilde{C}_{ni}|^2$ 
for the molecular orbitals which are located
right below and above the DFT semiconductor gap.
In particular, 
here results are shown for the molecular orbitals
with index $109 \leq n \leq 130$, 
which includes the states located in the intervals
$-10.22 \eV \leq \varepsilon \leq -7.77 \eV$
and
$-5.64 \eV \leq \varepsilon \leq -1.78 \eV$.
We are interested in how much Co($3d$) NAO weight 
these molecular orbitals contain.
We are also interested in how much NAO weight 
of the CN axial ligand is included
in these energy intervals. 
We will compare them with the HF+QMC data. 
It is for these purposes that the data in Table I
are presented.

The first and the second columns of Table I denote the location 
of a given molecular orbital with respect to the semiconductor gap
and the molecular orbital index $n$.
The third column shows the corresponding molecular energy $E_n$.
The following five columns show $|\tilde{C}_{ni}|^2$
for the Co($3d$) NAO states. 
The last two columns show the NAO weights summed over the three 
$2p$ orbitals, 
$\sum_{2p}  |\tilde{C}_{ni}|^2$,
for the C and N atoms of the CN axial ligand.
We note that the C and N atoms of the CN axial ligand are labelled as
30C and 31N following the notation of the Gaussian program\cite{Gaussian}.
 
Here, we observe that the molecular orbitals with the
index $n=115$ through 118 contain
significant amount of spectral weight from the 
Co $3d_{xz}$, $3d_{x^2-y^2}$
and $3d_{yz}$ NAO's. 
In comparison,
the HOMO state ($n=119$) and the LUMO state ($n=120$) contain 
smaller amount of Co $3d_{yz}$ NAO weight.
On the other hand,
the molecular orbitals $n=121$ and 122 contain significant amount of  
Co $3d_{xy}$ and $3d_{3z^2-r^2}$ NAO weights, respectively. 

Next, 
we compare these DFT results with the HF+QMC data from Section III. 
In the DFT results,
we see that in the interval $-10 \eV \ltsim \varepsilon \ltsim -8 \eV$ 
there is spectral weight originating from the 
Co $3d_{x^2-y^2}$,
$3d_{xz}$ and $3d_{yz}$ NAO states.
This agrees with the HF+QMC data.
However, 
the amounts of the single-particle spectral weight
contained in this interval are different.
According to the DFT results,
in this interval the Co $3d_{xz}$, $3d_{x^2-y^2}$
and $3d_{yz}$ NAO's 
accommodate 1.56, 1.84 and 1.88 electrons, respectively,
including spin. 
On the other hand, 
as seen in Fig. 6(a),
the HF+QMC approach finds that each of these orbitals contains 
in this interval
about 0.9 electrons including spin.
In this case,
the interval $-10 \eV \ltsim \varepsilon \ltsim -8 \eV$
is approximately equal to the value of 
$\tilde{\varepsilon}_{d\nu} + U$.
Hence, the states in this interval correspond mainly 
to the upper Hubbard states of the $t_{2g}$-like NAO's. 
This comparison shows that the DFT calculations
do not yield the splitting of the 
$t_{2g}$-like states into the lower and upper
Hubbard states found by HF+QMC.
 
Next, 
we discuss the DFT results for
the interval $-5 \eV \ltsim \varepsilon \ltsim -2 \eV$,
which contains spectral weight from the 
Co $3d_{xy}$ and $3d_{3z^2-r^2}$ NAO states.
The HF+QMC data also finds spectral weight from the 
Co $3d_{xy}$ and $3d_{3z^2-r^2}$ NAO's in this interval. 
However, the amounts of the spectral weight contained 
in this interval are different. 
According to the DFT results,
the Co $3d_{xy}$ and $3d_{3z^2-r^2}$ NAO's 
contain 1.0 and 1.14 electrons in this interval. 
The HF+QMC yields 0.2 electrons for each of these 
orbitals in the same interval. 

In the last two columns of Table I on the DFT results,
we observe that in the interval 
$-5.5 \eV \ltsim \varepsilon \ltsim -2 \eV$,
there is little amount of NAO spectral weight 
originating from the CN axial ligand. 
In the HF+QMC data, however, there is significant amount of 
spectral weight from the CN axial ligand in this energy interval.
This is because the CN axial ligand
hybridizes strongly with the Co $3d_{xy}$ and
$3d_{x^2-y^2}$ NAO's. 
Hence, in the DFT calculations the states in the interval 
$-5.5 \eV \ltsim \varepsilon \ltsim -2 \eV$
do not arise from an impurity bound state as found in the 
HF+QMC calculations. 

When we compare the DFT and the HF+QMC results,
we find good agreement for 
the magnitude of the HOMO-LUMO gap and
the total occupation number $\langle n_d\rangle$
of the Co($3d$) orbitals.
However, there exist important differences 
in the overall distribution of the single-particle
spectral weight near the semiconductor gap edges.
In particular, 
we see that the DFT calculation
yields much less spectral weight in the interval
$-5.6 \eV \ltsim \varepsilon \ltsim -1.8 \eV$ 
originating from the CN axial ligand. 
In addition, according to DFT , 
the Co($3d_{\nu}$) spectral weight in the interval
$-10 \eV \leq \varepsilon \leq -8 \eV$ 
does not arise from the 
split upper-Hubbard states of the Co $t_{2g}$-like 
orbitals. 

Furthermore, 
in Table I we see that,
in the interval $-10.22 \eV \leq \varepsilon \leq -1.78 \eV$,
there are 22 molecular orbitals, 
which can accommodate a total of 44 electrons. 
On the other hand,
in the HF+QMC data shown in Fig. 7(c),
we observe that there is spectral weight to contain 
only 5 electrons in the interval
 $-10 \eV \ltsim \varepsilon \ltsim -2 \eV$.
Hence, 
the single-particle spectrum in this interval 
is much denser according to the DFT results.

In addition to these, 
we have also performed DFT
calculations by using the B3LYP exchange-correlation functional 
\cite{Becke93,LYP}
instead of the BP86.
We do not present these results here
because of the length of the manuscript.
In this case, the energy gap is
$\approx 3.2 \eV$.
However, 
the distribution of the spectral weight 
near the HOMO and LUMO levels is similar to that 
shown in Table I for BP86.

Furthermore,
we have also performed calculations
by using the local spin density approximation \cite{LSDA} (LSDA) 
for Im-[Co$^{\rm III}$(corrin)]-CN$^+$
with the Gaussian program \cite{Gaussian}.
In that case, 
the energy gap is $\approx 2.1 \eV$, and 
the overall distribution of the single-particle spectral weight
is similar to the DFT results shown in Table I. 
As in the DFT calculations, 
the LSDA approach does not capture the Hubbard 
splitting of the $t_{2g}$-like orbitals,
and the states at the LUMO level 
do not arise from an impurity bound state.

In addition to metalloproteins and metalloenzymes,
the DFT technique is widely applied for studying 
transition-metal-containing organometallic molecules, 
which are used in fields ranging from solar-cell applications 
to hydrogen storage. 
The differences discussed here 
between the DFT and the HF+QMC approaches 
might simply be due to the fact that 
the present HF+QMC approach is too simple to describe the 
electronic state of CNCbl. 
If that is not the case, 
then
these comparisons suggest that caution is necessary 
in interpreting the results of the DFT technique
when used for studying
organometallic molecules containing transition metal atoms. 

In addition to these, 
the DFT and HF+QMC results differ on the 
nature of the electronic correlations.
The HF+QMC technique finds that antiferromagnetic correlations exist between 
the Co $e_g$ magnetic moments and 
the electronic spins localized at the CN axial ligand
depending on the filling of the impurity bound states.
However, the magnetic correlations are not contained
in the DFT results,
as it is well known.

\subsection{Comparison of the HF+QMC data with the photoabsorption spectrum of CNCbl} 

The photoabsorption spectrum of CNCbl is characterized by several
distinctive peaks\cite{Firth}. 
The lowest-energy peak in the spectrum occurs at
$\approx 2.25 \eV$ and it is associated with the so-called $S_1$ 
electronically excited state of CNCbl. 
The largest peak, on the other hand, is located at $\approx 3.4 \eV$.
There have been various DFT based studies 
\cite{Andruinov,Solheim,Reiq,Kornobis2011,Kornobis2013}
to identify the origin of the peaks in the spectra of 
the Cbl cofactors.

For a long time, the lowest-energy part of the photoabsorption 
spectrum of Cbl had been associated with a transition which is dominated
by the $\pi \rightarrow \pi^*$ excitations of the corrin ring \cite{Pratt},
which represent electronic transitions among 
the $2p\pi$ orbitals in the corrin ring. 
The reason for this assignment is that the metal-free corrinoids 
also exhibit a similar peak in the spectrum.
The time-dependent DFT calculations with the BP86 functionals 
on the truncated molecule Im-[Co$^{\rm III}$(corrin)]-CN$^+$
have supported the view that the $S_1$ state corresponds
mainly to $\pi \rightarrow \pi^*$ transitions 
in the corrin ring \cite{Kornobis2011}.
On the other hand, 
the ultrafast transient absorption spectroscopy 
measurements on CNCbl \cite{Shiang,Harris} have suggested a picture for the $S_1$ 
transition in which charge transfer from the $\pi$ orbitals of the corrin ring 
to the Co atom takes place.
In spite of these developments, 
today it is considered that much still remains to 
be understood about the nature of the excited states of Cbl cofactors. 
In particular, there is much interest to clarify the origin of the lowest 
excited $S_1$ state.

In order to compare the photoabsorption spectrum of CNCbl with the 
HF+QMC data on the electronic structure of Im-[Co$^{\rm III}$(corrin)]-CN$^+$,
we assume that the rigid band approximation can be used for 
describing the single-particle spectral weight 
distribution of the Haldane-Anderson model. 
According to the HF+QMC results shown in Fig. 7 for $U=36 \eV$,
when the chemical potential $\mu\approx -8.5 \eV$, 
the truncated system Im-[Co$^{\rm III}$(corrin)]-CN$^+$
has 238 electrons. 
In this case, 
the highest occupied states consist of the 
Co $t_{2g}$-like states which are 
the $3d_{xz}$, $3d_{x^2-y^2}$ and $3d_{yz}$
NAO's in our coordinate system.
Smaller amount of weight from the $2p\pi$ orbitals
of the corrin ring is also included.
In the dominant lowest-energy photoabsorption process,
an electron initially occupying a Co $t_{2g}$-like state 
would be transferred to the lowest excited state located at 
$\approx -5.5 \eV$.
We have seen in Section III.C that this state at $\approx -5.5 \eV$ 
corresponds to an impurity bound state and
consists mainly of the $2p\pi$ NAO's at the CN axial ligand. 
The impurity bound state
contains lesser amounts of spectral weight 
from the $2p\pi$ NAO's of the corrin ring as well as 
from the Co($3d_{xy}$) NAO.
Hence, according to HF+QMC,
the lowest excited state in the absorption spectrum corresponds 
mainly to a transition involving the transfer of an electron 
from the Co $t_{2g}$-like states
to the impurity bound state, 
which consists of
states at the CN axial ligand, 
the corrin ring and a Co $e_g$ NAO.

If the electron is excited to the second lowest energy of
Im-[Co$^{\rm III}$(corrin)]-CN$^+$,
then the in-gap states located at $\approx -4 \eV$ 
would become occupied.
This impurity bound state consists mainly of the $2p\sigma$ and $2s$ 
NAO's of the CN axial ligand.
Smaller amounts of spectral weight from the $2p\sigma$ orbitals of the 
corrin ring and of the Co($3d_{3z^2-r^2}$) NAO are also included.
Hence, 
this electronic transition would be dominated by the transfer of an electron from 
the Co $t_{2g}$-like states to 
the second impurity bound state.
The transitions to the $\approx -5.5 \eV$ and 
$\approx -4 \eV$ impurity bound states would be the main features
of the photoabsorption spectrum expected for 
Im-[Co$^{\rm III}$(corrin)]-CN$^+$
according to the HF+QMC data
for the Haldane-Anderson model.

In summary, 
according to the HF+QMC calculations 
the lowest-energy excitations are dominated by electron transfer 
from the Co $t_{2g}$-like states to the CN axial ligand. 
Now, the HF+QMC approach
might be too simple to correctly deduce the excitation spectrum
of vitamin B$_{12}$.
Nevertheless, the HF+QMC results presented here suggest that it would be 
useful to look for signatures of a possible 
$3d\rightarrow {\rm CN}$ ligand electron transfer process
in the excitation spectrum of CNCbl. 
According to the HF+QMC calculations, such $3d\rightarrow {\rm CN}$ ligand 
transitions would be accompanied by the vanishing of the antiferromagnetic correlations 
between the Co $e_g$ and the CN ligand NAO's 
along with the collapse of the magnetic moments at the CN ligand.
In addition, 
when an electron is removed from $t_{2g}$-like orbitals,
a magnetic moment would develop due to the lifting of the double occupancy 
in these orbitals. 
Hence,
the $3d\rightarrow$CN ligand electron transfer would be accompanied 
by the magnetic moment transfer 
in the reverse direction 
from the CN ligand to the Co $t_{2g}$ orbitals.
It would be useful to probe experimentally these magnetic correlations
predicted by HF+QMC.
In particular,
it would be interesting to see 
how the many-body effects discussed here would influence 
the magnetic circular dichroism spectrum of CNCbl \cite{Stich}.

\subsection{DFT+QMC approach} 

An alternative method for studying the electronic structure 
and correlations of organometallic molecules such as Cbl is 
the DFT+QMC approach. 
There are various differences between the HF+QMC and 
DFT+QMC approaches. 
For example,
in HF+QMC the correlations of the $s$ and $p$ orbitals
in the molecules are treated at the HF level rather than 
the more accurate DFT. 
In addition, 
in HF+QMC a bare unscreened Coulomb interaction is used in the QMC part. 
The bare Coulomb repulsion has a value of 
order 36 eV when evaluated in the basis of the atomic orbitals
with the 6-31G Gaussian functions.
In the DFT+QMC, 
the QMC part of the calculations
are performed by using a renormalized Coulomb interaction,
which is usually of order 4 eV.
This renormalization is due to screening by the $s$ and $p$ electrons 
in the system,
and the renormalized Coulomb interaction
can be obtained by the constrained local-density approximation
\cite{Dederichs,McMahan,Gunnarsson}. 
It is important to note that 
in both approaches
the choices used for the Coulomb interaction and the double-counting 
$\mu_{DC}$ represent different levels of approximations \cite{Held}.
Hence, it would be useful to compare them.
 
A transition-metal atom placed in an organic molecule, as in Cbl,
is one of the simplest strongly-interacting systems.
With the availability of spectroscopic data,
this is a case where the combined electronic-structure 
and many-body approaches can be tested. 
In this respect,
a thorough comparison of the HF+QMC and DFT+QMC approaches would be useful. 
Because of manuscript length considerations, 
this will be presented elsewhere \cite{Mayda2}. 

\subsection{Inter-orbital Coulomb interaction and the Hund's coupling}

In this paper, 
we have presented HF+QMC results,
where we used
a constant $U$ for the intra-orbital Coulomb interaction 
in the QMC part.
It is well known that the inter-orbital Coulomb interaction
is important in the cobaltates \cite{Maekawa}.
Here, 
the inter-orbital Coulomb interactions are taken into 
account only at the Hartree-Fock level. 
A more accurate approach would be to include the inter-orbital
Coulomb interaction in the QMC calculations.
This can be done by replacing the last term in Eq. (1) by
\begin{eqnarray}
\dfrac{1}{2} \sum_{\nu,\nu'} \sum_{\sigma}
& & \,
U_{\nu\nu'}
n_{\nu\sigma} n_{\nu,-\sigma} \\
& &+  
\dfrac{1}{2} \sum_{\nu\neq\nu'} \sum_{\sigma} \,
( U_{\nu\nu'} - J_{\nu\nu'} ) \,
n_{\nu\sigma} n_{\nu'\sigma}, \nonumber
\label{Hund's}
\end{eqnarray}
where $U_{\nu\nu'}$ and $J_{\nu\nu'}$
are the Coulomb matrix elements for the direct and the Hund's exchange couplings.
For consistency,
it is necessary to evaluate the Coulomb matrix elements 
in the NAO basis, 
since the one-electron parameters of the Anderson Hamiltonian
are obtained by using the NAO's.

The interaction, Eq.~(22), includes only density-density terms, 
hence it can be implemented in the Hirsch-Fye QMC algorithm.
This improvement would still neglect the pair-hopping and the spin-flip terms 
of the Hund's coupling,
which have been shown to be important in transition-metal oxides
\cite{Sakai,Belozerov}.
The double-counting correction $\mu_{\rm DC}$
would also need to be re-evaluated after including 
the inter-orbital Coulomb interactions 
\cite{Anisimov,Czyzyk,Kunes}.
 
We think that it is necessary to include the inter-orbital Coulomb interaction
before performing quantitative comparisons with the experimental data on the 
Cbl cofactors.
The Hund's coupling may influence the magnetic correlations and 
cause additional splitting of the energy levels. 
After these improvements,
the spin state of Co can be determined.
In addition,
the uniform magnetic susceptibility for electronic spins can be calculated 
to extract the effective magnetic moment
for the molecule. 

\section{Summary and conclusions}

In summary, 
we have studied the electronic structure and correlations 
of CNCbl from the perspective of the many-body physics.
For this purpose, 
we have used the framework of the multi-orbital single-impurity
Haldane-Anderson model of a transition-metal impurity placed in a 
semiconductor host.
First, 
we have constructed an effective Haldane-Anderson model by using the HF
approximation.
In particular, 
we determined the one-electron parameters of the Anderson Hamiltonian 
from the Fock matrix written in the basis of the natural atomic 
orbitals by using HF.
The double-counting of the Coulomb interaction within HF+QMC 
was taken into account by a chemical shift $\mu_{\rm DC}$ 
of the Co($3d_{\nu}$) levels. 
We used an orbital independent intra-orbital Coulomb interaction $U$
in the QMC calculations neglecting the inter-orbital terms.
We presented QMC results for various values $U$. 
However, 
we concentrated on the case $U=36 \eV$, 
since this yields values for the HOMO-LUMO gap and the total 
Co($3d$) electron number which are comparable to the experimental data. 

The QMC data on this effective model
showed how the energy gap found in the HF approximation
is reduced by the generation of new in-gap states. 
The correlated nature of the induced states is clearly seen.
In particular, 
states arising from the double-occupancy of the Co $t_{2g}$ orbitals are
induced near the gap edge above the HOMO level from HF, 
whereas impurity bound states 
are induced below the LUMO level. 
The impurity bound states arise from the strong hybridization of the Co $e_g$ orbitals 
with the host eigenstates
and contain significant amount of weight from the CN axial ligand and the corrin ring. 
We have also seen that magnetic moments can develop at the 
CN axial ligand and in the corrin ring, 
which are antiferromagnetically correlated with the 
Co $e_g$ moments. 
These disappear rapidly with electron filling of the impurity 
bound states.

We have also presented a comparison of the HF+QMC data with the DFT calculations on
Im-[Co$^{\rm III}$(corrin)]-CN$^+$.
While these two approaches yield similar values for the HOMO-LUMO gap
and the total Co($3d$) occupation number,
the results on the distribution of the single-particle spectral weights
can be very different. 
In particular,
the Hubbard-type splitting of the Co($3d_{\nu}$) states,
the impurity bound states and the magnetic correlations are not 
contained in the DFT results.
We have also presented the predictions of HF+QMC for the 
photoabsorption spectrum according to which
the lowest-energy photoabsorption excitations are dominated 
by electron transfer from the Co $t_{2g}$ orbitals to the 
impurity bound states. 

It will be interesting to perform similar calculations for AdoCbl and MeCbl
and compare their results with the results presented here on CNCbl.
In particular,
it would be useful to see whether it is possible to 
understand the catalytic functioning of these Cbl cofactors 
by using the HF+QMC calculations.
It would also be useful to use the same framework to study 
hemoglobin containing Fe. 

It is worth noting that  
the Co-C bond in Cbl is the first case of a metal-carbon bond 
to be found in biology. 
All of the known reactions of the Cbl-dependent 
enzymes involve the making and breaking of the Co-C bond
\cite{Krautler}.
In this respect,
it is interesting that the impurity bound state 
contains significant amount of spectral weight 
from the CN axial ligand attached to Co.
It would also be interesting to see whether 
the impurity bound state plays a role in the important Co-C bonding. 
If the making and breaking of the Co-C bond somehow involves 
the magnetic correlations around the Co atom, 
then this could be a way such that the occupancy 
of the impurity bound state becomes important.

The present HF+QMC approach contains various approximations.
For instance, 
the inter-orbital Coulomb interaction is not treated by QMC,
instead it is included at the Hartree-Fock level.
In addition,
the bare Coulomb repulsion $U$ and the double-counting
$\mu_{\rm DC}$ are used more as variables than as
true {\it ab initio} parameters. 
The geometrical parameters for the molecular structure 
are simple estimates since the actual atomic coordinates
for CNCbl are not known. 
In spite of these, 
the HF+QMC yields reasonable values for the HOMO-LUMO gap 
and the total Co($3d$) occupation number,
and interesting results for the photoabsorption spectrum. 
In particular, 
the HF+QMC results emphasize the many-body effects
such as the Hubbard splitting of the Co($3d_{\nu}$) states
and the possible existence of impurity bound states and 
magnetic correlations.

In conclusion,
according to the HF+QMC results presented here, 
there is a possibility that an impurity bound state exists 
in the electronic spectrum of CNCbl.
We think that it is necessary to seek
further experimental evidence for 
this prediction of the HF+QMC calculations.

\begin{acknowledgements}

We thank 
Tahir \c{C}a\u{g}{\i}n, Mehmet Sar{\i}kaya, Nuran Elmac{\i}, \"Ozg\"ur \c{C}ak{\i}r,
Devrim G\"u\c{c}l\"u, 
Jingyu Gan, Bo Gu, and Sadamichi Maekawa 
for valuable discussions and suggestions. 
The numerical calculations reported here were performed 
in part at the TUBITAK ULAKBIM, High Performance and Grid
Computing Center (TRUBA resources).
Financial support by the Turkish Scientific and Technical Research Council 
(TUBITAK grant numbers 110T387 and 113F242) is gratefully acknowledged. 

\end{acknowledgements}

All authors contributed equally to this paper. 

\bibliographystyle{apsrev4-1}


\end{document}